\documentclass[aip, jmp, amsmath,amssymb,reprint]{revtex4-2}

\usepackage{graphicx}
\usepackage{dcolumn}
\usepackage{bm}

\usepackage{amsmath,amssymb}
\usepackage{braket}
\usepackage{mathtools}
\usepackage{hyperref}
\usepackage{multirow}
\usepackage{color}
\usepackage[normalem]{ulem}
\usepackage{amsfonts}
\usepackage{float}

\newcommand{\dd}{\mathrm{d}}
\newcommand{\ii}{\mathrm{i}}

\def\bra#1{\mathinner{\langle{#1}|}}
\def\ket#1{\mathinner{|{#1}\rangle}}

\def\beq{\begin{equation}}
\def\eeq{\end{equation}}
\def\bea{\begin{eqnarray}}
\def\eea{\end{eqnarray}}

\begin{document}

\title{Superdiffusion in spin chains}


\author{Vir B. Bulchandani}
\affiliation{Princeton Center for Theoretical Science, Princeton University, Princeton NJ 08544, USA}

\author{Sarang Gopalakrishnan}
\affiliation{Department of Physics, Pennsylvania State University, University Park PA 16802, USA}
\author{Enej Ilievski}
\affiliation{Faculty of Mathematics and Physics, University of Ljubljana, Jadranska 19, 1000
Ljubljana, Slovenia}

\date{\today}

\begin{abstract}

This review summarizes recent advances in our understanding of anomalous transport in spin chains, viewed through the lens of integrability. Numerical advances, based on tensor-network methods, have shown that transport in many canonical integrable spin chains---most famously the Heisenberg model---is anomalous. Concurrently, the framework of generalized hydrodynamics has been extended to explain some of the mechanisms underlying anomalous transport.
We present what is currently understood about these mechanisms, and discuss how they resemble (and differ from) the mechanisms for anomalous transport in other contexts.
We also briefly review potential transport anomalies in systems where integrability is an emergent or approximate property.
We survey instances of anomalous transport and dynamics that remain to be understood.
%

%
\end{abstract}

\maketitle
\tableofcontents

\section{Introduction}

Many of the canonical models of condensed matter physics in one dimension are exactly or approximately \emph{integrable}, in the sense that their eigenfunctions can be written down exactly using the Bethe ansatz~\cite{Korepin_book, takahashi_book}. Integrable systems have infinitely many local conserved densities~\cite{prosen_drude, PhysRevLett.115.157201, 1742-5468-2014-9-P09037, QLreview}, and therefore strongly violate our expectations from conventional thermodynamics and hydrodynamics. In particular, they have well-defined, ballistically propagating quasiparticles, even in lattice models where total momentum is not a good quantum number~\cite{prosen_drude}. The existence of stable ballistically propagating quasiparticles might suggest that all quantities in such systems are transported ballistically (as opposed to nonintegrable systems, which exhibit diffusive transport), since one would expect the quasiparticles to carry charge. Surprisingly, however, this is not always the case: in one of the most studied integrable models, the anisotropic Heisenberg (or XXZ) model, spin transport in the absence of an external field can be ballistic, diffusive, or superdiffusive, depending on the parameters of the model~\cite{SD97, Zotos99, MarkoKPZ, Ljubotina_nature}. 

Understanding the origin of this rich transport phenomenology in models where the underlying degrees of freedom apparently behave so simply (ballistic motion plus forward scattering) has been an important challenge since the phenomenology was first numerically discovered. Although enormous progress has been made numerically, and more recently even experimentally~\cite{SchemmerExpt,malvania2020generalized,scheie2020detection}, exact calculations have proved challenging: even so simple a quantity as the linear-response a.c. conductivity relies on the matrix elements of local operators between the eigenstates of integrable systems, and the asymptotics of these matrix elements (``form factors'') is only understood in some simple cases~\cite{Korepin_book, de2015density, gohmann2017thermal, granet2020systematic, cubero2020generalized}.
Nevertheless, the past few years have seen enormous theoretical progress as well, spurred by the advent of \emph{generalized hydrodynamics} (GHD)~\cite{Doyon, Fagotti, DS17, Bulchandani17, PhysRevLett.120.045301, Bulchandani18, Doyon_notes}. 
GHD is believed to be an asymptotically exact theory of the long-wavelength dynamics of integrable systems: it treats each quasiparticle as a semiclassical object propagating (with nontrivial interactions) through a dense medium of other quasiparticles. This physical picture and the associated concrete computational approach have motivated many of the developments we will discuss below.
 
This review article summarizes the past few years of theoretical progress on anomalous transport in \emph{clean} spin chains. (Strongly disordered quantum spin chains also host regimes of anomalous transport, but the mechanisms involved there are different; for recent reviews, see, e.g., Refs.~\onlinecite{agarwal2017rare, gopalakrishnan2020dynamics}.) 
The numerical evidence for anomalous transport in the Heisenberg spin chain is at least a decade old; however, in the past three years, the nature of this phenomenon has begun to come into focus. Although important parts of the picture remain indistinct, it is now understood, e.g., which degrees of freedom are responsible for anomalous transport. 
We aim to lay out, as simply as possible, this emerging picture of the dynamics of integrable spin chains. 
Integrability---whether exact or approximate---is central to our analysis, because it guarantees the stability of quasiparticles in states of finite energy density. 
However, the specifics of Bethe ansatz and GHD calculations are not needed to understand the key results, which follow instead from much more general considerations.  

Correspondingly, the scope of our review is somewhat restricted. The topics we do not cover here are, however, amply addressed in other review articles. The framework of generalized hydrodynamics is laid out in Ref.~\onlinecite{Doyon_notes}. A much broader overview of transport in one-dimensional physical systems, including a thorough account of transport theory, integrable spin chains and experimental and numerical advances and results is presented in  Ref.~\onlinecite{Bertini_transport_review}.
%
Finally, a companion review~\footnote{B. Doyon et al., to appear in the same volume.} summarizes the current understanding of diffusion in integrable systems. We introduce the elements of all of these concepts and results that we will need to fix notation and present our results in a self-contained way. 


This review is organized as follows. We close this introductory section with a historical overview of numerical and analytical results concerning finite-temperature transport in integrable spin chains. In Sec.~\ref{background} we briefly review the background concepts---on integrable transport, conventional and generalized hydrodynamics, and quasiparticle diffusion in integrable systems---that we will assume in subsequent sections. This summary is meant to be self-contained, but the topics discussed there are addressed in more depth elsewhere in the literature, including in companion reviews\footnote{To appear in the same volume.}. In Sec.~\ref{anisoXXZ} we will encounter the simplest examples of ``anomalous'' (or at least non-ballistic) spin transport, in the context of the anisotropic XXZ model. The anisotropic XXZ model illustrates, in a simpler setting, many of the subtleties that are present in the canonical example of anomalous diffusion, viz. the isotropic Heisenberg model, which we cover in Sec.~\ref{isoXXZ}.  Our discussion of the Heisenberg model suggests that the key ingredient leading to anomalous diffusion in integrable spin chains is the presence of a global nonabelian Lie-group symmetry. This observation is solidified in Sec.~\ref{SUN}, where we assemble numerical and analytical evidence that both this anomalous diffusion phenomenon and its associated $z=3/2$ dynamical exponent are ``superuniversal'', in the sense that they occur in \emph{all} integrable spin chains with short-range interactions that possess global nonabelian Lie-group symmetries. In Sec.~\ref{domainwalls} we extend our considerations from linear response about thermal equilibrium to the dynamics of more general nonequilibrium initial states.

Finally, we turn our attention from integrable spin chains to \emph{chaotic} spin chains. (Here, ``chaotic'' means that the model in question exhibits the properties expected of a generic, thermalizing system, namely Wigner-Dyson level statistics and normal transport at non-zero temperature and asymptotically long times\cite{d2016quantum}.) In Sec. \ref{lowT}, we present some generic classes of chaotic spin chains in which robust signatures of anomalous transport arise at low temperature, due to the emergence of an integrable effective field theory at zero temperature. In Sec. \ref{undular}, we show how the interplay of Goldstone physics and diffusion in chaotic nonlinear sigma models can lead to complex effective diffusion constants, even at infinite temperature. We then conclude with a summary of the key open questions that remain to be settled in Sec.~\ref{conclusion}.

\subsection{Historical overview}

Since the bulk of this review will concern the spin-$1/2$ XXZ spin chain, it is useful to introduce its Hamiltonian at the outset:
\begin{equation}\label{xxzham}
H = J \sum_i (S^x_i S^x_{i+1} + S^y_i S^y_{i+1} + \Delta S^z_i S^z_{i+1}).
\end{equation}
In the rest of this work, we will usually set $J = 1$ for convenience. However, most of our discussion concerns physics at finite temperature, where there is no sharp distinction between the antiferromagnet and the ferromagnet (i.e., effectively $\beta \equiv (k_B T)^{-1}$ can take either sign). For some purposes below it will be more helpful to imagine a ferromagnet at low temperatures; we will be explicit when we are assuming this. Many of the results discussed here concern the high-temperature limit of transport, or transport at infinite temperature and finite chemical potential. These limits should be understood as follows: at infinite temperature, all transport coefficients vanish; however, in the limit of small $\beta$, transport coefficients like the conductivity are proportional to $\beta$. We are interested in this proportionality constant, i.e., in $\lim_{\beta \to 0} \sigma/\beta$, where $\sigma$ is a transport coefficient such as the conductivity. Also, when we work at infinite temperature with a finite net magnetization density. This involves computing transport coefficients in the density matrix $\rho \propto \exp(- \beta H - h S^z)$, in the $\beta \to 0$ limit. (Note that $h = \beta \mu$ where $\mu$ is the conventionally defined chemical potential.) Without loss of generality we specialize to $\Delta \geq 0$, as this parameter regime contains all the physics of interest.

At zero temperature this model has two phases: an easy-plane phase $\Delta < 1$ where the spectrum above the ground state is gapless, and an easy-axis phase $\Delta > 1$ with a gapped spectrum. The ground state in the easy-axis phase breaks the Ising symmetry; unlike the transverse-field Ising model, however, its dynamics is constrained by the $U(1)$ conservation law. Therefore, e.g., domain walls in the ferromagnet cannot move freely. The isotropic point $\Delta = 1$ is a quantum critical point separating these two phases. Depending on whether the couplings are ferromagnetic or antiferromagnetic, this critical point has dynamical critical exponent $z_0 = 1$ (antiferromagnet) or $z_0 = 2$ (ferromagnet)~\cite{Sachdev_book}. (To avoid confusion, throughout this review, we will use $z_0$ to denote the zero-temperature dynamical critical exponent and $z$ to denote the dynamical exponent that governs finite-temperature transport. We remind the reader that a dynamical critical exponent $z$ corresponds to space-time scaling of the form $t\sim x^z$.)

\subsubsection{Early history}

A vast literature exists on \emph{zero-temperature} transport in integrable systems; this is not directly relevant to our considerations, and we will not discuss it further. The general picture at zero temperature is the same for integrable and nonintegrable systems: transport is due to elementary excitations that (in clean lattice systems) propagate ballistically. That spin transport at \emph{nonzero} temperatures might exhibit richer behavior was first pointed out in the late 1990s and early 2000s~\cite{CZP95,Zotos1996,Zotos1997,SD97, DS98, Zotos99, narozhny1998transport, gros2002, saito2003strong, heidrich2003zero}. Two of these early works are particularly relevant to our considerations. Sachdev and Damle~\cite{SD97, DS98} explained the presence of normal diffusion in the easy-axis XXZ antiferromagnet at finite temperature \emph{despite} integrability using a semiclassical quasiparticle picture that is reminiscent of the GHD framework (see also Ref.~\onlinecite{sachdev1997low}). Meanwhile, various authors~\cite{Zotos1996,Zotos99, narozhny1998transport} found ballistic transport in the easy-plane regime, using numerical techniques as well as methods based on the thermodynamic Bethe ansatz, such as the Kohn formula (for a recent overview see Ref.~\onlinecite{KlumperDrude}). 

Taken together, these findings strongly suggested the existence of a finite-temperature ``phase transition'' in spin transport. The isotropic Heisenberg point was the natural critical point for this putative phase transition. It took further numerical advances, particularly the development of matrix-product methods for boundary-driven quantum spin chains, to clearly establish both the distinct transport behaviors on the easy-axis and easy-plane sides, and to find anomalous diffusion with the space-time scaling $x \sim t^{2/3}$ (i.e., $z = 3/2$) at the isotropic point~\cite{MarkoKPZ}. 

\subsubsection{Recent developments: superdiffusion at the isotropic point}

Superdiffusion of spin in the Heisenberg chain was discovered in Ref. ~\onlinecite{MarkoKPZ}, which obtained the steady state of an open XXZ chain coupled to a small magnetization gradient via magnetization or thermal~\cite{ZnidaricJSTAT11} Lindblad baths at its endpoints. The resulting steady state carries a linear-response spin current $j_S$ that depends on system size $L$ as $j_S \sim L^{-1/2}$. In the continuity equation, this implies the space-time scaling law $t \sim x^{3/2}$, or a dynamical exponent $z=3/2$. Subsequent tDMRG studies showed that this $z=3/2$ dynamical exponent could be probed by considering time evolution from ``weak domain wall'' initial conditions, of the form $\rho \propto (1+h S_z)^{L/2} \otimes (1-h S_z)^{L/2}$ with $h \ll 1$~\cite{Ljubotina_nature, Ljubotina19} (see also Ref.~\onlinecite{PhysRevB.89.075139}). Such states can be viewed as magnetic domain walls at very high temperature, and in the thermodynamic limit they simulate the dynamics of infinite-temperature spin autocorrelation functions through the identity~\cite{Ljubotina19}
\begin{equation}
\langle S_0^z(0) S_i^z(t)\rangle_{\beta=0} \propto \lim_{h \to 0} \frac{\langle S_{i-1}^z(t) \rangle_{h} - \langle S_i^z(t)\rangle_{h} }{h}.
\end{equation}
This result was used to fit numerically obtained scaling functions for $\langle S_0^z(0) S_i^z(t)\rangle_{\beta=0}$ against universal Kardar-Parisi-Zhang (KPZ) scaling functions\footnote{Here and throughout this review, we will refer to the Pr{\"a}hofer-Spohn scaling function~\cite{Prahofer2004}, obtained for the polynuclear growth model, as the Kardar-Parisi-Zhang scaling function. Thus, following Pr{\"a}hofer and Spohn, we assume that this scaling function is universal.}, leading to the conjecture that infinite-temperature spin dynamics in the spin-$1/2$ Heisenberg chain lies in the Kardar-Parisi-Zhang universality class~\cite{Ljubotina19}.
An important breakthrough in these numerical studies was the discovery of \emph{integrable Trotterizations} of the XXZ model~\cite{VZP18}, which allow one to simulate its dynamics for longer periods without worrying about errors in the Trotter decomposition.

These detailed numerical studies of the spin-$1/2$ Heisenberg chain have been complemented by numerical investigations of a plethora of other classical and quantum spin chains, both integrable and non-integrable and with various internal Lie group symmetries~\cite{MarkoKPZ,Bojan,Das19,DupontMoore19, PhysRevE.100.042116,Weiner2019,KP20,MatrixModels,Fava20,ssd}. From this body of work, an intriguing picture of ``superuniversal'' $z=3/2$ transport in isotropic spin chains has emerged, whose main empirical features are as follows:

\begin{enumerate}
    \item Both classical and quantum spin chains can exhibit anomalous, $z=3/2$ spin transport at half-filling.
    \item Integrability is a necessary condition for $z=3/2$ dynamical scaling of the spin density to persist to long times.
    \item Nonabelian Lie group symmetry is a necessary condition for $z=3/2$ spin transport to arise at all. This phenomenon is ``superuniversal'' in the sense that it seems to arise from global symmetry with respect to \emph{any} compact nonabelian Lie group $G$ and irreducible representation thereof.
    \item Spin chains with robust $z=3/2$ spin transport seem to exhibit scaling collapse of their spin autocorrelation functions to Kardar-Parisi-Zhang scaling functions at long times. The numerical evidence is most convincing for classical spin chains with pure $SU(2)$ symmetry and less convincing for quantum spin chains, for which access to asymptotically long times is limited.
\end{enumerate}

We now summarize the main developments in the theoretical understanding of $z=3/2$ transport in spin chains. Strikingly, the first attempts at a theoretical explanation did not appear until some seven years after the numerical result was reported, in part owing to the lack of a theory of generalized hydrodynamics in integrable systems. The initial breakthrough was a demonstration that generalized hydrodynamics predicts a divergent spin diffusion constant for the half-filled Heisenberg and Hubbard chains~\cite{Ilievski18} (building on a rigorous lower bound on diffusion constants~\cite{PhysRevLett.119.080602}). An analysis of the finite-size scaling of this divergence in terms of the microscopic kinematics of quasiparticles subsequently revealed that the unique self-consistent dynamical exponent for spin transport in the half-filled Heisenberg chain is $z=3/2$~\cite{GV19}. While these studies allowed for a fairly detailed understanding of the microscopic kinetics giving rise to superdiffusion in the Heisenberg chain (see also Refs. ~\onlinecite{NMKI19, GVW19}), they did not explain the apparent universality of $z=3/2$ dynamics in spin chains, as observed in numerical simulations~\cite{DupontMoore19}. An alternative approach based on studying dynamical fluctuations of the Bethe pseudovacuum was proposed in Ref. ~\onlinecite{Vir20}. This provided a macroscopic argument for the emergence of KPZ universality (and thence a $z=3/2$ exponent) in the Heisenberg chain, which was subsequently adapted to a variety of other classical and quantum spin chains~\cite{NMKI20}, including the Hubbard model~\cite{Fava20}. More recently, the ``microscopic'' and ``macroscopic'' descriptions of $z=3/2$ transport in spin chains have been unified through the fundamental observation that mean-field vacuum dynamics, which underpins the theoretical discussion of Ref.~\onlinecite{Vir20}, emerges from the scattering phase-shifts of infinitely large quasiparticles in the quantum Heisenberg chain~\cite{NGIV20}, and more generally of large solitons in the Goldstone sector of spin chains with global continuous nonabelian symmetries~\cite{ssd}.

Finally, we mention that the theoretical prediction of $z=3/2$ transport in the Heisenberg chain was recently experimentally verified for the first time~\cite{scheie2020detection} by scattering neutrons off a single crystal of the near-ideal Heisenberg spin chain material $\mathrm{KCuF}_3$. Further evidence for superdiffusion, as well as KPZ universality in the growth of transport fluctuations, was recently found in ultracold atomic systems using quantum gas microscopy~\cite{zeiher2021}.

\subsubsection{Drude weight and corrections in the easy-plane phase}

The study of finite-temperature transport in the spin-$1/2$ Heisenberg chain has a rather long history. The study of transport properties and Drude weights in integrable models was initiated in Refs.~\onlinecite{CZP95,Zotos1996,Zotos1997,Naef1998}, which sparked further interest in this area. A striking early observation~\cite{Zotos1997} was that energy transport in many interacting integrable models, including the Heisenberg chain, is purely ballistic, implying that Fourier's law of heat conduction does not hold in such models. The problem of characterizing spin transport proved to be more subtle. The finite-temperature, easy-plane spin Drude weight was first evaluated analytically from Bethe-ansatz techniques in Ref.~\onlinecite{Zotos99}, based on earlier related papers \cite{FujimotoKawakami98,Fujimoto99}. This result was afterwards successfully reobtained in Ref.~\onlinecite{Benz05} (the same study however also reported a failed attempt of reproducing the result within an alternative spinon-basis computation). One enduring puzzle, that remained unresolved for some time, was how the spin Drude weight could possibly assume a non-zero value in the apparent absence of local conservation laws with odd parity under spin reversal (other than the magnetization density itself), that are in principle necessary to protect the spin current from dissipating away~\cite{Zotos1997,Sirker2009,Sirker2011}.

This was partially resolved in Ref.~\onlinecite{prosen_drude}, which identified a suitable, hitherto unknown, \emph{quasilocal} conservation law of the easy-plane Heisenberg chain (see Ref.~\onlinecite{QLreview} for a more detailed discussion).
The associated quasilocal charges were used to construct a lower bound on the Drude weight (known as a Mazur bound, see Sec.~\ref{drudesec} below); this bound nonetheless appeared to be loose, in the sense that it omitted spectral weight compared to the TBA result~\cite{Zotos99}. Moreover, the origin of this novel conservation law remained elusive.
A striking prediction of these works was that the conjectured lower bound had a discontinuous fractal dependence on the anisotropy (which we will return to in Sec.~\ref{anisoXXZ}). Soon after the advent of GHD, an explicit calculation in the GHD framework suggested that this lower bound was \emph{exact}, so the true Drude weight is indeed discontinuous~\cite{IN17}---this conclusion was since also reached by other means~\cite{KlumperDrude}.

Since various theoretical approaches have now converged on the conclusion of a discontinuous Drude weight, a natural question to ask is how the spectral weight gets redistributed between the Drude peak and the low-frequency regular part of the conductivity as one continuously varies $\Delta$. The first major progress in addressing this question was the demonstration~\cite{Ilievski18} that the d.c. limit of the conductivity is infinite for irrational values of the anisotropy. The nature of this divergence was recently addressed in Ref.~\onlinecite{Agrawal20} using GHD (see also Refs.~\onlinecite{PhysRevB.97.081111, PhysRevB.102.180409}): it was found that $\sigma(\omega) \sim 1/\sqrt{\omega}$ at irrational values of the anisotropy, and the crossover scales between rational and irrational values were identified. This nontrivial $1/\sqrt{\omega}$ scaling is consistent with numerical results~\cite{LZP19} on the approach of the finite-time current-current correlator to its long-time limit (which sets the Drude weight). 

\subsubsection{Spin helices and out-of-equilibrium dynamics}

The revival of interest in integrable dynamics was driven by developments in ultracold atomic gases, where it is natural to study the real-time dynamics of isolated quantum systems far from equilibrium~\cite{bloch_review, d2016quantum}. However, the most natural experiments in ultracold settings are somewhat different from equilibrium linear-response transport (note, though, that equilibrium spin transport has been measured~\cite{nichols2019spin}, as has charge transport for interacting fermions~\cite{PhysRevX.8.011053}). 
Thermal equilibrium states are hard to prepare reliably in cold-atom experiments (because equilibration is slow and thermometry is hard), whereas far-from-equilibrium product states are relatively simple. Thus, one series of influential experiments has studied the relaxation of an initial ``spin helix''~\cite{hild2014far, brown2015two, jepsen2020spin}. In Ref.~\onlinecite{jepsen2020spin} a helix is created as follows: one realizes the spin model by trapping two distinct spin states of ultracold atoms. One initializes all the atoms in the same spin state, then applies a radio-frequency pulse to rotate them into the equator of the Bloch sphere. A magnetic field gradient then causes each atom to precess at a position-dependent rate, so after some wait time the spins form a helically modulated state. The contrast of the helix is measured as a function of time, e.g., by in-situ imaging. 

Since these initial states are not thermal, the theoretical framework developed in the bulk of this work does not directly apply. The linear-response diffusion constant should describe the relaxation of a weak spin modulation of wave-vector $q$ created on top of a thermal state; we expect that the contrast of such a modulation at later times should decay as $\rho_q(t) \propto \exp(-D(q) q^2 t)$. By contrast, the cold-atom experiments create a large-amplitude modulation on top of a vacuum state. 
One might naively have expected the late-time decay to be the same in both types of experiments, since (intuitively) when the initial helix has mostly relaxed the system can be regarded as a weakly modulated thermal state.
Experimental and numerical evidence suggests that this intuition is incorrect: while ballistic transport persists in the easy-plane case, it seems that isotropic systems exhibit diffusion while easy-axis systems exhibit subdiffusive spin transport. 
At present there is no detailed understanding of these results within the GHD framework or any other framework, although various approximate treatments exist, using either nonequilibrium field-theoretic methods~\cite{PhysRevX.5.041005} or short-time series expansions~\cite{jepsen2020spin}.
However, a simpler version of this setup---consisting of a domain wall between two regions of opposite spin polarization---has been studied theoretically, and we return to this in Sec.~\ref{domainwalls}.

\section{Background}\label{background}

\subsection{Hydrodynamics: a reminder}
\label{reminder}
Constructing a microscopic theory of transport in strongly interacting systems at finite temperature is, in general, an intractable task. However, the main qualitative aspects of transport in this regime are captured by conventional hydrodynamics. The framework of hydrodynamics posits a separation of timescales between fast degrees of freedom (i.e., any variable that can relax locally) and slow degrees of freedom, which correspond to long-wavelength fluctuations of conserved quantities, Goldstone modes associated with broken continuous symmetries, and other similar fluctuations that are constrained to be long-lived. Assuming such a separation of scales, one can carve a system up into mesoscale hydrodynamic ``cells,'' which are large compared with the microscopic scales that govern ``fast'' dynamics but small compared with the density fluctuations, Goldstone modes, etc. of interest. One can decompose the Hamiltonian $H = \sum_{\mathbf{x}} H(\mathbf{x})$, where each term acts on the state space of a cell centered at $\mathbf{x}$; boundary terms can be neglected because of the separation of scales between the range of the Hamiltonian and the size of a cell. A similar decomposition holds for other conserved charges. Note, however, that the Hamiltonian and the conserved charges are translation invariant.

Each cell is described by a thermal equilibrium state with a local temperature, local chemical potentials, and (in the case of broken continuous symmetries) a local orientation for the order parameter. A system that can be partitioned into local equilibrium states in this way is said to be in local equilibrium. As a specific example, a generic quantum system with only energy and particle-number conservation has local reduced density matrices of the form $\rho(\mathbf{x}) \propto \exp[-\beta(\mathbf{x})(H(\mathbf{x}) - \mu(\mathbf{x}) N(\mathbf{x}))]$. Here, as elsewhere in this review, the label $\mathbf{x}$ is implicitly coarse-grained over a mesoscale hydrodynamic cell, and we are also implicitly assuming that the chemical potentials vary smoothly in space. 


Under the assumption of local equilibrium, the dynamics of a system can be reduced to the dynamics of a small number of conserved densities (and Goldstone modes, though we will mostly not be concerned with these below). Assuming, further, that the dynamics is spatially \emph{local}, one can write continuity equations for the conserved quantities: these relate the time derivative of each conserved density to its current. (We will not consider problems where the microscopic dynamics is spatially nonlocal; for a recent discussion of these, see, e.g., Ref.~\onlinecite{PhysRevB.101.020416}, as well as a recent experimental verification~\cite{roos2021}.) To close this system of equations, we have to relate the currents back to the densities. We can achieve this by the logic of hydrodynamic projections~\cite{pomeau, forster2018hydrodynamic}: since the only long-lived variables are conserved quantities, their products and derivatives, the part of the current that is long-lived enough to have interesting consequences must itself be made up of these ingredients. Thus, the slow part of the current is in general some arbitrary function of the conserved charges and their low-order spatial derivatives. If we assume further that the fluctuations of conserved charges are not too large, we can expand the current of the $\mu$th charge
\begin{equation}\label{consrel}
j_\mu = A_{\mu\nu} q_\nu + D_{\mu\nu} \partial_x q_\nu + G_{\mu\nu\lambda} q_\nu q_\lambda + f_\mu \ldots, 
\end{equation}
where $q_\mu$ is the $\mu$th conserved charge and the matrices are left general for now. Finally, the ``rest'' of the current (i.e., the part of the current operator that consists of typical rapidly fluctuating degrees of freedom) is incorporated as the noise term $f_\mu$, which is usually taken to be white noise. Eq.~\eqref{consrel} is called a \emph{constitutive relation}. Here, and below, we specialize to one-dimensional systems.

This generic procedure (called hydrodynamics, or sometimes fluctuating hydrodynamics) leads in general to a set of nonlinear partial differential equations. To make further progress, one typically linearizes the theory, leading to a solvable ``fixed point,'' and then includes nonlinearities using some combination of self-consistent and renormalization-group approaches (see, e.g., Refs.~\onlinecite{RevModPhys.54.195, PhysRevLett.54.2026,  KPZ, PhysRevLett.108.180601}). When the nonlinearities are RG-irrelevant, they can still give rise to non-analytic long-time tails~\cite{pomeau, PhysRevA.1.18, ernst1984long, kirkpatrick2002long, delacretaz2020heavy}; when they are relevant, they can destabilize the fixed-point theory and give rise to anomalous transport, as discussed in Sec.~\ref{nlfhd}. Also, if one considers systems with quenched spatial randomness, the coefficients that enter the constitutive relation can vary strongly in space. If these variations are strong enough, transport is anomalous even at the linear level, as we will discuss in Sec.~\ref{griffiths}.

The phenomenology implied by the hydrodynamic framework above strongly depends on whether the matrix $A_{\mu\nu}$ is nonzero: i.e., whether any currents themselves are truly conserved. \emph{Generically}, currents are conserved only in systems with Galilean invariance, where momentum is conserved. In such Galilean systems, a density fluctuation will propagate ballistically. This ballistic propagation is accompanied by spreading, which the linear theory would predict to be diffusive, but which is in fact superdiffusive with the KPZ exponent $z = 3/2$ (see Sec.~\ref{nlfhd}). However, the lattice systems we are concerned with in this review do \emph{not} have momentum as a legitimate slow mode, since it can relax locally via umklapp scattering. (In particular, the anomalous transport phenomena discussed in this review are unrelated to proximate momentum conservation\cite{RoschAndrei,Rosch05}.) In generic one-dimensional lattice systems, therefore, the current is \emph{not} conserved, and to leading order it is given by $j_\mu = D_{\mu\nu} \partial_x \rho_\nu$, i.e., Fick's law. A standard analysis shows that there are no RG-relevant corrections to diffusion. Therefore, one expects that in one-dimensional lattice models all conserved charges diffuse.

\subsection{Integrable systems}

Integrable systems in one dimension are a special class of interacting system in which the scattering among particles is ``non-diffractive,'' in the sense that any scattering process can be factored into a sequence of two-body scattering processes~\cite{takahashi_book}. Since two-body scattering in one dimension can only permute momenta among particles, the dynamics of integrable systems preserves (in some intuitive sense) all the information about the momentum distribution of a generic initial state. This feature of integrable dynamics can be understood in two ``dual'' ways: (i)~integrable systems have stable, ballistically propagating quasiparticles; and (ii)~integrable systems have infinitely many conservation laws. (The relation between these perspectives is sometimes called ``string-charge duality'' and will be addressed further in Sec.~\ref{drudesec}.) From either perspective, the conventional hydrodynamic framework is inappropriate to describe integrable systems. We will turn next to a recently developed alternative, the framework of \emph{generalized} hydrodynamics (GHD). 

Before introducing GHD, though, it is worth clearing up an important strategic point. Since integrable models are famously ``exactly solvable'' via the Bethe ansatz, why does one need \emph{any} coarse-grained framework at all? The answer is that although \emph{some} properties of integrable models, such as their thermodynamics, are straightforward to access, the computation of dynamical correlation functions in the thermodynamic limit is notoriously difficult~\cite{gohmann2004integral}. Thus, even the simplest nonequilibrium quantities like linear response transport coefficients are daunting; if one insists on working with exact form factors, far-from-equilibrium dynamics is restricted to modest system sizes~\cite{caux2009correlation}.

\subsection{Generalized hydrodynamics}

Before turning to GHD, we very briefly review some properties of quasiparticles in integrable systems. This overview is mostly to fix notation; for a pedagogical introduction to this topic, see, e.g., Ref.~\onlinecite{Doyon_notes}. In a finite system, the Bethe ansatz yields a discrete set of eigenstates, each labeled by the quantum numbers of the occupied quasiparticles. In the thermodynamic limit, the allowed quantum numbers form a continuum, and one specifies a state by specifying the \emph{density} of quasiparticles with each set of quantum numbers. We denote this quantity (often called the ``root density'') $\rho_{a,\lambda}$, where $\lambda$ is a continuous index and $a$ is a discrete index or set of indices. In the simplest integrable models, such as the repulsive Lieb-Liniger model, there is only one quasiparticle species and the index $a$ can be suppressed, but in the lattice models we are concerned with the elementary quasiparticles generically form bound states (which are treated as separate species within GHD). Two other basic quantities are the density of states $\rho^{\mathrm{tot}}_{a,\lambda}$ (which is related to the root density via the Bethe equations, $\rho^{\mathrm{tot}}_{a,\lambda} + \sum_b \int d\eta \, \mathcal{K}_{a\lambda,b\eta} \rho_{b,\eta} = \partial_\lambda p_{a,\lambda}/(2\pi)$, where $p_{a,\lambda}$ is the bare momentum of the state parameterized by $(a,\lambda)$, and $\mathcal{K}$ is a matrix that captures the scattering phase shifts in a particular integrable model) and the filling factor $n_{a,\lambda} \equiv \rho_{a,\lambda}/\rho^{\mathrm{tot}}_{a,\lambda}$. An equilibrium state may be specified in terms of either the set of root densities $\{ \rho_{a,\lambda} \}$ or the set of filling factors $\{ n_{a,\lambda} \}$; the two descriptions are related by a nonlinear transformation. An advantage of working with filling factors is that in an equilibrium state, their fluctuations $\delta n_{a,\lambda}$ are uncorrelated, $\langle \delta n_{a,\lambda} \delta n_{b,\eta} \rangle \propto \delta_{ab} \delta_{\lambda\eta}$; the fluctuations of root densities are not diagonal in this sense, since each quasiparticle affects the available state space for all other quasiparticles via scattering~\cite{PhysRevB.54.10845}.

One way to understand GHD is that it starts from a local equilibrium state, in which each hydrodynamic cell is specified by some set of variables as above, and then uses standard hydrodynamic logic to write down equations of motion for these vectors $\{ \rho_{a,\lambda} \}$ or $\{ n_{a,\lambda} \}$. In an integrable system the number of quasiparticles of each type is conserved (since there is only forward scattering); this yields the family of continuity equations $\partial_t \rho_{a,\lambda} + \partial_x j_{a,\lambda} = 0$. The second fundamental equation is a constitutive relation, which says that $j_\alpha = v_{a,\lambda}^{\mathrm{dr}} \rho_{a,\lambda}$, i.e., each quasiparticle propagates ballistically with an effective group velocity set by its ``dressed'' dispersion relation (in which the energies and momenta of each state are computed using the thermodynamic Bethe ansatz~\cite{takahashi_book}). (Note that this constitutive relation is only valid at the Euler scale; the neglected terms are higher order in derivatives but have important physical consequences, giving rise to quasiparticle diffusion.)
Combining these two equations and re-expressing the GHD equations in terms of the filling factors $\{ n_{a, \lambda} \}$, one arrives at a particularly convenient form of these equations:
\begin{equation}\label{bbe}
\partial_t n_{a,\lambda} + v^{\mathrm{eff}}_{a,\lambda}[\{ n \}] \partial_x n_{a,\lambda} = 0.
\end{equation}
This equation describes the advection of the filling factors $n_{a,\lambda}$, which are also called ``normal modes'' of GHD. In a spatially homogeneous system, at Euler scale, the propagator $\langle n_{a,\lambda}(x,t) n_{a,\lambda}(0,0) \rangle = \delta(x - v^{\mathrm{eff}}_{a,\lambda} t)$~\cite{Doyon_notes}.

Thus, given an initial state specified in terms of root densities or filling factors in each cell, GHD gives a prescription to propagate this data forward in time. To compute the dynamics of some conserved charge density $q$, one must work out how much $q$-charge each normal mode transports. This ``dressed charge'' is denoted $q^{\mathrm{dr}}_{a, \lambda}$. It is operationally defined as follows: in the presence of a potential $V$ coupling to $q$, we expect that the filling factor $n_{a,\lambda} \propto (1 + \exp(V q^{\mathrm{dr}}_{a,\lambda}))^{-1}$. The dressed charge is \emph{defined} in terms of $n_{a,\lambda}(V)$ by inverting this relation.

%

%

Instead of regarding Eq.~\eqref{bbe} as a hydrodynamic equation, an alternative and in some ways more natural perspective is that GHD is instead a kind of \emph{kinetic theory}~\cite{Bulchandani18}, analogous to that of transport in Fermi liquids. Like a quasiparticle in an integrable system, a quasiparticle near the Fermi energy in a two-dimensional Fermi liquid is kinematically blocked from scattering in any direction but forward~\cite{pines2018theory, RevModPhys.66.129}. Such a quasiparticle is characterized by its direction along the Fermi surface. Collisions with other quasiparticles will \emph{dress} its properties but cannot change its direction. Quasiparticles at each point on the Fermi surface therefore propagate ballistically along their (conserved) direction, at the Fermi velocity (which interactions can renormalize). 
An appealing visual representation of the kinetic theory is the ``flea-gas'' picture~\cite{PhysRevLett.120.045301}. In this picture, one regards the quasiparticles of an integrable system as rigidly propagating bodies, which recoil from one another whenever they undergo a collision. We will revisit the flea-gas picture below when we address diffusion in integrable systems. 

From either the flea-gas perspective or the perspective of hydrodynamics with many conservation laws, one can see that ballistic transport is natural (though not inevitable) in interacting integrable systems. From the flea-gas picture, indeed, it is not obvious why the leading transport behavior would ever \emph{fail} to be ballistic, since the quasiparticles are always ballistic. We will return to this in Sec.~\ref{anisoXXZ}. At an even more elementary level, the existence of infinitely many conservation laws makes it natural for any particular operator (like the current) to have a large overlap with one or more conserved quantities, giving rise to a ballistic transport. In fact, the earliest results firmly establishing ballistic transport in integrable systems used precisely this approach, explicitly constructing the overlap of the current operator with known conserved operators\cite{Zotos1997}. 

\subsection{The Drude Weight}\label{drudesec}

At the linear-response level, ballistic transport in integrable lattice systems is characterized by the Drude weights $\mathcal{D}_{\mu\nu}$.
The most common and widespread definition of the Drude weight is the long-time limit of the connected current-current autocorrelation function,
\begin{equation}\label{kubodrude}
\mathcal{D}_{\mu\nu} = \lim_{t \rightarrow \infty} \int \dd x \, \langle j_\mu(x,t) j_\nu(0,0) \rangle^c,
\end{equation}
where the superscript $c$ denotes the \emph{connected} part of the correlator, and $j_\mu$ denote densities of current operators $J_\mu = \int \dd x \, j_\mu(x)$. For a system with a nonzero Drude weight, the frequency-dependent conductivity $\sigma_{\mu\nu} = \beta \pi \mathcal{D}_{\mu\nu} \delta(\omega) + \ldots$, where the ellipsis denotes regular contributions, to which we will return below. It is straightforward to see that a persistent current~\eqref{kubodrude} implies ballistic transport. 




\emph{Mazur-Suzuki bound}.---We now formalize the intuition that the Drude weight~\eqref{kubodrude} should be nonzero in integrable systems using the method of hydrodynamic projections. Suppose we have a set of $n$ conserved charges $Q_k$, spanning a vector space of operators orthogonal (but not normalized) under an appropriate inner product. For generalized Gibbs ensembles~\cite{rigol2008thermalization, vidmar2016generalized}, the suitable choice is
$(a,b)=\int \dd x \, \langle a(x) b(0)\rangle^{c}$, under which any two densities $a$ and $b$ of conserved charges are time-invariant (see e.g. Ref.~\onlinecite{Doyon2019diffusion} for formal treatment). In the late-time limit, the non-conserved of currents $J_\mu$ average out, and what remains is the projection onto the conserved subspace,
\beq\label{mazurbound}
    \mathcal{D}_{\mu\nu} \geq \sum_{k = 1}^n \frac{(J|Q_k)^2}{(Q_k|Q_k)}
\eeq
known as the Mazur--Suzuki inequality \cite{Mazur1969,Suzuki1971}. For a finite set of charges, the above formula provides a lower bound on the Drude weights~\cite{Zotos1997}. Upon including all the relevant charges $Q_k$, the inequality turns into a strict equality. However, identifying a complete basis of conserved charges is a difficult task in general, even for the relatively tractable case of integrable lattice models. (The subtleties of constructing such a basis in finite-dimensional classical systems are discussed in Ref.~\onlinecite{DHAR2021110618}, while Ref.~\onlinecite{QLreview} provides a review in the context of quantum integrable lattice models.)

\medskip

\emph{String-charge duality}.---The Mazur--Suzuki inequality is convenient for \emph{lower-bounding} Drude weights; however, if one wants compact and exact expressions for the Drude weight, it is more practical to adopt the quasiparticle perspective.
One is allowed to use either perspective because of a principle known as ``string-charge duality''~\cite{StringCharge}. This reconciles earlier predictions from the thermodynamic Bethe ansatz with the more recent discovery of quasilocal charges, demonstrating that the thermodynamic Bethe ansatz formalism implicitly encodes these charges, despite preceding their discovery by some forty years.
This means that one can specify an equilibrium state in one of two equivalent ways: as a generalized Gibbs ensemble with a separate chemical potential for each charge, or equivalently via the thermodynamic Bethe ansatz, by specifying all the root densities. 

Let us briefly motivate the notion of quasilocal conservation laws. In practice, the notion of locality in lattice systems can be rather subtle. In particular, the traditional infinite set of local conservation laws introduced by integrability textbooks, which are obtained by series expanding the logarithm of the fundamental transfer matrix to yield conservation laws with densities supported on a compact region of space, turns out to be \emph{insufficient}~\cite{Wouters2014,MTPW15}. Due to the presence of bound states in lattice models, one can only fully specify a macrostate upon including quasilocal conservation laws that derive from fused transfer matrices with auxiliary higher-dimensional irreducible representations~\cite{IMP15,PhysRevLett.115.157201}. 

We will return to the issue of quasilocal charges when we discuss the XXZ model below. However, we emphasize that these are not peculiar to the XXZ model, but are a generic feature of quantum integrable lattice models (see, for instance, an explicit construction in the $SU(3)$-invariant chain~\cite{Feher2020}). The core idea behind their existence, which motivates the string-charge duality, can be justified as follows. As we have argued above while introducing GHD, the root densities $\{ \rho_{a,\lambda} \}$ are separately conserved for every species. Preservation of the whole distribution requires the existence of extensive conserved charges with quantum numbers that are additive in the Bethe roots. Suppose for simplicity that quasiparticle excitations are only labelled by their rapidities, as e.g. in the \emph{repulsive} Lieb--Liniger Bose gas (or sinh-Gordon model). Then the traditional, local conserved quantities will suffice to ensure that the rapidity distribution remains invariant in time. Integrable models however generically accommodate bound states; this includes typical integrable spin chains and integrable QFTs that possess internal degrees of freedom that are exchanged upon elastic collisions (known as non-diagonal scattering). In this case, additional conservation laws with additive spectrum are required to ensure conservation of all the rapidity densities for the entire quasiparticle spectrum (as a matter of fact, one per species).

\medskip
\emph{Drude weights from GHD}.---We now express the Drude weight using TBA data, which is justified by the string-charge duality sketched above. Quasiparticles (or more precisely normal modes) with quantum numbers $(a,\lambda)$ propagate with a dressed velocity $v^{\mathrm{eff}}_{a,\lambda}$, which depends on the background state but is dynamically conserved. Each quasiparticle also carries a dressed charge $q^{\mathrm{dr}}_\mu$, so it carries a current $j_{a,\lambda} = q^{\mathrm{dr}}_\mu(a,\lambda) v^{\mathrm{eff}}_{a,\lambda}$, which (at the Euler scale) does not change with time. 
The probability that the quasiparticle state $(a,\lambda)$ will be occupied in thermal equilibrium is $n_{a, \lambda}$. The variance in the occupation is therefore $n_{a, \lambda}(1-n_{a,\lambda})$. Putting these expressions together we arrive at the formula
\beq\label{hydrodrude}
\mathcal{D}_{\mu\nu} = \sum_a \int \dd \lambda\, \rho^{\mathrm{tot}}_{a,\lambda} n_{a,\lambda}(1-n_{a,\lambda}) (v^{\mathrm{eff}}_{a,\lambda})^2 q^{\mathrm{dr}}_{a,\lambda,\mu} q^{\mathrm{dr}}_{a,\lambda,\nu},
\eeq
i.e., the variance in the current carried by a macroscopic state comes entirely from the variance in the occupation of its normal modes, since each occupied normal mode carries (at this level of analysis) some fixed current.

The full matrix of Drude weights is positive semi-definite. Moreover, since the total state density and Fermi occupation functions are strictly positive (at non-zero temperature), a diagonal Drude weight $\mathcal{D}_{\nu \nu}$ becomes zero \emph{if and only if} the corresponding dressed charge vanishes for the entire quasiparticle spectrum. 
For generic (quasi)local conservation laws of the model this does not happen (irrespective of chemical potentials associated with the equilibrium state). An important exception are quasilocal conservation laws from the easy-plane regime of the XXZ chain which possess odd parity under spin reversal~\cite{QLreview}. Conserved charges associated with global Lie symmetries play a distinguished role; their equilibrium dressed values, which are set by the $U(1)$ chemical potentials, can be deduced from the dressed dispersion relations.

\medskip

\emph{Other approaches}.---We now mention, for completeness, some alternative but physically helpful ways of defining the Drude weight.
It was noted in Refs. \onlinecite{VasseurKarraschMoore,Karrasch_2017} that a hydrodynamic Riemann problem (also known as the ``bipartitioning protocol'' or ``two-reservoir quench'' in related literature) consisting of two adjacent, semi-infinite thermal regions with infinitesimally different chemical potentials $\pm \delta h/2$ coupling to the charge $Q$ of interest, could be viewed as a trick for computing the (diagonal) Drude weight, via the relation~\footnote{An alternative definition, similar in spirit, had been proposed earlier in Ref.~\onlinecite{Ilievski12} using the formalism of operator $C^{*}$-algebras.}
\begin{equation}
\label{eq:DrudeTrick}
\mathcal{D} = \lim_{\delta h \to 0}\lim_{t\to \infty}\lim_{L\to \infty}\frac{1}{t \delta h}
\int_{-L}^L \dd x \, j(x,t),
\end{equation}
with $j$ the current density associated with the charge $Q$. This trick was subsequently applied to the specific problem of computing spin Drude weights in the easy-axis XXZ chain from generalized hydrodynamics~\cite{Bulchandani17, IN_Drude}.

A second useful relation is that Drude weights correspond to the second moment of dynamical correlation functions at late times
\begin{equation}
\mathcal{D}_{\mu \nu} = \int \dd x \, x^2 \langle q_{\mu}(x,t)q_{\nu}(0,0) \rangle
\simeq \mathcal{D}_{\mu \nu}t^{2}.
\end{equation}
This is essentially a version of the Einstein relation between conductivities and diffusion constants. It can be derived from the Kubo formula~\eqref{kubodrude} using the continuity equation.

Finally, we note that the Drude weight of a system on a ring can equivalently be defined, following Kohn~\cite{Kohn1964}, in terms of the response of energy levels to twisting the boundary conditions. This approach was applied to the Heisenberg model at $T=0$ in Ref.~\onlinecite{Shastry1990}, and later extended to finite temperatures \cite{FujimotoKawakami98,Fujimoto99,Zotos99}. Kohn's approach works naturally only for conservation laws associated with $U(1)$ global symmetries (such as e.g. spin/charge in quantum spin chain). We will not take this perspective in what follows; for reviews see Ref.~\onlinecite{Sirker_lecture}.

\subsection{Diffusion in integrable systems}

\begin{figure}[tb]
\begin{center}
\includegraphics[width = 0.5\textwidth]{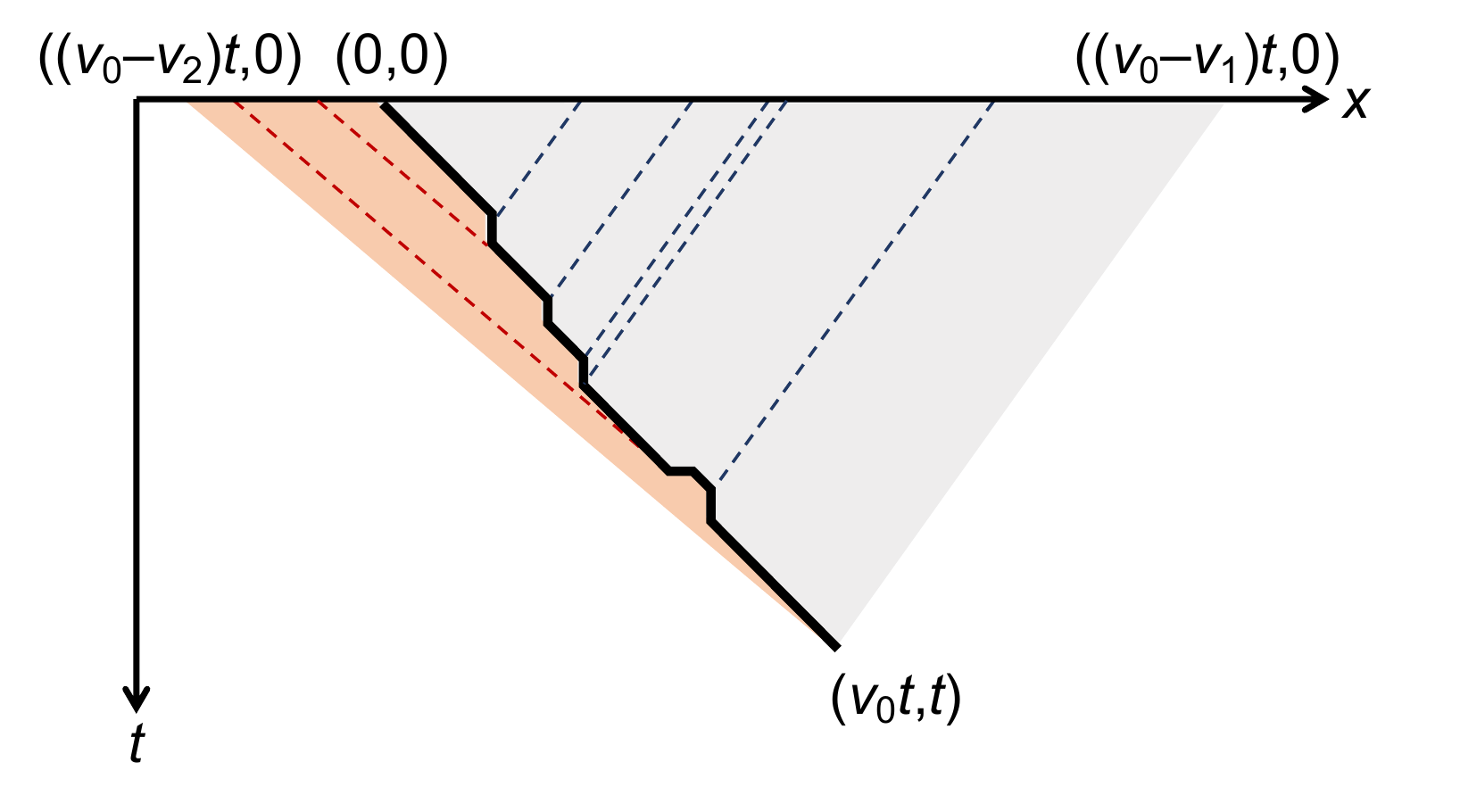}
\caption{Kinematics of forward scattering leading to velocity dressing and diffusion in integrable systems. On average, a tagged quasiparticle traveling to the right encounters more left-movers than right-movers, giving some net velocity renormalization. The fluctuations in the number of collisions in a time window cause diffusion (see text). Reproduced from S. Gopalakrishnan et al., \textit{Phys. Rev. B} {\bf 98}, 220303(R) (2018).}
\label{figdiff}
\end{center}
\end{figure}

We briefly review the next-order dynamical process in integrable systems, which is quasiparticle diffusion~\cite{DeNardis2018, DeNardis_SciPost, Gopalakrishnan18, Medenjak19, Doyon2019diffusion}.
This topic is discussed extensively in a companion review~\footnote{To appear in the same volume.}; here, we briefly summarize a ``kinetic'' derivation~\cite{Gopalakrishnan18} of the main result that will be helpful to us in the rest of this review.
To leading order, quasiparticles in integrable systems move with a fixed velocity. However, the soliton-gas picture of GHD makes it clear that this is not an exact statement. Rather, the displacement of a quasiparticle over a time interval $\Delta t$ has two components: \emph{free propagation} and \emph{collisional shifts}. In a general soliton gas, each collisional process between a pair of quasiparticles is associated with its own collisional shifts. 
The key observation is that the collisional shifts a quasiparticle experiences depends on the number of collisions it has in a given time interval; this in turn depends on the density of other quasiparticles of each type in the interval, which is a quantity that experiences Gaussian thermal fluctuations. Thus the effective velocity of a quasiparticle also has thermal fluctuations, which as we will see give rise to diffusion.

To be more quantitative, we consider a specific normal mode of quantum numbers $(a, \lambda)$ initialized at the origin of spacetime, and consider its motion over a time $t$. At the Euler scale, this normal mode will deterministically propagate to $v^{\mathrm{eff}}_{a,\lambda} t$ and its position will have no variance. Once collisional fluctuation effects are included, the variance can be expressed as
\beq
\delta x_{a,\lambda}^2 = [(\delta v^{\mathrm{eff}}_{a,\lambda})^2] t^2 = t^2 \sum_{b} \int \dd \eta \left(\frac{\delta v^{\mathrm{eff}}_{a,\lambda}}{\delta n_{b,\eta}} \right)^2 [\delta n_{b, \eta}^2], 
\eeq
where the square brackets denote an average over the trajectory of the quasiparticle in the time interval $t$, and we have used the fact that fluctuations of different occupation factors are uncorrelated~\cite{PhysRevB.54.10845}. In a high-temperature thermal state, fluctuations are spatially uncorrelated, so we can write $[\delta n_{b,\eta}^2] = C_{b,\eta}/\ell$ over any region $\ell$. The main subtlety here is computing the number of fluctuations correctly. One might naively want to average over a fixed spatial window, but this is inaccurate: normal modes that are moving almost parallel to the one of interest will rarely collide with it, whereas those moving in the opposite direction will have many more collisions. In general, a quasiparticle moving at velocity $v_{a,\lambda}$ will collide with one moving at velocity $v_{b,\eta}$ \emph{if} the second quasiparticle started out in a window of size $|v^{\mathrm{eff}}_{b,\eta} - v^{\mathrm{eff}}_{a,\lambda}| t$ (Fig.~\ref{figdiff}). Since the density fluctuations in a high-temperature state are Gaussian and uncorrelated across hydrodynamic cells, we see that
\bea\label{intdiffusion}
\delta x_{a,\lambda}^2 & = & t \sum_b \int \dd\eta \left(\frac{\delta v^{\mathrm{eff}}_{a,\lambda}}{\delta n_{b,\eta}} \right)^2 \frac{C_{b,\eta}}{|v^{\mathrm{eff}}_{a,\lambda} - v^{\mathrm{eff}}_{b,\eta}|} \\ & & \quad = t \left[ \frac{1}{(\rho^{\mathrm{tot}}_{a,\lambda})^2} \sum_\beta \int \dd\eta\, |v^{\mathrm{eff}}_{a,\lambda} - v^{\mathrm{eff}}_{b,\eta}| \rho^{\mathrm{tot}}_{b,\eta} n_{b,\eta} (1-n_{b,\eta}) [\mathcal{K}^\mathrm{dr}_{a\lambda;b\eta}]^2 \right]. \nonumber
\eea
%
%
The first equality shows that each quasiparticle trajectory undergoes diffusive broadening with a diffusion constant determined (implicitly) by GHD and TBA data; the second equality presents an explicit expression, which we will not derive here; see Ref.~\onlinecite{Gopalakrishnan18}. The expression in square brackets is the ``quasiparticle diffusion coefficient,'' i.e., the rate at which the propagator for each normal mode broadens. 

To summarize, the propagator for each normal mode in an integrable system broadens as in Eq.~\eqref{intdiffusion}. While, at Euler scale, its propagator would read $\langle n_{a,\lambda}(x,t) n_{a,\lambda}(0,0) \rangle \propto \delta(x - v^{\mathrm{eff}}_{a,\lambda} t)$, at the diffusive scale this delta function is broadened to a Gaussian of width $\sqrt{D_{a,\lambda} t}$, where the diffusion constant is given in Eq.~\eqref{intdiffusion}. This broadening was explicitly checked for the integrable Rule 54 cellular automaton~\cite{Gopalakrishnan18, klobas2019time, PhysRevB.98.060302}. One can transform this result for quasiparticle diffusion back into the basis of conserved charges, giving an ``Onsager matrix'' of diffusion constants~\cite{DeNardis2018, DeNardis_SciPost}; we will not cover this here in detail. 

For completeness we note two alternative approaches to deriving diffusion in integrable systems, which we briefly mention here. The first is to explicitly evaluate the Kubo formula using thermodynamic form factors~\cite{DeNardis2018, DeNardis_SciPost}. The second is to write the current operator $j_\mu = A_{\mu \nu} q_\nu + G_{\mu\nu\lambda} q_\nu q_\lambda + \ldots$. Evaluating the contribution of the term quadratic in $q$ yields a regular term in the d.c. conductivity~\cite{Medenjak19, Doyon2019diffusion}. 

One important point remains to be addressed. As we remarked above, in general a ballistically propagating mode in one dimension broadens \emph{anomalously}, according to the KPZ equation, rather than simply broadening diffusively. Why this does not happen in the present context was first explained in Ref.~\onlinecite{Vir20}: KPZ broadening would take place if the dressed velocity $v_{a,\lambda}^{\mathrm{eff}}$ depended linearly on the filling factor of the \emph{same} type of quasiparticle, i.e., on $n_{a,\lambda}$. Since the velocity dressing is due to scattering off \emph{other} quasiparticle species, this direct linear dependence is not present, so one has diffusion rather than KPZ broadening for each individual quasiparticle. Intuitively, the hydrodynamic evolution Eq. \eqref{bbe} cannot form shocks~\cite{Bulchandani17}, which means there is no dynamical mechanism for roughening~\cite{KPZ} within GHD for an integrable system with a finite number of quasiparticle species. 

\subsection{Generic mechanisms for anomalous transport}

In this section, we summarize some well-known and generic mechanisms for anomalous transport in classical systems. One of the unexpected findings from studies of quantum many-body systems over the past decade is that all these distinct forms of anomalous transport seem to be realized in one-dimensional spin chains, as we will see in subsequent sections.

\subsubsection{L\'evy flights and fractional diffusion}\label{griffiths}

L{\'e}vy flights are ubiquitous in nature, and can most simply be understood as the limiting stochastic process of ``fat tailed'' random walks, whose step lengths are so unpredictable that their variance is infinite. To motivate the notion of L{\'e}vy flights, it is helpful to recall the probabilistic understanding of normal transport. Classically speaking, ``normal transport'' is a prediction of the central limit theorem, in the following sense~\cite{BOUCHAUD1990127}. Consider a random walker in one dimension, who at a series of discrete and regularly spaced time steps $t_n = n\Delta t$, undergoes jumps of length $l=l_n$. Here, the lengths $l_n$ are assumed to be independent, identically distributed random variables, drawn from a continuous distribution with some density function $p(l)$. The total distance travelled at time $t=t_n$ is given by $X_{t} = \sum_{j=1}^n l_j$. At long times, the ``typical'' behaviour of $X_t$ is ballistic with diffusive corrections,
\begin{equation}
\langle X_t \rangle = vt, \quad \langle X_t^2\rangle - \langle X_t \rangle^2 = 2D t,
\end{equation}
with drift velocity and diffusion constant $v = \langle l \rangle/\Delta t$ and $D = (\langle l^2 \rangle - \langle l \rangle^2)/(2\Delta t)$ respectively. By the central limit theorem applied to the scaling variable $W=(X_t-vt)/(2Dt)^{1/2}$, the asymptotic probability density function $u(x,t)$ for $W=(x-vt)/(2Dt)^{1/2}$ is Gaussian and satisfies the Fokker-Planck equation
\begin{eqnarray}
\partial_t u = -v \partial_x u + D \partial_x^2 u.
\end{eqnarray}
This behaviour is generic insofar as the random jumps $l_n$ satisfy the hypotheses of the central limit theorem. Suppose, however, that the asymptotic behaviour of the jump distribution $p(l) \sim l^{-(1+z)}$ as $l \to \infty$, with $1<z<2$. Then $\langle l \rangle$ is finite but $\langle l^2 \rangle$ exhibits an infrared divergence, leading to an infinite variance of $p(l)$ and invalidating the above analysis. Nevertheless, once the correct scaling variable $W = (X_t-vt) / t^{1/z}$ is identified, a limiting probability density function for $W$ is found that is no longer Gaussian, but instead a so-called \emph{L{\'e}vy $z$-stable distribution function} $L_{z,\beta}(w)$, where $z$ denotes the dynamical exponent and $\beta$ quantifies the asymmetry in $w$ about $w=0$. Explicit formulas for these distribution functions are complicated in general; here we merely quote the result~\cite{BOUCHAUD1990127} for the symmetric ($p(l)=p(-l), \, v=\beta=0$) case, which yields (up to rescaling) the probability density function
\begin{eqnarray}
L_{z,0}(w) = \int \frac{\dd k}{2\pi}\, e^{ikw} e^{-|k|^z}.
\end{eqnarray}

Now $u(x,t) = L_{z,0}(x/t^{1/z})/t^{1/z}$ is a source function for the so-called \emph{fractional diffusion equation}
\begin{equation}
\label{eq:FDE}
\partial_t u = -(-\nabla^2)^{z/2} u,
\end{equation}
which therefore stands in the same relation to L{\'e}vy flights as the ordinary diffusion equation stands in relation to Brownian motion. The ``fractional Laplacian'' $(-\nabla^2)^s$ appearing in this equation is mostly simply defined by its action in Fourier space, for example
\begin{equation}
(-\nabla^2)^s f(x) = \int \frac{\dd^dk}{(2\pi)^d}\,|k|^{2s} \tilde{f}(k)e^{ikx},
\end{equation}
where $\tilde{f}(k) = \int d^dx \, e^{-ikx} f(x)$ is the Fourier transform of $f(x)$. The space-time scaling of solutions to Eq. \eqref{eq:FDE} can be read off to be $t \sim x^{z}$, and normal diffusion is recovered at the point $z=2$.

The scaling invariance of Eq. \eqref{eq:FDE} is reflected in the microscopic distribution of jump sizes. To see this, it is helpful to consider the behaviour of the ``largest jump'', $l^*(t)$, at time $t=t_n$. We define $l^*(t)$ such that a jump with length greater than or equal to $l^*(t_n)$ occurs exactly once in $n$ time steps, so that $n \int_{l^*(t_n)}^\infty dl \, p(l) = 1$. This yields the scaling law
\begin{equation}
l^*(t) \sim t^{1/z},
\end{equation}
at large times. Notice that $l^*(t)$ defines a natural infrared cut-off for regulating the divergence in $\langle l^2 \rangle$. One can then estimate the variance of the total path length as
\begin{eqnarray}
\langle X_t^2 \rangle - \langle X_t \rangle^2 \sim t\int^{l^*(t)} \dd l\, l^2 p(l) \sim t^{2/z} \sim [l^*(t)]^2.
\end{eqnarray}
The key point is that the mean square length of the path $X_t$ at time $t$ is dominated by the largest jump up to time $t$. This provides an intuitive sense in which the L{\'e}vy flight is self-similar, and can be interpreted more precisely as a continuously varying fractal dimension $z$ for the resulting trajectories~\cite{Seshadri4501}.

Finally, we note that fat-tailed distributions in microscopic transport typically arise from L{\'e}vy \emph{walks}, which are random walks with a finite average velocity and power-law distributed times-of-flight (see Ref. \onlinecite{DharLW} for a recent review in the context of heat transport). In contrast, for the anomalous quasiparticle transport of interest here (cf. Sec. \ref{gaplessxxzsec}), scattering off large quasiparticles is a highly non-local process and allows for an unbounded effective quasiparticle velocity in the limit of infinite system size. The resulting dynamics is most naturally interpreted as a L{\'e}vy flight, rather than a L{\'e}vy walk.

\subsubsection{Nonlinear diffusion}\label{NLD}
The nonlinear diffusion equation is given by
\begin{equation}
\label{eq:NLDE}
\partial_t u = \nabla^2 u^m,
\end{equation}
i.e. the effective diffusion constant is nonlinear,
\begin{equation}
\label{eq:NLDEDiff}
D_{\textrm{eff}}[u] = mu^{m-1}.
\end{equation}
In physical applications it tends to be the case that $u \geq 0$, which will henceforth be assumed. For $m=1$, Eq. \eqref{eq:NLDE} recovers normal diffusion. For $m>1$, Eq. \eqref{eq:NLDE} is known as the ``porous medium equation'' while for $m<1$, it is instead called the ``fast diffusion equation''. Nonlinear diffusion equations are useful for modelling diverse physical systems, ranging from groundwater flow to high-temperature heat transfer in plasmas.

In contrast to the other two models for anomalous transport considered in this section, the dynamics of Eq. \eqref{eq:NLDE} in the linear response regime, $u(\mathbf{x},t) = u_0 + \delta u(\mathbf{x},t)$, $|\delta u| \ll u_0$ is normal diffusion, with effective diffusion constant $D_{\mathrm{eff}}[u_0]$ as in Eq. \eqref{eq:NLDEDiff}. However, the linear response approximation breaks down as $u_0 \to 0$, with $D_{\mathrm{eff}}[u_0] \to 0$ for the porous medium equation, indicating subdiffusive dynamics, while $D_{\mathrm{eff}}[u_0] \to \infty$ for the fast diffusion equation, indicating superdiffusive dynamics.

\begin{figure}[t]
    \centering
    \includegraphics[width=.6\columnwidth]{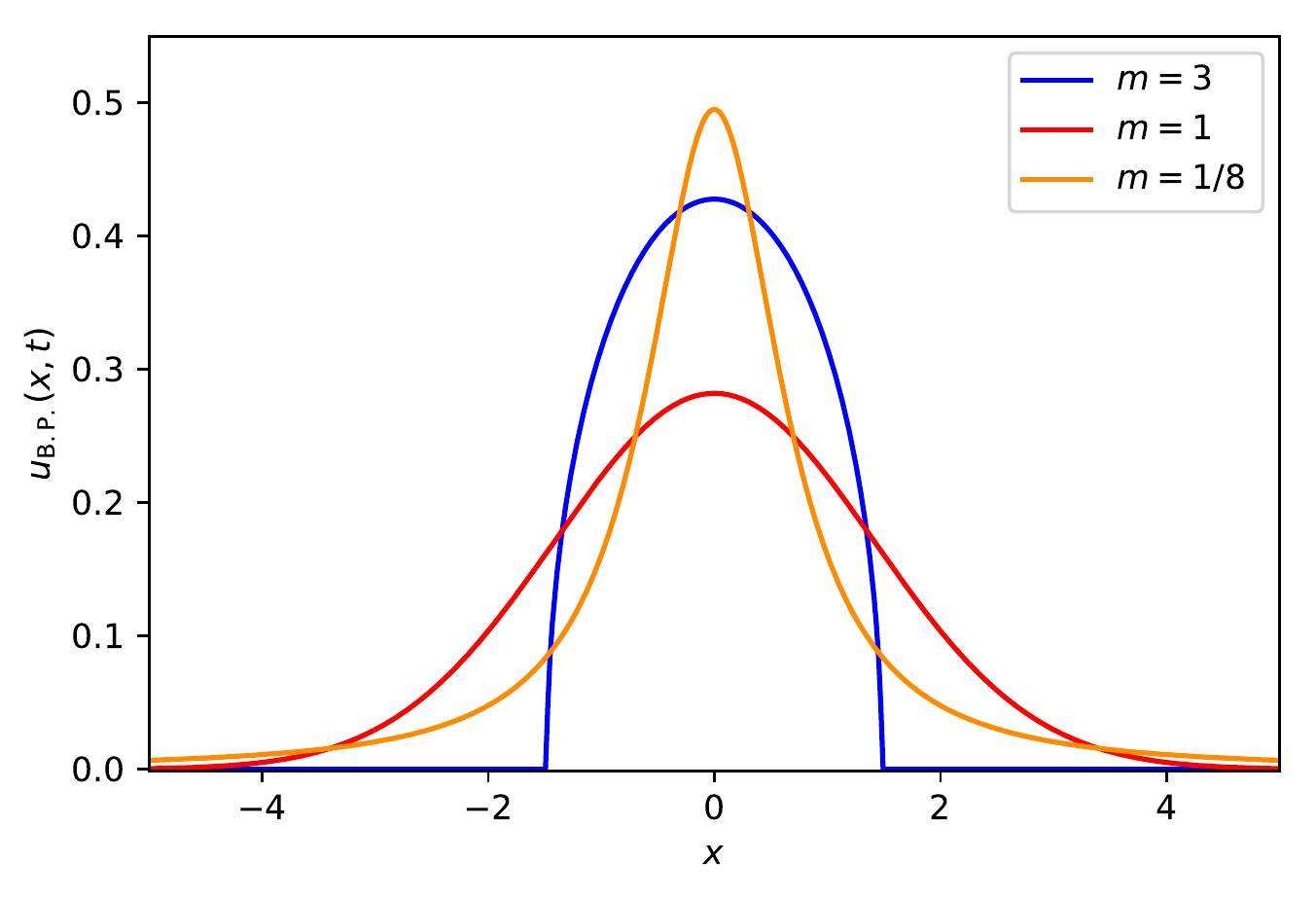}
    \caption{Fundamental solutions to the nonlinear diffusion equation with the same initial area $M=1$ and varying nonlinear exponents $m=3$, $m=1$ and $m=1/8$, at time $t=1$. These profiles exhibit subdiffusion, normal diffusion and superdiffusion respectively, with associated space-time scaling exponents $\alpha = 1/4$, $\alpha=1/2$ and $\alpha=8/9$.}
    \label{fig:BPsolns}
\end{figure}

These peculiarities are explained by observing that although the linear response of Eq. \eqref{eq:NLDE} is diffusive for bulk densities $u_0>0$, its fundamental solutions for $m \neq 1$, which have the physical interpretation of free expansion into the $u_0=0$ ``vacuum'', are non-Gaussian. Instead, the nonlinearity and scaling invariance of Eq. \eqref{eq:NLDE} gives rise to fundamental solutions with anomalous space-time scaling, known as ``Barenblatt-Pattle profiles''. In one spatial dimension, the case relevant to this review article, the Barenblatt-Pattle profiles have the form~\cite{VazquezPME}
\begin{equation}
\label{eq:BarenplattPattle}
u_{\mathrm{B.P.}}(x,t) = t^{-\alpha}\mathrm{max}[(C-k(x/t^\alpha)^2)^{-\frac{1}{m-1}},0], \quad \alpha = 1/(m+1),
\end{equation}
with $k=k(m)$ constant and $C(m,M)$ fixed by the initial area $M$. Representative examples of these profiles are depicted in Fig. \ref{fig:BPsolns}. The Barenblatt-Pattle profiles for nonlinear subdiffusion have a discontinuous edge (in this sense, they define ``weak'' solutions to the underlying PDE), while those for nonlinear superdiffusion are characterized by a power-law decay in space as $|x|\to\infty$.

Barenblatt-Pattle solutions exist for the porous medium equation for all $m>1$ and (in one dimension) for the fast diffusion equation in the range $0<m<1$. They can be obtained by substituting the scaling ansatz $u_{\mathrm{B.P.}}(x,t) = t^{-\alpha} F(x/t^{\alpha})$ into Eq. \eqref{eq:NLDE} and solving for $F$. These solutions are ``fundamental'' in the sense that they originate from delta functions, with
\begin{equation}
\lim_{t\to 0^+} u_{\mathrm{B.P.}}(x,t) = M \delta(x)
\end{equation}
as a distribution. For $m \leq 0$ the fast diffusion equation becomes increasingly ill-posed, exhibiting phenomena such as ``extinction'' in finite time, whereby diffusion occurs so fast that fundamental solutions instantly start to lose mass at infinity. A detailed discussion of such pathologies can be found in Ref. \onlinecite{VazquezSmoothing}.

For the purposes of this review, we will be concerned with the superdiffusive regime $0 < m < 1$; an interesting open question is how far other regimes of Eq. \eqref{eq:NLDE} can be realized in the collective dynamics of many-body quantum systems.

\subsubsection{Nonlinear fluctuating hydrodynamics}\label{nlfhd}

The two models for anomalous transport considered above can be formulated as deterministic PDEs, that emerge as a suitable scaling limit of random, microscopic processes. The models that we consider in this section have a fundamentally different character, and take the form of stochastic PDEs, with microscopic fluctuations included at the macroscopic scale in the form of coupling to white noise {\`a} la Langevin.
Such equations arise naturally in fluid dynamics and we briefly summarize their derivation (see Ref. \onlinecite{spohn2016fluctuating} for a fuller exposition). Suppose we are given a one-dimensional Hamiltonian system with $N$ local conserved charges, $Q_a = \int_0^L \dd x \, q_a(x)$, satisfying microscopic conservation laws $
\partial_t q_a + \partial_x j_a = 0$. We take as our starting point the closed system of Euler-scale hydrodynamic equations
\begin{equation}
\partial_t \langle q_a \rangle  + \partial_x \langle j_a \rangle = 0, \quad a=1,2,\ldots,N,
\end{equation}
in the space of local Gibbs ensembles $\rho = Z^{-1} e^{-\sum_{a=1}^N \int dx \, \beta_a(x)q_a(x)}$ that were discussed in the introduction. Linearizing the hydrodynamic equations about an equilibrium state $\rho = \rho_0 + \delta \rho$ with $\rho_0 = Z^{-1} e^{-\sum_{a=1}^N \beta_a Q_a}$ and defining $A_{ab} = \partial \langle j_a \rangle_0 / \partial \langle q_b \rangle_0$ yields a linear equation
\begin{equation}
\partial_t u_a + \sum_{b=1}^N A_{ab} \partial_x u_b = 0, \quad a=1,2,\ldots,N,
\end{equation}
for small perturbations $u_a = \langle q_a \rangle - \langle q_a \rangle_0$ about $\rho_0$. The generalized ``speeds of sound'' for this system in the state $\rho_0$ are then given by the eigenvalues $c_a$ of $A_{ab}$. Under mild assumptions~\cite{spohn2016fluctuating}, the $c_a$ are real and there is a basis $\phi_a = \sum_{b=1}^N R_{ab} u_b$ of eigenmodes of $A_{ab}$, which satisfy
\begin{equation}
\partial_t \phi_a + c_a \partial_x \phi_a = 0, \quad a=1,2,\ldots,N.
\end{equation}
It follows by characteristics that a small initial perturbation $\phi_i(x,0)$ will spread ballistically as $\phi_a(x,t) = \phi_a(x-c_a t,0)$, unless an eigenvalue $c_a = 0$, in which case the leading behaviour of the mode $\phi_a$ is sub-ballistic.

Realistic physical systems at non-zero temperature exhibit dissipation, but treating this within hydrodynamics inevitably requires introducing some form of approximation, since purely deterministic time evolution can never give rise to irreversible behaviour. More precisely, the strength of dissipation (and its associated fluctuations) in a given hydrodynamic description is sensitive to the choice of coarse-graining length scale $\ell$. To the best of the authors' knowledge, there is no systematic way to treat such ``mesoscopic'' effects. This state of affairs can be summarized by the observation that dissipative hydrodynamics is an ``effective stochastic field theory'', a point of view that animates both earlier studies of fluctuating hydrodynamics~\cite{pomeau} and the more explicitly field-theoretical formulation of dissipative hydrodynamics that was achieved recently~\cite{Crossley2017}.

The idea behind fluctuating hydrodynamics is to couple the Euler-scale hydrodynamic equations to noise and dissipation by hand, to yield
\begin{equation}
\label{eq:NLFH}
\partial_t u_a + \partial_x \sum_{b=1}^N (A_{ab} u_b - D_{ab}\partial_x u_b + B_{ab}\zeta_b ) = 0, \quad a=1,2,\ldots,N,
\end{equation}
where the $\zeta_a$ are unit normalized, uncorrelated white noise variables $\langle \langle \zeta_a(x,t) \zeta_b(x',t') \rangle \rangle = \delta_{ab}\delta(x-x')\delta(t-t')$ and $\langle \langle \ldots \rangle \rangle$ denotes the average over noise realizations. Demanding stationarity of initial thermal correlations $\langle \langle u_a(x,0) u_b(x',0) \rangle \rangle = C_{ab} \delta(x-x')$ imposes the fluctuation-dissipation relation $DC + CD^T = BB^T$.

So far, Eq. \eqref{eq:NLFH} merely \emph{describes} the expectation that hydrodynamic fluctuations of a finite temperature system should be diffusive. In fact, this is only true in $d \geq 3$. In lower dimensions, non-linear corrections to Eq. \eqref{eq:NLFH} become important. In $d=2$ such corrections are marginal; they lead to a logarithmic divergence in transport coefficients that may be rephrased as a weak finite-size effect. In $d=1$, the leading non-linear correction to Eq. \eqref{eq:NLFH} is relevant, and alters the universality class of hydrodynamic fluctuations~\cite{spohn2016fluctuating}. Working in the normal mode basis and restoring leading non-linearities in the expansion of the currents $\langle j_a \rangle$ yields an equation of the form
\begin{equation}\label{nlb}
\partial_t \phi_a + \partial_x \left(c_a \phi_a + \sum_{b=1}^N (-D'_{ab}\partial_x \phi_b + B'_{ab}\zeta_b) + \sum_{b,\,c=1}^N G^{a}_{bc} \phi_b\phi_c\right) = 0, \quad a=1,2,\ldots,N.
\end{equation}
When the sound velocities $c_a$ are distinct\footnote{This condition guarantees that hydrodynamic normal modes are well-separated at long times. Mode-coupling theory can of course be applied when sound velocities coincide, but the resulting theoretical analysis is more complicated~\cite{Ertas}.}, the dynamical scaling exponents $z_a$ of fluctuations of the normal modes $\phi_a$ are determined by the sets $I_a = \{ b: G^a_{bb} \neq 0\}$. Specifically, a one-loop calculation in stochastic field theory yields the dynamical exponents\cite{Popkov15}
\begin{equation}
\label{eq:expcriterion}
z_a = \begin{cases} 2 & I_a = \emptyset \\ 3/2 & a \in I_a \\ \min_{b\in I_a} \left(1 + \frac{1}{z_b}\right) & \mathrm{else} \end{cases}.
\end{equation}
Although it treats dissipation phenomenologically, this strategy of ``nonlinear fluctuating hydrodynamics'' has proved to be invaluable for explaining the emergence of anomalous transport in one-dimensional classical fluids~\cite{spohn2016fluctuating}. In $d>1$, this approach is less informative because it merely demonstrates consistency of the assumption of diffusive fluctuations, that was put in by hand.

Before we close this section, we briefly mention one simple physical way of understanding the exponent $z = 3/2$ in the classical KPZ context. This approach was hinted at in Ref.~\onlinecite{KPZ}. The argument runs as follows. Consider the noisy Burgers equation $\partial_t \rho + \rho \partial_x \rho = D \partial_x^2 \rho + \xi$, where $\xi$ is some noise term. This is essentially Eq.~\eqref{nlb} with indices suppressed. Let us anticipate that the stationary measure under this equation has equilibrium thermal fluctuations. Now consider the propagation of a particle over a  region of length $L$. The typical fluctuation of $\rho$ over this length-scale will scale as $L^{-1/2}$; thus, so will the typical velocity. The time it takes to cross this region is therefore $t(L) \sim L/v \sim L^{3/2}$, giving the KPZ exponent.

\section{Anisotropic Heisenberg chain}\label{anisoXXZ}

We begin our exploration of anomalous transport in integrable models with two examples that are in some ways simple, since the anomalous transport is inherently \emph{linear}: namely the easy-axis and easy-plane regimes of the Heisenberg XXZ model, 
\beq\label{hamxxz}
H_{\mathrm{XXZ}} = \sum\nolimits_i (\sigma^x_i \sigma^x_{i+1} + \sigma^y_{i} \sigma^y_{i+1} + \Delta \sigma^z_i \sigma^z_{i+1}). 
\eeq
The easy-axis case $|\Delta| > 1$ and the easy-plane case $|\Delta| < 1$ will be dealt with separately, since they exhibit very different physics. 

From a general statistical mechanics perspective it might seem unexpected that the finite-temperature dynamics of a one-dimensional spin chain should change discontinuously as a parameter is tuned. \emph{Thermodynamically}, indeed, all the states we will consider are paramagnets. Further, the dynamics at finite times must also evolve continuously with $\Delta$. However, the late-time limit of the dynamics can change discontinuously because the structure of the exact conserved quantities is extremely sensitive to the precise value of $\Delta$. 

Some intuition as to why this is so can be gleaned from considering the ground states in the two cases. In the easy-plane regime, the ground state consists of spins pointing along the equator of the Bloch sphere, and breaks a continuous $U(1)$ symmetry. The low-energy excitations are Goldstone modes. In the easy-axis regime, the ground state consists of spins pointing along the $z$ axis and breaking an Ising symmetry. The excitations above this state are magnetic domains of various sizes. Because the XXZ model is integrable, these elementary excitations remain infinitely long-lived, and thus the difference in ground states gets promoted to a difference in the entire spectrum. (A similar phenomenon happens with ``strong zero modes'' in the transverse-field Ising model~\cite{fendley2016strong}.)

\subsection{The easy-axis XXZ model}

We begin with the easy-axis case. To develop an intuition for the dynamics of this model it is helpful to consider systems for which $\Delta \gg 1$, so that one can treat the hopping terms in the Hamiltonian as a perturbation. In the purely ``classical'' limit $H = \sum_i \sigma^z_i \sigma^z_{i+1}$, the eigenstates are simply product states (which we will call ``configurations'') in which each spin is either up or down. The energy of an eigenstate is set by the number of domain walls, i.e., the number of anti-aligned neighboring spins. (Depending on the sign of the interaction, anti-alignment is either favored or penalized; either way, it changes the energy and is not an on-shell process.) Starting from this trivially solvable point, we now introduce the flip-flop terms as a weak perturbation, and ask what pairs of configurations are hybridized by this perturbation. To hybridize, two configurations must have the same energy (i.e., the same number of domain walls); however, because of $U(1)$ symmetry, the two configurations must also have the same number of $\uparrow$ spins. Thus (unlike, e.g., the transverse-field Ising model) domain walls are not freely propagating quasiparticles. Rather, the propagating quasiparticles are \emph{entire domains}. 

Let us start with the ferromagnetic vacuum state; evidently this has infinitely many different species of quasiparticles, corresponding to sequences of $s$ flipped spins. These species are referred to as $s$-strings; an $s$-string has a dispersion with a bandwidth $\propto \Delta^{1-s}$. Because the model is integrable, these $s$-string labels continue to label stable quasiparticles even at finite excitation density; however, $s$-strings are strongly dressed by collisions with one another, and cannot be simply identified with domains.

\subsubsection{Absence of ballistic transport at half filling}\label{gappedsec}

\begin{figure}[tb]
\begin{center}
\includegraphics[width = 0.375\textwidth]{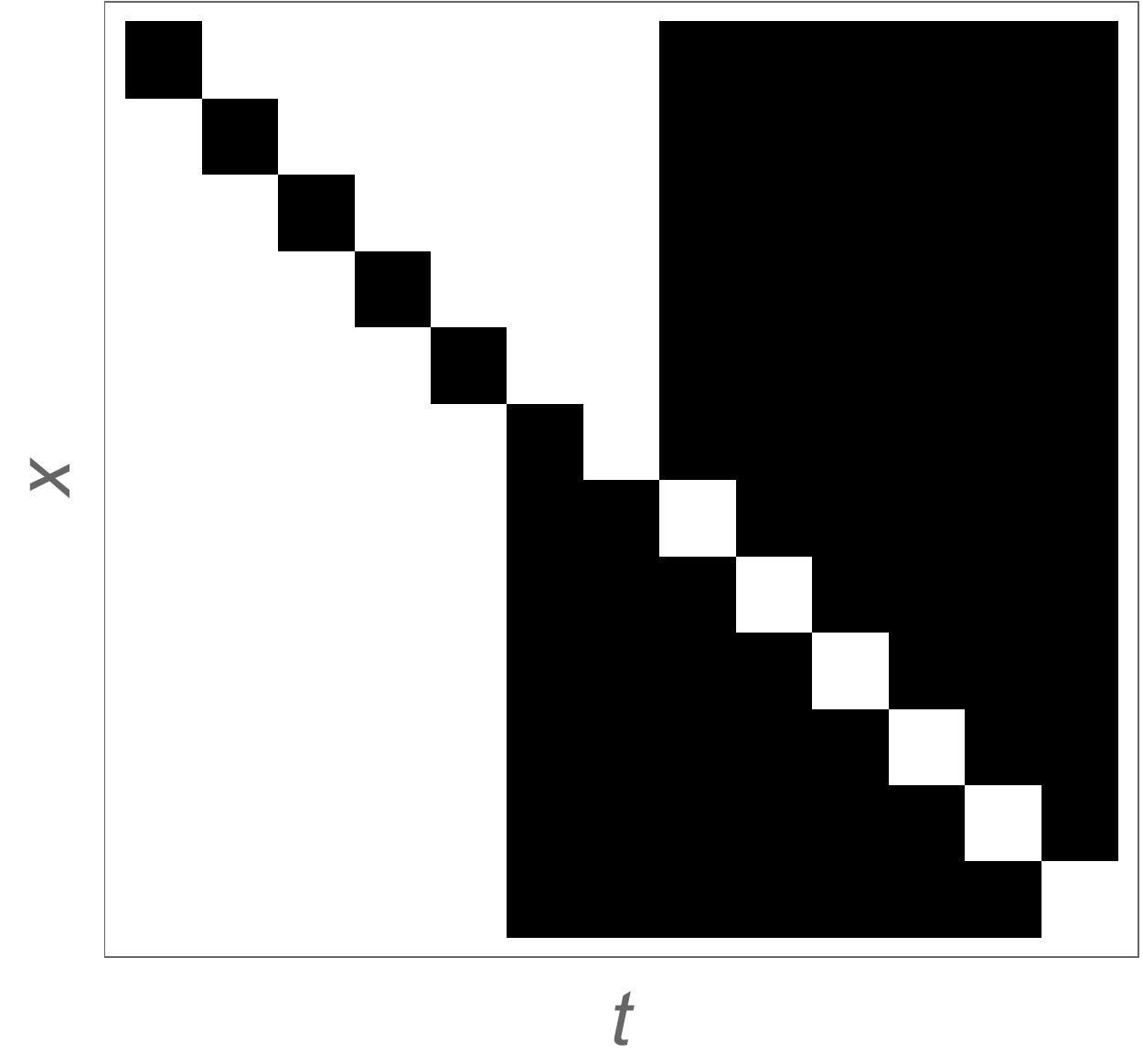}
\includegraphics[width = 0.35\textwidth]{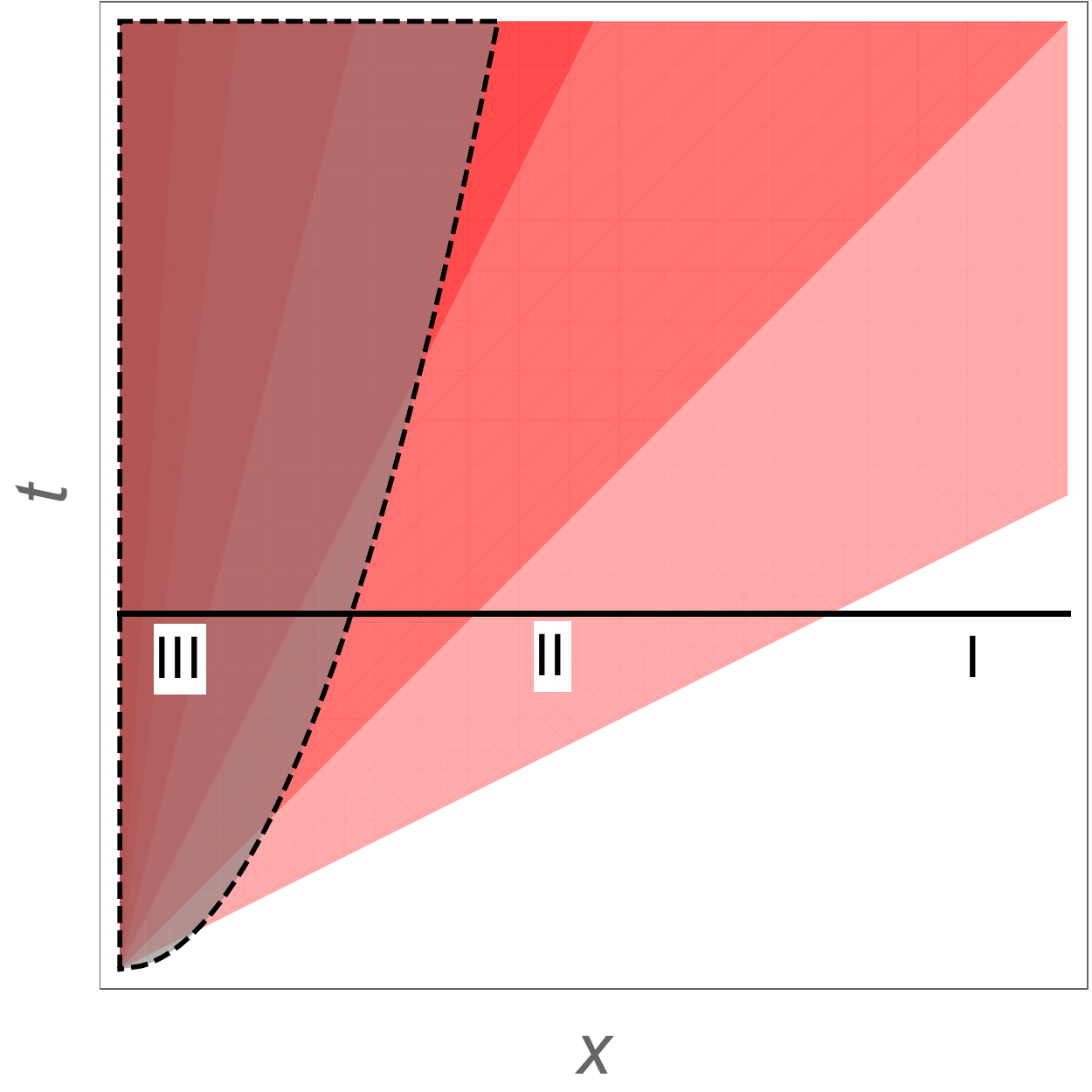}
\caption{Left: Passage of a magnon through a domain wall in the easy-axis XXZ model. Note that the magnon is ``stripped'' of its magnetization as it goes through. Right: Dynamical spin correlation function of the XXZ model at generic filling, showing three regimes: (I)~outside the light-cone, the correlator vanishes, (II)~there is a large ``Eulerian'' window in which the correlator has anomalous exponents that are computable within Eulerian GHD, and (III)~diffusive regime, in which diffusive corrections are important. In the limit of zero magnetization, regime~(II) loses all of its spectral weight to regime~(III). Right panel adapted from S. Gopalakrishnan et al., \textit{Proc. Nat. Acad. Sci. USA} {\bf 116}, 16250 (2019).}
\label{gappedxxz}
\end{center}
\end{figure}

The easy-axis XXZ model is the simplest example of an integrable model in which some conserved densities (in this case the magnetization) do not undergo ballistic transport. The reason ballistic transport is absent is easy to understand in the $\Delta \gg 1$ limit. To simplify matters we consider ferromagnetic interactions and work at low but nonzero temperature. A typical equilibrium configuration has randomly distributed domain walls at some low density set by $T$; the domain sizes are distributed as $P(\ell) = \xi^{-1} \exp(-\ell/\xi)$, where $\xi \sim e^{\Delta/T}$ is the typical domain size, i.e., the equilibrium correlation length. In this large-$\Delta$ limit, the only domains that are mobile are the rare domains (``magnons'') consisting of a single $\uparrow$ spin in a sea of $\downarrow$ spins (or vice versa). Transport is dominated by these magnons. Let us consider an $\uparrow$ magnon moving through a large $\downarrow$ domain. Eventually it will reach the end of the domain. At this point, there are two kinematically allowed processes: either it can reflect back, or it can propagate through the next ($\uparrow$) domain as a minority $\downarrow$ spin (see Fig.~\ref{gappedxxz}). Integrability implies that only the latter process has a matrix element. Total magnetization is conserved because the new domain ``swells'' by two sites to accommodate the magnon. The upshot is that when a magnon crosses domain walls it is \emph{stripped} of its magnetization~\cite{PhysRevB.92.214427}; thus, over large distances the magnon is a magnetically neutral quasiparticle, transporting energy but not magnetization. Away from half filling the magnon transports some spin, which is proportional to the average magnetization of the state.

This basic intuition survives away from the large-$\Delta$ limit. At infinite temperature, a closed-form solution exists for the dressed quasiparticle magnetization, i.e., the amount by which the energy of the magnon state changes in response to an infinitesimal field. It will turn out (for reasons that we discuss below) that it is useful to work at finite chemical potential $h$ (or equivalently at finite net magnetization) and take the $h \rightarrow 0$ limit at the end. We quote the asymptotic behavior of the dressed magnetization, occupation factor, and dressed velocity of an $s$-string~\cite{Ilievski18}:
\beq\label{gappedxxzdata}
m_s^{\mathrm{dr}} = \left\{\begin{array}{lr} \tfrac{1}{3}h(s+1)^2 & sh \ll 1 \\ s & sh \gg 1 \end{array}\right. \qquad n_s = \left\{\begin{array}{lr} \frac{1}{(s+1)^2} & sh \ll 1 \\ e^{-sh} & sh \gg 1 \end{array}\right. \qquad v_s^{\mathrm{eff}} = e^{-\eta s},
\eeq
where in the last expression $\Delta = \cosh \eta$. Note that the anisotropy $\Delta$ shows up only in the expression for the dressed velocity: otherwise, the high-temperature thermodynamics of the easy-axis XXZ model is the same for all $\Delta > 1$.

Two instructive quantities to evaluate are the static spin susceptibility and the Drude weight. The former is given by the expression 
\beq\label{susceptibility}
\chi = \sum_s n_s (1 - n_s) (m^{\mathrm{dr}}_s)^2 \int \dd\lambda\, \rho^{\mathrm{tot}}_{s,\lambda} = \frac{1}{4}.
\eeq
This result can, of course, be computed by elementary methods at infinite temperature. 
However, it is instructive to consider the structure of the sum over strings~\eqref{susceptibility}. Since $n_s$ is exponentially suppressed for $s h \geq 1$, we can cut off the sum at $s h \approx 1$. Crucially, this sum \emph{cannot} depend on $h$ although each of the summands scales as $h^2$. This is only workable if the summands diverge as $s$, so that $\chi = h^2 \sum_{s \leq 1/h} s = O(1)$. 
Thus the expression in terms of strings reveals an important feature: in the limit $h \rightarrow 0$, the static susceptibility sum rule must be dominated by strings with $s h \approx 1$; much smaller strings effectively become neutral, as our intuitive argument suggested.

We now turn to the Drude weight. This is given by a very similar expression, but with added factors of the dressed velocity:
\beq
\mathcal{D} = \sum_s n_s (1 - n_s) (m^{\mathrm{dr}}_s)^2 \int d\lambda \rho^{\mathrm{tot}}_{s,\lambda} \left|v^{\rm dr}_{s,\lambda}\right|^2 \propto h^2.
\eeq
Because of the exponentially decaying velocities, the sum over $s$ in the Drude weight converges, so the Drude weight vanishes as $h \rightarrow 0$ within GHD. The physics of this is that spin is almost entirely transported by large-$s$ strings, which are immobile at the Euler scale.

As the susceptibility calculation showed, the $h \rightarrow 0$ limit is delicate: a naive application of GHD exactly at half-filling would yield the (obviously erroneous) conclusion that $\chi = 0$. This reflects a certain formal difficulty with applying GHD for the XXZ model at zero magnetization, which will arise again in the context of nonabelian spin chains. The issue is as follows: GHD formally splits the system up into cells of a certain size $\ell$, and takes the system to be described by a typical thermal Bethe eigenstate inside this cell. For many models, the state is fully specified given a quasiparticle distribution. However, in the XXZ model the quasiparticle distribution does not uniquely fix the state: rather, there are two Bethe vacua (all up or all down) from which one could have started, so to fully specify the state in a cell one needs to specify both the vacuum and the quasiparticle distribution. However, GHD as usually formulated gives equations of motion only for the quasiparticle distribution, \emph{not} for the local vacuum. Thus, if one fixes a cell size $\ell$ at the outset, GHD must be supplemented by additional data. 

A more mundane way of explaining this obstruction is that since strings come in all sizes, if we chop our system into $\ell$-sized blocks we will disrupt the structure of strings with $s > \ell$. The apparent ``vacuum'' of an $\ell$-sized region has dynamics because of the motion of such ``giant strings'' through it. 
Formally, we will largely avoid this issue within GHD by working at fixed nonzero magnetization $\propto h$, and assuming $\ell \gg 1/h$, so that the density of giant strings is exponentially suppressed and they do not matter for dynamics. 

\subsubsection{Diffusion constant}

Since the Drude weight vanishes, the leading transport behavior at half filling is diffusive. This diffusive behavior is, again, easiest to understand in the limit of low $T$ and very large $\Delta$. Suppose the correlation length (typical domain size) is $\xi$. Then a magnon carries a current of order unity for a time $\sim \xi$, while it is propagating through this domain. At later times, the magnon is moving through a domain whose magnetization is uncorrelated with the magnetization at $(0,0)$, so the magnon does not carry any net magnetization. The d.c. conductivity therefore goes as $\int \dd t \langle J(t) j(0,0) \rangle^{c} \propto \xi$. By the Einstein relation this gives a finite diffusion constant.

This argument can be made much more general, as follows~\cite{PhysRevLett.119.080602,spohnintnonint}. Consider the finite-time Drude weight, defined as $\mathcal{\tilde{D}}(t) \equiv \frac{1}{2t} \int_{-t}^t dt' \langle J(t') j(0,0) \rangle^{c}$. Clearly $\mathcal{\tilde{D}}(t) \rightarrow \mathcal{D}$ in the infinite-time limit. Evidently, $\mathcal{\tilde D}(t)$ closely resembles the finite-time d.c. conductivity $\tilde\sigma(t) = (\beta/2) \int_{-t}^{t} dt' \langle J(t') j(0,0) \rangle^{c} = 2t \, \mathcal{\tilde D}(t)$. We will now use this relation to bound the d.c. conductivity.

Since the operator $J$ commutes with total magnetization we can write $\mathcal{\tilde D}(t) = \sum_x P(h = x) \mathcal{\tilde D}(t, x)$, where $P(h = x)$ is the probability that the system is in a sector of total magnetization $m(h)$---re-expressed in terms of the conjugate chemical potential $h$---and $\mathcal{\tilde D}(t, h)$ is the finite-time Drude weight in that sector. Note that $\mathcal{\tilde D}(t,h)$ is positive-semidefinite, so the full sum is lower-bounded by any of its partial sums. The crucial next step in this argument is to note that, for all dynamical purposes, one can truncate the system to a region of size $v_{\rm LR} t$, where $v_{\rm LR}$ is the Lieb-Robinson velocity. Thus, we have $\mathcal{\tilde D}(t) = \sum_h (v_{\rm LR} t)^{1/2} \exp(-h^2 v_{\rm LR} t) \mathcal{\tilde D}(x,t)$. Assuming (plausibly) that $\mathcal{\tilde D}(t,h) \agt \mathcal{D}(h)$ (i.e., that the time-averaged current-current correlator monotonically decays with averaging time), and writing $\mathcal{D}(h) = h^2 \partial_h^2 \mathcal{D}(h)$, we end up with the bound
\beq
\sigma(\beta, h=0) \geq \beta \, \partial_h^2 \mathcal{D}(h)\vert_{h = 0}.
\eeq
%
Note that this bound is loose, because the Drude weight is a sum over quasiparticle species, and the Lieb-Robinson speed grossly overestimates how far most quasiparticles have traveled. One can parametrically improve this bound as follows~\cite{GV19}. Let us decompose 
\beq
\mathcal{D}(h) = \sum_s \int \dd\lambda \, \mathcal{D}(h, s, \lambda).
\eeq
We now follow the previous argument up to the step where we averaged physical quantities over a region of size $L = v_{\rm LR} t$. Here, we tighten the estimate by introducing a quasiparticle-specific distance scale $L_{s, \lambda} = v_{s, \lambda} t$. Plugging this in, we find
\beq
\mathcal{\tilde D}(t) \geq \sum_{h, s} \int d\lambda (v_{s, \lambda} t)^{1/2} \exp(-h^2 v_{s,\lambda} t) \mathcal{D}(h, s, \lambda). 
\eeq
Combining this with our expression for the Drude weight, we get the final result
\beq\label{lbdiff}
\sigma \geq \beta \sum_s \int d\lambda \rho^{\mathrm{tot}}_{s,\lambda} n_{s,\lambda} (1 - n_{s,\lambda}) (m^{\mathrm{dr}}_{s,\lambda})^2 |v^{\mathrm{eff}}_{s,\lambda}|.
\eeq
As we have derived it, this result is a lower bound, since it allows for the possibility of other mechanisms for diffusion that we did not include. 

Remarkably, this bound is saturated in the XXZ model, as one can verify by an explicit calculation~\cite{NMKI19}. We will very briefly outline the logic of this calculation. There are two crucial observations: (1)~at small $h$, magnetization is transported entirely by very heavy strings $s h \sim 1$, since they saturate the magnetic susceptibility sum rule, and (2)~in the easy-axis regime, these strings have exponentially suppressed velocities. Thus they move essentially via Brownian motion from collisions with lighter strings. It follows that to obtain the full conductivity one must simply compute the Brownian motion of each quasiparticle, which is given by Eq.~\eqref{intdiffusion} above. The conductivity can be written as
\beq\label{dndiff}
\sigma = \beta \chi \sum_{s'} \int \dd \lambda' \, |v^{\rm eff}_{s', \lambda'}| \rho^{\mathrm{tot}}_{s',\eta} n_{s',\lambda'}(1-n_{s',\lambda'}) \lim_{s \rightarrow \infty} \left[\mathcal{K}^{\mathrm{dr}}_{s;s'\lambda'}/\rho^{\mathrm{tot}}_{s}\right]^2.
\eeq
Comparing Eqs.~\eqref{lbdiff},~\eqref{dndiff} we see that they would be identical if the following equality held: 
\beq
\lim_{s \rightarrow \infty} \left[\mathcal{K}^{\mathrm{dr}}_{s;s'\lambda'}/\rho^{\mathrm{tot}}_{s}\right]^2 = (1/\chi) \lim_{h \rightarrow 0} (m^{\mathrm{dr}}_{s}(h)/h)
\label{eqn:magic_formula}
\eeq 
Perhaps surprisingly, this identity (termed the ``magic formula'' in Ref.~\onlinecite{NMKI19}) is exact. Thus the \emph{only} mechanism for spin diffusion in the easy-axis XXZ model is the quasiparticle screening mechanism outlined above.

%


\subsubsection{Dynamic correlation function at finite net magnetization}\label{finitemag}

We now turn to the behavior away from half filling, focusing on the dynamical density correlation function~\cite{GVW19}, $C(x,t) = \langle S^z(x,t) S^z(0,0) \rangle^{c}$, which one can write as
\beq
C(x,t) = \sum_s \int \dd \lambda \, \rho^{\mathrm{tot}}_{s,\lambda} n_{s,\lambda} (1-n_{s,\lambda}) (m^{\mathrm{dr}}_{s,\lambda})^2 \sqrt{D_{s,\lambda} t} \exp\left[-\frac{(x - v_{s,\lambda} t)^2}{2 D_{s,\lambda} t}\right].
\eeq
To simplify our discussion we specialize to magnetization $h \approx 1$ and to $T = \infty$, so we can replace all TBA quantities with their large-$s$ asymptotics. This reduces the above expression to
\beq\label{sfactor}
C(x,t) = \sum_s s^2 e^{-h s} \int d\lambda \, \rho^{\mathrm{tot}}_{s,\lambda} \sqrt{D_{s,\lambda} t} \exp\left[-\frac{(x - v_{s,\lambda} t)^2}{2 D_{s,\lambda} t}\right].
\eeq
%
%
Spin transport away from half filling is dominated by the ballistic spreading of light strings. However, this is not necessarily true for the \emph{local} autocorrelator, $C(0,t)$ (or more generally for $C(x,t)$ at fixed $x$ in the limit $t \rightarrow \infty$). 

To understand the asymptotics of this quantity, we note that light (small-$s$) strings spread ballistically, so their probability of remaining at the origin scales as $\sim 1/(v_s t) \sim e^{\eta s} / t$. Thus, the summand in Eq.~\eqref{sfactor} contains a factor $e^{(\eta - h) s}$, leading to two sharply distinct regimes of behavior. If $\eta < h$, heavy strings are too rare to affect the leading decay of the local autocorrelator, which instead goes as $1/t$ from the light strings. However, if $\eta > h$, the autocorrelator is dominated by very heavy strings. As discussed in the previous section, these very heavy strings move diffusively, with a diffusion constant $D$ that is $s, \lambda$-independent. However, if one waits long enough any string's motion is primarily ballistic. The crossover from primarily ballistic to primarily diffusive motion happens when $e^{-\eta s} t = \sqrt{D t}$, i.e., $s^* = \frac{1}{2} \eta$. In this $\eta > h$ regime, at late times the autocorrelator near the origin is dominated by diffusive strings, and scales as 
\beq
C(0,t) \simeq \frac{1}{\sqrt{Dt}} \sum_{s > s^*} e^{-h s} \sim t^{-\frac{1}{2} - \frac{h}{2\eta}}.
\eeq
Thus, the autocorrelation function has multiple nontrivial spatio-temporal regimes (Fig.~\ref{gappedxxz}), with ballistic motion near the light-cone coexisting with anomalous decay of local autocorrelation functions. This type of ``mixed'' behavior, with a ballistic front coexisting with a large slow tail, has been seen in multiple other contexts recently, including random integrable spin chains~\cite{PhysRevB.99.174203, PhysRevB.98.024203} and the transverse-field Ising model with random hyperuniform couplings~\cite{PhysRevB.100.134206}. 

\subsection{The easy-plane XXZ model}\label{gaplessxxzsec}

The structure of conductivity in the easy-plane axis is superficially more natural, as there is ballistic transport even in the half-filled sector. However, as discussed in the Introduction, the Drude weight is a highly nontrivial, nowhere-smooth, function of the anisotropy $\Delta$. In what follows we will discuss the reasons for such a seemingly pathological behavior of the Drude weight, and its implications for finite-time dynamics. We will find that the finite-time response is anomalous, and appears to be described by a quasiparticle L\'evy flight~\cite{Agrawal20}.



\subsubsection{Fractal spin Drude weight}


A key advance in our understanding of the easy-plane XXZ model was the discovery of a one-parameter analytic family of quasilocal conservation laws (called nowadays the ``$Z$-charges''), including the charge of Ref.~\onlinecite{prosen_drude}, which were constructed in Ref.~\onlinecite{PhysRevLett.111.057203} (afterwards adapted to periodic boundary conditions~\cite{ProsenNPB,1742-5468-2014-9-P09037}).
Optimizing a Mazur-Suzuki bound (Sec.~\ref{drudesec}) using these charges yields the following striking prediction.
%
%
Parameterizing $\Delta = \cos{(\gamma)}$ with $\gamma/\pi \equiv m/\ell \in [0,\pi)$, for $m$ and $\ell$ being co-prime integers, one obtains the following Mazur--Suzuki bound \cite{PhysRevLett.111.057203}
\begin{equation}
\mathcal{D}_{\rm spin} \geq \frac{\beta}{16}\frac{\sin^{2}(\pi m/\ell)}{\sin^{2}(\pi/\ell)}
\left[1-\frac{\ell}{2\pi}\sin{\left(\frac{2\pi}{\ell}\right)}\right].
\label{eqn:Drude_bound}
\end{equation}
This bound is in fact in precise agreement with an earlier analytical TBA calculation\cite{Zotos99} specialized to a set of isolated primitive roots of unity ($m=1$). It nonetheless remained unclear at the time whether the computed ``fractal dependence''~\footnote{To be precise, the dependence on $\Delta$ is not truly fractal (nor is it a Weierstrass function) but rather what is known as a nowhere continuous function.} was actually tight and if so, whether such a result is even physical. As we explain next, this uncertainty has since been settled by reobtaining and reconciling the result with an underlying microscopic description~\cite{IN_Drude}.

\medskip

\paragraph*{Spin Drude weight from generalized hydrodynamics.}
As described in Sec.~\ref{background}, the GHD framework allows one to compute Drude weights directly. For the easy-plane XXZ model, this calculation was first undertaken in Refs.~\onlinecite{IN_Drude, Bulchandani18} and refined in a number of subsequent works \cite{IN17,DS17,KlumperDrude}. Since the GHD approach relies essentially on long-established results of the thermodynamic Bethe ansatz, it is \emph{a priori} surprising that it seems to capture the spin Drude weight to within numerical accuracy~\cite{Bulchandani18,Karrasch_2017,IN17}. That GHD (equivalently, TBA) can capture the necessary quasilocal charges was demonstrated in Ref.~\onlinecite{IN_Drude}, by invoking the notion of ``string-charge duality'' (Sec.~\ref{drudesec}). The approach of Refs.~\onlinecite{IN_Drude,Bulchandani18} provides an independent numerical check on Eq.~\eqref{eqn:Drude_bound}, thus demonstrating that discontinuities persist at any finite temperature.
These formulae have been afterwards further improved and eventually superseded by compact closed-form expressions~\cite{DS17,IN17}.
A closed-form solution to the associated Riemann problem in the high-temperature limit was afterwards derived in Ref.~\onlinecite{PhysRevB.97.081111}, thus recovering the anticipated exact result \eqref{eqn:Drude_bound}. The same result has later been also rederived within the Quantum Transfer Matrix formalism~\cite{KlumperDrude}, which offers an alternative, but equivalent~\cite{IQ19}, framework for performing thermodynamic computations in Bethe Ansatz solvable models.


\paragraph*{Quasiparticle spectrum.}
To understand where the subtleties of the easy-plane (gapless) regime come from, we now take a closer look at the quasiparticle spectrum. The latter has a much richer structure compared to the easy-axis phase, which can be traced back to the representation theory of the quantum algebra $\mathcal{U}_{q}(\mathfrak{su}(2))$ with a unimodular quantum deformation parameter $q=e^{\ii \gamma}$. The complete description of the thermodynamic quasiparticle spectrum goes all the way back to Ref.~\onlinecite{Takahashi1971}, which singled out a stability condition for formation of bound magnon excitations (using the fusion relations for the scattering amplitudes), finding that both the total number and internal structure of allowed bound states depend on the interaction parameter $\gamma$ in a rather intricate fashion. By expanding the interaction parameter $\gamma$ as the continued fraction $\gamma/\pi = 1/(\nu_{1} + 1/(\nu_{2} + 1/(\nu_{3}+\ldots)))$, the number of distinct quasiparticle species is obtained as $\nu = \sum_{i}\nu_{i}$. Therefore, at roots of unity, only \emph{finitely} many species are protected against decay.

The stability criterion indeed changes in a discontinuous manner, causing an abrupt reorganization of the quasiparticle spectrum upon varying the interaction parameter $\gamma$.~\footnote{It is essential to respect the correct \emph{order of limits}: by first fixing $\Delta$, one has to take the thermodynamic limit prior examining the dependence of thermodynamic quantities on $\Delta$. In contrast, in any finite (but arbitrarily large) system, both energy levels and eigenstates obviously depend continuously upon $\Delta$.}
To give a glimpse, let us here focus specifically on the vicinity of commensurate points at primitive roots of unity $\gamma/\pi = 1/\nu_{1}$, where discontinuities are most pronounced. The simplest way to approach these isolated points is to follow a sequence $\gamma=1/\nu_{1}+1/\nu_{2}$ by increasing $\nu_{2}$. For large values of $\nu_{2}$, the thermodynamic excitations will mostly comprise ``very heavy'' bound states. For example, for $\nu_{1}=3$ (corresponding to $\Delta = 1/2$) the heaviest bound states at, say, $\nu_{2}\in \{10,20,40,100\}$ are made out of $\{28,58,118,298\}$ magnetization quanta, in respective order. On the other hand, in a chain of finite length $L$, magnetization carried by quasiparticles cannot exceed $L/2$, indicating that such heavy strings do not even fit into systems of size $L\sim 30$ (roughly within reach of present-day exact diagonalization). Whether this has any significance for thermodynamic quantities or physical observables is difficult to estimate a priori. This is essentially controlled by the discarded integrated spectral weight attributed to such heavy strings. We briefly remind the reader that in the case of the Drude weights, individual quasiparticle contributions are weighted by their static susceptibilities $\rho_{s,\lambda}(1-n_{s,\lambda})$. It may happen that the net weight carried by such heavy strings is small. It is instructive to add here that all quasiparticle contributions to the free energy density can be explicitly resummed, yielding a final expression that exhibits a continuous dependence on $\Delta$. A more detailed exposition of the mathematics behind these results (including Baxter $Q$-operators, Wronskian relations and Fabricius--McCoy complete strings), which further corroborates the above findings, can be found in
Ref.~\onlinecite{Miao20}.

\medskip

\paragraph*{Dressed magnetization and velocity.} We now fix some generic rational value of $\gamma$, with a large but finite denominator $\ell$, and summarize the properties of the quasiparticles that are present for that specific $\gamma$. For a generic rational number with denominator $\ell$, there are $\sim \log \ell$ different quasiparticle species. At half filling, one finds (see, e.g., Ref.~\onlinecite{Agrawal20}) that all but two of these quasiparticles have zero dressed magnetization, while the dressed magnetization of the two largest strings is unrenormalized, and scales as $\ell$. The \emph{velocity} of these large strings, meanwhile, saturates to a constant, $\ell$-independent value. Since current is carried by large but ballistic quasiparticles, the Drude weight is nonzero. (This is in contrast to the isotropic and easy-axis regimes, in which the velocities of large quasiparticles vanish; we comment further on this point in Sec.~\ref{simplesec}.)

\medskip

\paragraph*{Numerical evidence.}
The outlined exceptional structure of the quasiparticle spectrum in the easy-plane (gapless) phase renders numerical computation of the spin Drude weight a particularly delicate issue.  The task proves very challenging even for current state-of-the-art methods, due to intrinsic limitations tied to finite-time (or finite-size) effects~\cite{PhysRevB.84.155125, PhysRevB.77.161101, PhysRevB.84.155125, PhysRevLett.108.227206, PhysRevB.96.245117, Ljubotina_nature, PhysRevB.97.081111, Agrawal20,Marcin_ballistic}.
Several discernible features of discontinuous behavior have
nevertheless been successfully demonstrated in recent DMRG simulations~\cite{PhysRevB.96.245117, Ljubotina_nature}. We would like to specifically highlight Ref.~\onlinecite{LZP19}, which demonstrates that upon increasing the time-averaging window in an integrable Trotterization of the XXZ Heisenberg model, a smeared-out spin Drude weight gradually collapses towards the discontinuous analytical prediction.

\subsubsection{Anomalous diffusive corrections and a.c. conductivity}

The Drude weight in the easy-plane regime changes by an $O(1)$ amount for an infinitesimal change in the anisotropy $\Delta$. This feature might seem unphysical at first sight, but it is not: for example, the Drude weight similarly drops by an $O(1)$ amount, all the way to zero, if one takes an integrable spin chain and adds an infinitesimal integrability-breaking perturbation. However, \emph{at finite times} the system cannot ``resolve'' $\Delta$ to arbitrary accuracy, so two points in the easy-plane phase with very similar $\Delta$ must have similar values of the correlator $\langle J(t) J(0)\rangle^{c}$ up to some crossover time $t$. By systematically analyzing this crossover time $t$, Ref.~\onlinecite{Agrawal20} was able to obtain the asymptotic behavior of the low-frequency conductivity. We now briefly summarize this logic.

The argument of Ref.~\onlinecite{Agrawal20} is built on the assumption that the correlator $\langle J(t) J(0) \rangle^{c}$ is monotonically decreasing with time. This assumption is consistent with the numerical evidence. We consider a generic irrational number $\varphi$---for concreteness chosen to be the Golden Ratio---and approximate it by a series of rational approximants $\varphi_\ell$ with increasingly large denominators $\ell$. (For the Golden ratio these approximants are ratios of Fibonacci numbers.) The expression~\eqref{eqn:Drude_bound} shows that the Drude weight decreases for larger denominators, as $1/\ell^2$. Now consider the approximant $\varphi_\ell$ with denominator $\ell$. Until a crossover time $t^*(\ell)$, the current-current correlator $\langle J(t) j(0,0) \rangle^c_\varphi \approx \langle J(t) j(0,0) \rangle^c_{\varphi_\ell} \geq \mathcal{D}(\varphi_\ell)$. Thus the correlators for rational approximants to some irrational number $\varphi$ must follow the irrational behavior until some time $t^*(\ell)$, which increases with $\ell$, and then saturate to their Drude-weight value.

\begin{figure}[tb]
\begin{center}
\includegraphics[width = 0.7\textwidth]{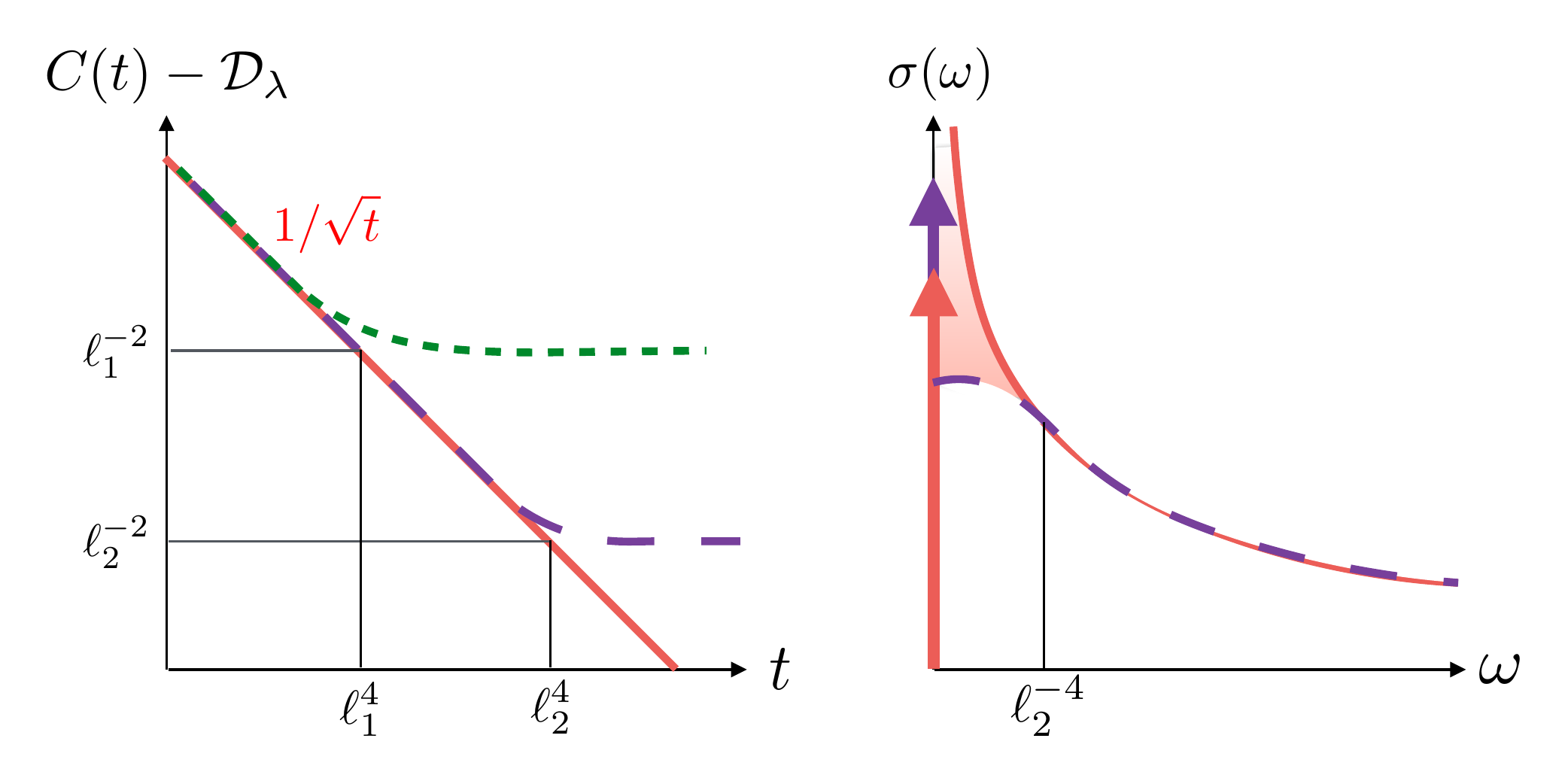}
\caption{Behavior of the current-current correlator (left) and the a.c. conductivity (right) as one tunes to the irrational point through a sequence of rational approximants with increasing denominators. At rational points, there are finitely many quasiparticles and the conductivity divergence is cut off at small $\omega$ leading to an enhancement of the Drude weight (by the sum rule). Reproduced from U. Agrawal et al., \textit{Phys. Rev. B} \textbf{101}, 224415 (2020).}
\label{gaplessxxzfig}
\end{center}
\end{figure}

\begin{figure}[tb]
\begin{center}
\includegraphics[width = 0.7\textwidth]{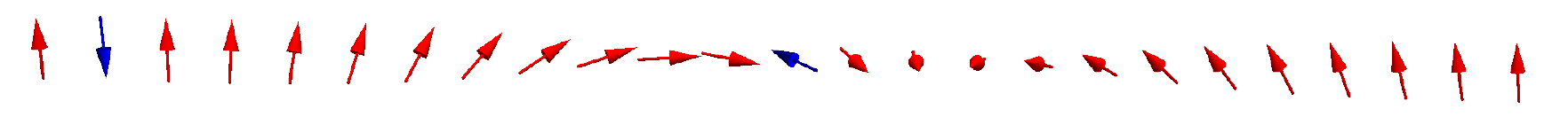}
\caption{Schematic illustration of two magnons (blue) and a large string in the Heisenberg model. The string looks locally like a ferromagnetic vacuum; the magnon is always oriented opposite to the local vacuum, and thus changes its magnetization as it  collides with large strings.}
\label{magnonsketch}
\end{center}
\end{figure}

We can use these very general observations to derive a series of scaling relations. Let us anticipate that $\langle J(t) j(0,0) \rangle^c_\varphi - \mathcal{D}_\varphi \sim 1/t^{1-\alpha}$, where $\alpha$ is an exponent that remains to be determined. By continuity, the correlators at a rational approximant $\varphi_\ell$ must follow this curve until $t^*(\ell)$, and then saturate to the Drude value. Since the curves are continuous and (by assumption) monotonic, we find that $(t^*(\ell))^{\alpha-1} \sim 1/\ell^2$, so $t^*(\ell) \sim \ell^{2/(1-\alpha)}$. Also, since the correlator for $\varphi_\ell$ followed the power law until time $t^*(\ell)$, we can find the d.c. limit of the conductivity as $\int_0^{t^*(\ell)} \dd t/t^{\alpha-1}$. This then yields that
\begin{equation}
\sigma^{\mathrm{d.c.}}_\ell \sim \ell^{2\alpha/(1-\alpha)}.
\end{equation}
The practical significance of this last expression is that, purely by computing the d.c. limit of the regular conductivity as a function of the approximant $\varphi_\ell$ (a straightforward calculation within GHD, using the results of Ref.~\onlinecite{DeNardis_SciPost}), one can fix the exponent $\alpha$ that sets the \emph{a.c. conductivity} at the irrational point. This allows us (surprisingly) to compute a.c. transport properties purely in terms of GHD data. 
Finally, a straightforward GHD calculation fixes $\alpha = 1/2$ in the relations above; thus, $\sigma(\omega) \sim \omega^{-1/2}$. The spectral ``bookkeeping'' between the Drude peak and the finite-frequency conductivity is illustrated in Fig.~\ref{gaplessxxzfig}. 

The mechanism underlying this anomalous correction turns out to be a quasiparticle L\'evy process, which one can understand as follows. As we noted above, only the largest two strings in this model carry a dressed magnetization at half filling. Diffusion is due to the broadening of these ``charged'' quasiparticles~\eqref{intdiffusion} due to their collisions with the smaller, neutral quasiparticles. For larger denominators, there are increasingly many species of neutral quasiparticles to scatter off. The larger species of neutral quasiparticle are rarer, but their scattering shifts are also larger, and the latter effect dominates: if we think of the charged quasiparticle as undergoing (in addition to its ballistic propagation) Brownian motion due to its scattering off the lighter quasiparticles, the \emph{step size} has a fat-tailed distribution, so the charged quasiparticle undergoes a L\'evy process as discussed in Sec.~\ref{griffiths}. This L\'evy-process interpretation has been used to independently compute the crossover time vs. denominator $\ell$ as $t^*(\ell) \sim \ell^4$ (confirming the general scaling relations above)~\cite{Agrawal20}. 

Three further comments on this result are in order. First, $\alpha = 1/2$ is \emph{not} the result one would get if one made the simplest assumption about crossover timescales: namely, that for a perturbation $\delta$, the crossover time should be $t \sim 1/\delta$ by the uncertainty theorem. Since one can achieve an error $\sim 1/\delta$ with a rational approximant that has denominator $\ell \sim 1/\sqrt{\delta}$ (by Dirichlet's approximation theorem), this simple theory would predict $\alpha = 0$, which is \emph{not} the GHD result. Rather, it seems that perturbing the anisotropy by an amount $\delta$ gives a crossover timescale $t \sim 1/\delta^2$. This quadratic behavior is suggestive of Fermi's Golden Rule, but no concrete connection has yet been shown and finding a general \emph{a priori} argument for this exponent remains an open question.

Second, it is interesting to consider how this argument is modified if, instead of approximating a generic irrational point in the easy-plane regime, we attempt to approximate the Heisenberg point via a series of roots of unity. At finite $\ell$, the Drude weight $\mathcal{D}_\ell \sim 1/\ell^2$~\eqref{eqn:Drude_bound}. However, calculating the d.c. conductivity yields $\sigma(\ell) \propto \ell$, yielding (through the scaling relations) $\alpha = 1/3$, which is the correct superdiffusive exponent at the isotropic point.

We remark, finally, that the exponent $\alpha = 1/2$ has also been confirmed numerically, in particular in Ref.~\onlinecite{LZP19}, which computed the finite-time current-current correlator vs. $\Delta$ and $t$. It was shown that the finite-time Drude weight (integrated up to a time $t$) has peaks at small-denominator rationals, with width $\sim t^{-1/2}$. This width follows immediately from our argument above that the crossover timescale changes with a perturbation of $\delta$ as $t \sim 1/\delta^2$: consequently if one fixes $t$, the width in $\delta$ scales as $1/\sqrt{t}$, as observed. Establishing whether the front indeed broadens as a L\'evy process is outside the scope of current numerical techniques, but is an interesting question for future numerical work.

\subsection{Beyond spin transport}

In contrast to spin transport, the transport behavior for other ($\mathbb{Z}_2$-invariant) conserved charges in the XXZ model is conventional, i.e., characterized by a Drude weight that in general varies smoothly with anisotropy, with no special features at $\Delta = 1$. In general, the transport of these charges is dominated by magnons and small strings, which do not change their character with $\Delta$.  Although the general formulae are known, the diffusion constants for these charges have not been explicitly computed to our knowledge, but there seems no reason to expect that these exhibit any anomalous features. For the same reasons, the growth of entanglement after a quench should remain linear in time across the transition~\cite{alba2017entanglement}. One point that merits clarification is the broadening of operator fronts (as captured, e.g., by the out-of-time-order correlator) discussed in Ref.~\onlinecite{Gopalakrishnan18}. The shape of operator fronts is governed by the diffusive broadening of the fastest quasiparticles, i.e., maximum-velocity magnons. One might wonder if these magnons are anomalously broadened by collisions with large strings. This is not, however, the case~\cite{DeNardis_SciPost}: the scattering phase shift for a collision between a magnon and an $s$-string does not diverge with $s$. Therefore, one expects conventional diffusive broadening of fronts~\cite{Gopalakrishnan18, GV19}, in both phases and also at the isotropic point. 

It is still possible to detect the effects of giant magnons using $\mathbb{Z}_2$-even observables, at least in the easy-axis phase, by computing the behavior of the corresponding dynamic correlation functions along spacetime rays where $x = \alpha t$, $\alpha \to 0$~\cite{GVW19}. These will be dominated by contributions from large, slow-moving strings, as discussed in Sec.~\ref{finitemag}. An information-theoretic quantity that will similarly be dominated by large slow strings is the mutual information between two disjoint regions, in the same scaling limit~\cite{PhysRevB.100.115150}. 

\section{Isotropic Heisenberg chain}\label{isoXXZ}

We now finally turn to the isotropic point $\Delta = 1$ of the Heisenberg spin chain. From the results of the previous section it is clear that something nontrivial must happen here: for $\Delta < 1$ spin transport is ballistic at half filling, for $\Delta > 1$ it is diffusive. Thus the isotropic Heisenberg point is a sort of \emph{dynamical critical point} between two ``phases'' of spin transport. At the same time, from the point of view of the Bethe ansatz quasiparticle structure, it is natural to think of this point as merely a special case of the easy-axis regime in the limit $\eta \to 0$. Although away from half filling one indeed finds regular behavior
(i.e., finite diffusion constants accompanied by finite Drude weights), one encounters \emph{superdiffusion} precisely at half filling. We already discussed how superdiffusion was discovered (see Introduction). In this section we provide a unified theoretical account of the origin of superdiffusion. We present, in turn, two different approaches to the problem---a \emph{microscopic} approach based on taking the $\eta \to 0, h \to 0$ limit in a self-consistent way~\cite{Ilievski18, GV19, NMKI19, GVW19}---and a \emph{macroscopic} approach that avoids explicit Bethe ansatz methods in favor of symmetry considerations~\cite{Vir20}. These approaches have complementary advantages, since the macroscopic approach gives more insight into the scaling functions and the relevance of the nonabelian symmetry, while the microscopic approach allows one to perform quantitatively accurate calculations. We finally discuss how these approaches can be unified, by taking an appropriate scaling limit of the thermodynamic Bethe ansatz equations~\cite{NGIV20}. 

\subsection{Quasiparticle content}

As we already saw in the easy-axis regime of this model, the quasiparticle content of a state does not completely specify local properties of thermodynamic eigenstates. One additional piece of information is needed, namely the vacuum on top of which the quasiparticles were created. (A different way to say this is that the many-body spectrum splits up into irreducible representations of $SU(2)$; within each $SU(2)$ multiplet, the Bethe ansatz describes only one particular state, which is the highest-weight state.) In both the easy-axis regime and at the isotropic point, the magnetization (and therefore the nature of spin transport) depends on the choice of vacuum, i.e., identical quasiparticle distributions above different vacua correspond to physically distinct states with different magnetization. When the Hamiltonian and/or the thermal equilibrium state breaks the symmetry, i.e., at nonzero magnetization and/or external field, this ambiguity is resolved. We address this general situation first, and then turn to the $SU(2)$ symmetric limit.


\subsubsection{Quasiparticles at finite magnetization}

We first consider the spectrum above the ferromagnetic ground state in the presence of a field. If we set the quantization axis to line up with the field, the ground state has the exact product form $|\downarrow\downarrow\ldots\rangle$, and there is a residual $U(1)$ symmetry for rotations around the quantization axis. (It is important to begin with the \emph{ferromagnetic} state here, regardless of the sign of the couplings in the microscopic Hamiltonian, since the Bethe ansatz vacuum---which we will also call the ``pseudovacuum''---is always a simple product state.) The elementary excitations above this state are \emph{magnons}, which can be understood as a single flipped spin, propagating with a quadratic dispersion; one can regard a magnon as a Goldstone mode with dynamical exponent~\cite{Sachdev_book} $z_0 = 2$.
In addition to magnons there exists an infinite series of bound states or \emph{$s$-strings}, each carrying magnetization $s$ (the magnetization here is the conserved charge under the $U(1)$ symmetry). 

Below, we will be concerned with the behavior of $s$-strings at large $s$. As we will see, these giant strings have a semiclassical interpretation as solitons made up of Goldstone modes. 
Their bare properties (which one can also explicitly compute using the Bethe ansatz~\cite{takahashi_book}) follow from this identification: an $s$-string is a slow vacuum rotation that takes place over a length-scale $\sim s$ and rotates the vacuum by an angle of order unity (Fig.~\ref{magnonsketch}). Thus an $s$-string carries bare magnetization $s$ relative to the vacuum. Its energy scales as $1/s$, because its size scales as $s$ while each bond along the soliton carries phase twist $1/s$ and therefore (because $z_0 = 2$) the soliton has energy density $1/s^2$. Finally, its characteristic group velocity scales as $1/s$, since it is made up of Goldstone modes with momenta $\leq 1/s$. 

Because of integrability, these magnons and $s$-strings persist as stable quasiparticles even when the system is at finite temperature, but their properties are heavily renormalized or ``dressed.'' The thermodynamics of the isotropic point at finite field are very similar to those of the easy-axis regime: in particular, small strings get depolarized by the process of passing through large strings (cf. Sec.~\ref{gappedsec}). The only major change is that the velocities now no longer fall off exponentially but as $1/s$.

At infinite temperature and fixed nonzero magnetization, one can express the Drude weight and the d.c. limit of the conductivity as follows:
\begin{eqnarray}
\mathcal{D} & = & \sum_s (m^{\mathrm{dr}}_{s})^2 n_s (1 - n_s) \int \dd\lambda \, \rho^{\mathrm{tot}}_{s,\lambda} \left|v^{\rm eff}_{s,\lambda}\right|^2 \label{heis_drude}. \\
\sigma & = & \beta \sum_s \left(\partial_h m^{\mathrm{dr}}_{s} \vert_{h \to 0} \right)^2 n_s (1 - n_s) \int \dd\lambda\, \rho^{\mathrm{tot}}_{s,\lambda} |v^{\rm eff}_{s,\lambda}| \label{heis_diff}.
\end{eqnarray}
Here, the thermodynamic quantities $n_s, m_s^{\mathrm{dr}}$ behave identically to the easy-axis phase (Sec.~\ref{gappedsec}), but the velocities decrease algebraically with $s$. Thus, the quantities $\int d\lambda \rho^{\mathrm{tot}}_{s,\lambda} |v^{\rm eff}_{s,\lambda}|^2 \sim 1/s^3$ and $\int d\lambda \rho^{\mathrm{tot}}_{s,\lambda} |v^{\rm eff}_{s,\lambda}| \sim 1/s^2$. At $h \neq 0$ we can cut off the sum over $s$ at $s \approx 1/h$, and use the $sh \ll 1$ scaling in Eq.~\eqref{gappedxxzdata} for all quantities. Then we see that $D \sim h^2 |\log h|$ and $\sigma \sim 1/h$. These nonanalyticities at $h = 0$ signal the onset of anomalous transport. In the next section we will use these scaling forms to estimate the transport exponent self-consistently.

\begin{figure}[tb]
\begin{center}
\includegraphics[width = 0.43\textwidth]{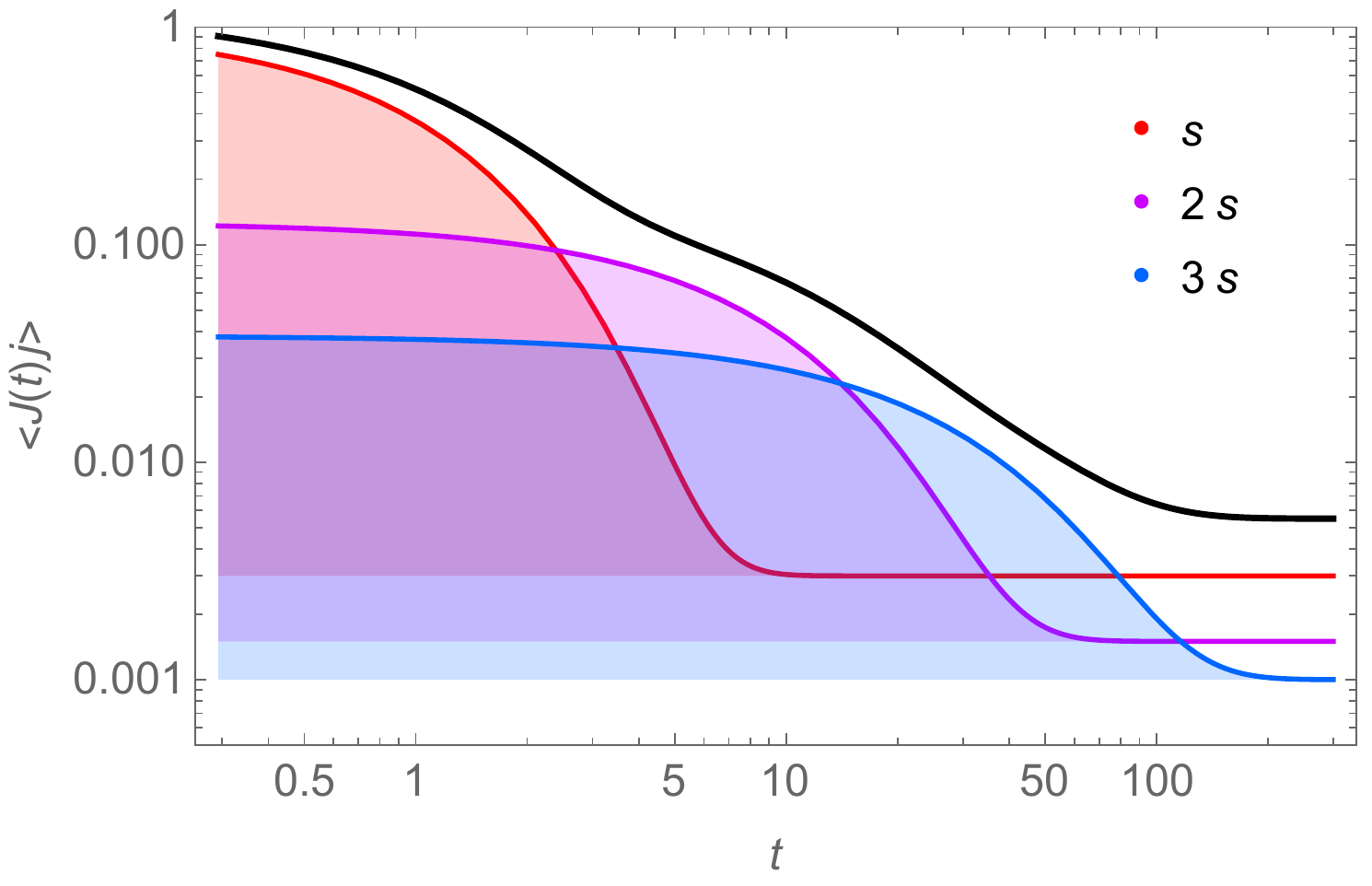}
\includegraphics[width = 0.38\textwidth]{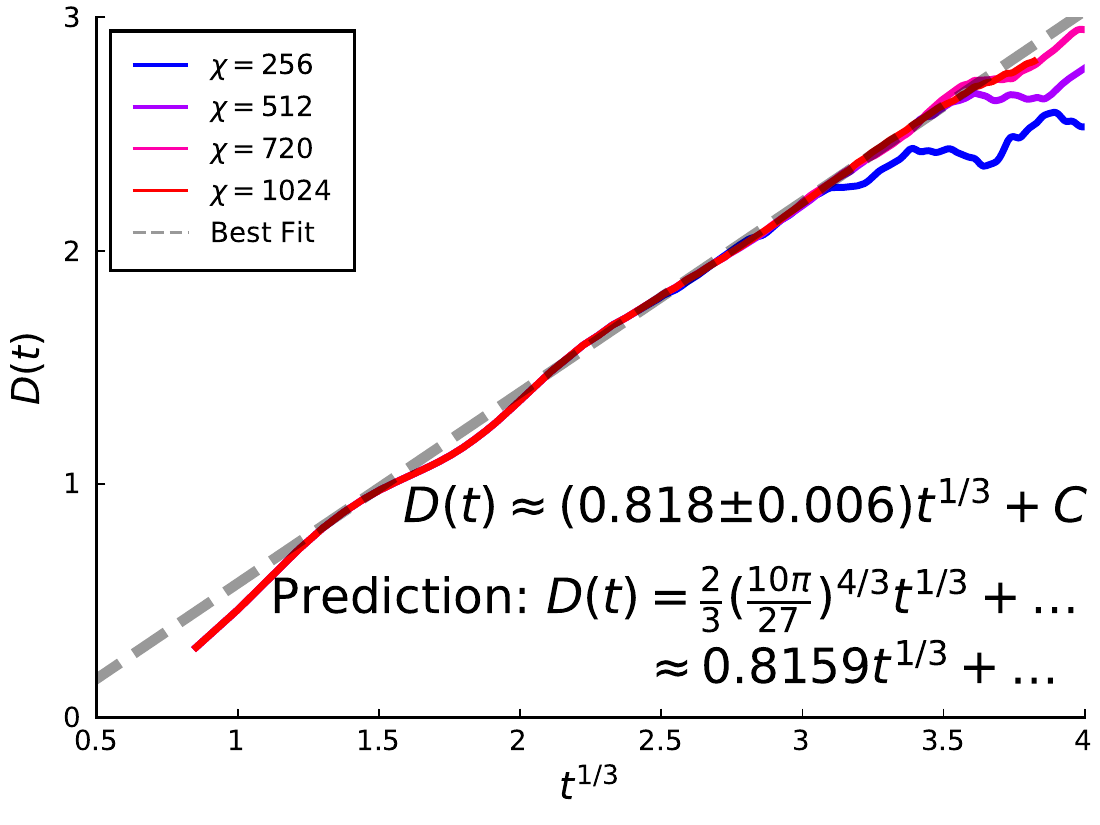}
\caption{Left: schematic illustration of the contributions to the current-current correlator of three large strings, of size $s, 2s, 3s$ respectively, at small finite $h$. Bigger strings are present at lower density, but carry current for longer. The late-time saturation value sets the Drude weight, while the area under the curve (shaded region) for each string is its contribution to the regular d.c. conductivity. Although each string has a finite relaxation time, summing over the strings gives a curve (solid black line) with nontrivial relaxation behavior. Units on both axes are arbitrary. Right: Comparison of $D(t)$~\eqref{dtdeff} extracted from TEBD simulations (at various choices of bond dimension) with the exact theory prediction (shown in figure). Reproduced from J. De Nardis \emph{et al.}, arXiv:2102.02219.}
\label{kubo_illustration}
\end{center}
\end{figure}

One can understand these expressions in a unified way by thinking about the Kubo formula, $\sigma(\omega) = \beta \int dt'\,e^{\ii \omega t'} \langle J(t) j(0,0) \rangle^{c}$. The current-current correlator can be written as a sum over strings. At short times, we expect each string to contribute as if it were not dressed (since it has not undergone any collisions yet). At long times, its contribution is constant and corresponds to the Drude weight. The d.c. conductivity is the area under the curve (Fig.~\ref{kubo_illustration}), which depends on the short-time value, the saturated value, and the saturation \emph{timescale}, which we will estimate in the next section. The conductivity diverges at low fields because of the increased prevalence of large strings, which carry currents for increasingly long times.

\subsubsection{Quasiparticles at zero magnetization}

We briefly comment now on the case of zero magnetization. Here, any finite-$s$ string is neutral (i.e. paramagnetic), and does not contribute to transport. Formally, transport is due entirely to the motion of the $s = \infty$ string. This raises a delicate conceptual issue: in the construction of GHD, one assumes that the state of the system can be described in terms of local generalized Gibbs ensembles on a mesoscopic length-scale $\ell$. Clearly this restriction is incompatible with the $s \to \infty$ limit since $\ell$ acts as a cutoff. Thus, to describe the dynamics of systems directly at zero magnetization, one must augment GHD with additional ``gauge'' degrees of freedom that describe fluctuations of the GHD vacuum. We turn to these gauge fluctuations in Sec.~\ref{pseudovacuum}. For linear response we are able to avoid directly confronting this issue, by working at finite $h$ and taking the $h \to 0$ limit at the very end. Note, however, that if one is to describe, e.g., the dynamics of nonequilibrium spin profiles with a magnetization profile that goes through zero, there does not seem to be any way to avoid explicitly invoking gauge degrees of freedom.

\subsection{Two simple arguments for superdiffusion}\label{simplesec}

We first present two simple arguments, constructed out of the GHD analysis above, that immediately recover the $z = 3/2$ exponent for superdiffusion (although not the KPZ scaling form). These arguments avoid subtleties related to the gauge degrees of freedom by working entirely at finite $h$. The former argument (made in Refs.~\onlinecite{NGVW21, Agrawal20}) is less quantitative at present, but perhaps more physically transparent; the latter argument~\cite{GV19, NGIV20} has the advantage of making accurate quantitative predictions for the value of the KPZ coupling constant. 

\subsubsection{Argument from the Kubo formula}

One can argue directly for $z = 3/2$ by analyzing the Kubo formula, as follows. The physics of quasiparticle scattering in the (isotropic and easy-axis) XXZ models is such that when any string encounters a larger string, it loses its magnetization. (On the other hand, collisions with shorter strings do not have this effect.) On short timescales, no such collision has happened and the string carries a current $v_{s,\lambda} m_s$ where these are the \emph{bare} values. This current is carried until a timescale $\tau_{s,\lambda}$ on which such a collision has taken place. We will simplify by ignoring the $\lambda$-dependence and focusing on the scaling with $s$. Thus the current carried by each string on a short timescale scales as $j_s \sim v_s m_s \sim 1/s \times s = O(1)$. Meanwhile, the time taken by an $s$-string to encounter a bigger string scales as the spacing between such strings, $\left(\sum_{s' > s} \rho_{s'}\right)^{-1} \sim s^2$, divided by the velocity difference between the two strings (which we can approximate as the velocity of the faster string, $\sim 1/s$). Thus the timescale on which the current carried by an $s$-string decays scales as $\tau_s \sim s^3$. 

Equipped with this information we can write the Kubo formula as
\begin{equation}\label{kuboheis}
\langle J(t) J(0) \rangle^{c} \sim \sum_s \rho_s e^{-t/\tau_s} \sim t^{-2/3}.
\end{equation}
Thus the dynamical exponent $z$ falls out of these elementary considerations. Since the velocity and magnetization follow from the way large $s$-strings are defined, the only model-specific consideration in this argument was the density scaling $\rho_s \sim 1/s^3$. As we will argue below, this scaling itself is not free but is fixed by the static susceptibility sum rule. 

\subsubsection{Argument from self-consistent field fluctuations}

We now present a slightly more complicated, but also more quantitative, argument for superdiffusion. 
As noted above this self-consistent theory involves working at finite field or chemical potential $h$, and taking the $h \to 0$ limit self-consistently. When $h$ is small enough, the majority of transport is diffusive, so one can assume the spreading of the strings that dominate transport to be diffusive with a diffusion constant $D(h) = D_0/h$ (Eq.~\eqref{heis_diff}). To complete the argument we estimate $h$ self-consistently as a function of $D$, as follows. Suppose quasiparticles have traveled over a distance $\ell$. The apparent field over this scale due to thermal fluctuations is $h^2 = 1/(4\chi \ell)$. However, this scale $\ell$ is related to time by $\ell^2 = 2 D(h(\ell)) \, t(\ell) = 4 D_0 \, \sqrt{\chi \ell} \, t(\ell)$. Rearranging this expression yields:
\begin{equation}
\ell^2 = (4 D_0 \sqrt{\chi} t)^{4/3}.
\end{equation}
The time-dependent apparent diffusion constant can be defined, in terms of the variance of the dynamical correlation function, as~\cite{NGVW21}
\begin{equation}\label{dtdeff}
D(t) \equiv \frac{\chi}{2} \frac{d}{dt}\left[  \int_{-\infty}^{\infty} dx \, x^2 \langle S^z(x,t) S^z(0,0) \rangle \right] = \frac{4}{3} (4 D_0 \sqrt{\chi} t)^{1/3}.
\end{equation}
Plugging in the exact analytic expression~\cite{NGIV20} for $D_0$ at infinite temperature, we finally arrive at the prediction
\begin{equation}
\label{eq:numKPZ}
D(t) = \frac{2}{3} \left(\frac{10\pi}{27} \right)^{4/3} t^{1/3} + \ldots
\end{equation}
where $\ldots$ denotes subleading terms. This last prediction is in very good agreement with numerical simulations of the Heisenberg model~\cite{NGIV20, NGVW21} (see Fig.~\ref{kubo_illustration}). 

\subsubsection{Inevitability of $z = 3/2$}

Superficially, both arguments above seem to use multiple pieces of Bethe ansatz data to establish the properties of quasiparticles. It is natural to ask how universal these properties are, and if they can be deduced in some more general way. As we discussed, some properties of the quasiparticles (such as their bare velocity and magnetization) follow from identifying them as solitonic packets of $z_0 = 2$ Goldstone modes. The fact that the dressed magnetization goes to zero at half filling also follows from elementary considerations about the dressing of solitons. Finally, the dressed magnetization and the filling factor $n_s$ at finite field are related by definition: in a finite chemical potential $h$, the filling factor $n_s \sim e^{-h m^{\mathrm{dr}}_s}$. 
However, the \emph{specific} scaling of Eqs.~\eqref{heis_drude},~\eqref{heis_diff} does not seem to be fixed, nor does the scaling with $s$ of the dressed density of states. In fact these quantities are also highly constrained by analyticity properties and the static susceptibility sum rule, as we now discuss. 

First, let us fix the scaling of the dressed magnetization. At small $h$, analyticity forces it to be linear in $h$. Further, excitations with $h s \gg 1$ are exponentially suppressed at the bare level and therefore do not get screened (since screening is due to collisions with yet larger quasiparticles). Continuity between these limits then forces $m^{\mathrm{dr}}_s \sim h s^2$ for $h s \ll 1$. 

The remaining data are fixed by the static susceptibility sum rule, which we present in a simplified scaling form
\begin{equation}
\chi \sim \sum_{s < 1/h} \rho_s (h s^2)^2 = O(1),
\end{equation}
and this scaling immediately fixes $\rho_s \sim 1/s^3$. As the argument from the Kubo formula showed, $z = 3/2$ immediately follows from this scaling of the quasiparticle density $\rho_s$.

Since essentially all scalings in Eqs.~\eqref{heis_drude}, \eqref{heis_diff} are fixed by thermodynamics, the only place where the dynamics of the system appears is the scaling of $v_s$ with $s$. Recall that $s$-strings are made up of Goldstone modes with a dispersion $\omega \sim q^{z_0}$, where so far we have taken $z_0 = 2$ for the Goldstone theory of a ferromagnet. More generally, $z_0$ is an integer by analyticity. When $z_0 = 1$, $v_s$ does not scale with $s$ and we simply have ballistic transport. (This is what happens in the easy-plane phase discussed above, where the Goldstone bosons have $z_0 = 1$.) When $z_0 = 3$, $v_s \sim 1/s^2$, in which case the arguments above give a logarithmic divergence of the conductivity at small $h$, and therefore logarithmic corrections to diffusion. Finally for $z_0 > 3$ the relevant integrals converge at large $s$ and one recovers diffusion~\cite{ssd, NMKI20}.

\subsection{KPZ from the pseudovacuum}\label{pseudovacuum}

We now discuss an approach to superdiffusion in the Heisenberg chain that sidesteps the complexities of Bethe strings. The basic idea is to treat the quasiparticle vacuum (``Bethe pseudovacuum'') as a dynamical degree of freedom. The fluctuating hydrodynamics of the resulting long-wavelength vacuum excitations---``soft gauge modes''---provides an intuitive physical explanation for why KPZ universality, with its associated $z=3/2$ dynamical exponent, can arise in spin chains with isotropic symmetry.

It is worth stressing that such fluctuations of the Bethe pseudovacuum have very little to do with the low-temperature physics of the Heisenberg chain: they instead provide one route to capturing the dynamics of symmetry-breaking excitations at \emph{all} temperatures. The reason we have to invoke the pseudovacuum at all is that there exists (at present) no other language for explicitly describing $SU(2)$-symmetry-breaking dynamics in the Heisenberg chain. This is a peculiarity of the manner in which Bethe's ansatz is formulated, as discussed at length below and in Ref. \onlinecite{Vir20}.

For example, ``pseudovacuum fluctuations'' in the Heisenberg antiferromagnet at high temperature correspond to soft fluctuations of the local magnetization vector, which are unrelated to the model's ground state physics. In particular, the fact that the Bethe pseudovacuum is ferromagnetic is incidental; what decides the possibility of superdiffusive transport is whether the space of physically distinguishable reference states is discrete or continuous. For the Heisenberg model, the space of physically distinct Bethe pseudovacua is continuous only at the isotropic point; more generally, it is continuous in integrable models with global non-Abelian Lie group symmetries.

\subsubsection{Domain walls and a gauge principle}

To see why KPZ physics in spin chains must arise from vacuum fluctuations, and what these have to do with gauge invariance, it suffices to consider the weak domain wall initial condition, $\hat{\rho}_{h} \propto (1+h S_z)^{L/2} \otimes (1-h S_z)^{L/2}$ with $0 < h \ll 1$, that was discussed in the introductory survey.

The ``two-reservoir'' initial condition $\hat{\rho} = \hat{\rho}^L \otimes \hat{\rho}^R$, with $\hat{\rho}^{L/R}$ both thermal states, is a touchstone for theories of non-equilibrium dynamics. Indeed, the very first studies of generalized hydrodynamics~\cite{Fagotti,Doyon} sought to solve this initial value problem and the technique is well-established by now. One first specifies a global quasiparticle vacuum, say $|\Omega\rangle = |\ldots\uparrow\uparrow\ldots\rangle$ for the XXZ chain, upon which quasiparticle states can be constructed via Bethe's ansatz. The thermal states $\hat{\rho}^{L/R}$ are then uniquely characterized by their quasiparticle root densities $\rho^{L/R}_{s,\lambda}$, and a hydrodynamical initial condition that models $\rho$ is given by 
\begin{equation}
\label{eq:twores}
\rho_{s,\lambda}(x) = \begin{cases} \rho^L_{s,\lambda} & x\leq 0\\ \rho^R_{s,\lambda} & x>0
\end{cases}.
\end{equation}

Thus, in principle, modelling time evolution from a weak domain wall $\hat{\rho}_h$ in the XXZ chain is simple: one merely solves the initial value problem Eq. \eqref{eq:twores} in generalized hydrodynamics. For anisotropies $|\Delta| < 1$, this procedure is unproblematic: stationary states with opposite signs of their magnetization $\langle S^z \rangle$ and $-\langle S^z \rangle$ are described by distinct sets of root densities, and ballistic transport that matches the prediction of Eq. \eqref{eq:twores} is observed. For anisotropies $|\Delta| \geq 1$, we apparently find that stationary states with opposite signs of their magnetization are described by the same sets of root densities, so that Eq. \eqref{eq:twores} predicts \emph{no bulk transport at all}. This is in manifest contradiction with numerical results. In the easy-axis phase, $|\Delta| > 1$, the problem was addressed to some extent~\cite{PhysRevB.96.115124} by introducing an additional $\mathbb{Z}_2$-valued variable corresponding to the sign of the local magnetization, but even in this case, no generalized hydrodynamic initial value problem has been written down that correctly models the diffusive relaxation of $\hat{\rho}_{h}$.

The origin of this confusion is a hidden gauge principle in generalized hydrodynamics, that was pointed out in Ref. ~\onlinecite{Vir20}. Let us first recall the essence of the gauge principle: suppose a physical system has a global symmetry that is not a local symmetry. Then the global symmetry can be promoted to a local (``gauge'') symmetry at the price that the symmetry becomes dynamical. To see the relevance to generalized hydrodynamics, observe that an arbitrariness in the choice of direction for the Bethe pseudovacuum $|\Omega\rangle = |\ldots\uparrow\uparrow\ldots\rangle$ implies a global $G=O(3)$ (resp. $G=\mathbb{Z}_2$) symmetry for $|\Delta|=1$ (resp. $|\Delta|>1$). Suppose that one now allows $G$ to act locally in space, i.e. one ``gauges'' the pseudovacuum symmetry $G$ of generalized hydrodynamics. Then homogeneity of the initial condition Eq. \eqref{eq:twores} for $\hat{\rho}_h$ is precisely the statement that the weak domain wall initial condition is gauge equivalent to a stationary state.

The problem with how internal symmetries are treated in generalized hydrodynamics is now apparent: the erroneous prediction of ``no dynamics" for weak domain walls $\Delta \geq 1$ occurs because we gauged the pseudovacuum symmetry $G$ without making it dynamical, in contradiction with the gauge principle. At this point it should be obvious that gauge dynamics is responsible for the KPZ phenomenon in spin chains with nonabelian internal symmetry: we now fill in the details of this argument for the Heisenberg chain.

\subsubsection{Dynamics in the gauge sector}
For simplicity, we focus on hydrodynamic initial states in the Heisenberg chain that exhibit dynamics purely in the gauge sector. Such states are gauge equivalent to thermal states, in the sense of the above discussion, and can be constructed as follows. Consider taking an infinite Heisenberg chain and coarse-graining it into fluid cells $\Gamma$ of length $\ell$ sites. Thermal states in each cell $\Gamma_i$ are specified by the expectation values of $O(\ell)$ $G$-invariant local charges, together with the direction of the local magnetization vector $\mathbf{\Omega}_i = \langle \mathbf{S}_i \rangle /\|\langle \mathbf{S}_i\rangle \|$. Let us therefore fix the expectation values of all $G$-invariant local charges $\{S,E,Q_3,Q_4,\ldots,Q_n\}_{n\sim\ell}$ to be the same in each fluid cell, i.e. set the local quasiparticle occupation numbers $\{\rho^i_{s,\lambda}\}_{s=1}^{\sim\ell}$ to be constant throughout the system (here $S=\|\langle \mathbf{S}_i \rangle \|$ denotes the scalar part of the local magnetization and $E$ the local energy density). At the same time, we allow the direction of the magnetization vector at each site, $\langle \mathbf{S}_j\rangle = g(j)\mathbf{e}_z$, to fluctuate, with $g: \mathbb{R} \to SO(3)$ a smooth function that varies on some length scale $\ell_\Omega \gg \ell$.

In the hydrodynamic limit $\ell,\ell_\Omega \to \infty, \, \ell/\ell_{\Omega} \to 0$, this state is gauge equivalent to the thermal state $\rho_{s,\lambda}$ via the local gauge transformation $g: \mathbb{R} \to SO(3)$. In this scaling limit, the gauge and quasiparticle degrees of freedom cannot interact at any finite time and are dynamically decoupled. Thus the field $g$ behaves as if the quasiparticle background were absent. By definition of $g$, its dynamics in the absence of quasiparticles is given by the Hamiltonian dynamics of the Heisenberg chain projected onto the space of gauge vacua, $|\mathbf{\Omega}(g)\rangle = \bigotimes_{x} |g(x) \mathbf{e}_z\rangle$. This is nothing but the mean-field dynamics of the Heisenberg chain and implies Landau-Lifshitz dynamics
\begin{equation}
\label{eq:LL}
\partial_t \mathbf{\Omega} = \lambda \mathbf{\Omega} \times \partial_x^2 \mathbf{\Omega},
\end{equation}
for the local pseudovacuum vector $\mathbf{\Omega}(x) = g(x)\mathbf{e}_z$ (we have restored the coupling strength $\lambda = J/2$, kept time implicit and set the lattice spacing $a=1$ to facilitate direct comparison with the Heisenberg chain).

In order to make contact with spin transport, we must first obtain the Euler-scale hydrodynamics of Eq. \eqref{eq:LL}. This is deceptively simple, and one soon finds that standard parameterizations of $\mathbf{\Omega}$ (e.g. spherical polar coordinates, or the usual canonical coordinates on $S^2$) are ineffective. The reason is that such coordinates are not invariant under global $G$ transformations; by Gauss's \emph{Theorema Egregium}, any local coordinate system on the sphere must exhibit some $G$-dependent singular point. To obtain $G$-invariant hydrodynamic equations, one should instead work in the tangent space of $G$. An ingenious and geometrically rather natural approach~\cite{lakshmanan1976} is to treat the instantaneous local gauge transformation $g : \mathbb{R} \to SO(3)$ as defining the Frenet-Serret frame of a space curve parameterized by its arc-length $x$. The curvature and torsion associated with the section $g$ are then invariant under global $G$ transformations by definition, and furnish suitable coordinates for obtaining the Euler-scale hydrodynamics of Eq. \eqref{eq:LL}.

The essence of this idea has been rediscovered at least three times in the past century; we refer the reader to Ref. ~\onlinecite{Ricca1991} for a survey of its fascinating history. We shall follow the treatment of Ref.~\onlinecite{lakshmanan1976} and let $\mathbf{\Omega}(x)$ be the tangent vector of a fictitious space curve in the arc-length parameterization. The Frenet-Serret equations for the tangent $(\mathbf{\Omega})$, normal $(\mathbf{n})$ and binormal $(\mathbf{b})$ vectors of this curve then read
\begin{align}
\nonumber \partial_x \mathbf{\Omega} &= \kappa \mathbf{n}, \\
\nonumber \partial_x \mathbf{n} &= - \kappa \mathbf{\Omega} + \tau \mathbf{b},\\ 
\nonumber \partial_x \mathbf{b} &= - \tau \mathbf{n},
\end{align}
where the curvature and torsion satisfy
\begin{align}
\kappa &= (\partial_x \mathbf{\Omega} \cdot \partial_x\mathbf{\Omega})^{1/2}, \\
\tau &= \kappa^{-2} \mathbf{\Omega} \cdot \partial_x\mathbf{\Omega}\times \partial_x^2 \mathbf{\Omega}.
\end{align}
Upon expressing the Landau-Lifshitz dynamics in terms of the fields $\mathcal{E} = \kappa^2/2$ (the local energy density) and $\tau$ and discarding a dispersive term in passage to the Euler scaling limit\footnote{The full hydrodynamic equation for $\tau$ is given by $\partial_t \tau + \partial_x[\lambda (\tau^2 - \mathcal{E} - \partial_x^2\kappa/\kappa)] = 0$. However, at the Euler scale for $\mathcal{E}$ and $\tau$, the dispersive term $\sim \partial_x^2 \kappa/\kappa$ is a third-order derivative correction and therefore vanishes in the scaling limit.}, we obtain the Euler-scale hydrodynamics of the Landau-Lifshitz evolution in the form
\begin{align}
\label{eq:curv}
\partial_t \mathcal{E} + \partial_x[\lambda (2\mathcal{E} \tau)] &= 0, \\
\label{eq:tors}
\partial_t \tau + \partial_x[\lambda (\tau^2 - \mathcal{E})] &= 0.
\end{align}

So far our treatment of the dynamics Eq. \eqref{eq:LL} has been exact. It may therefore seem surprising that although the Landau-Lifshitz equation is integrable in the sense of possessing an infinite number of local conserved densities, of which $\mathcal{E}$ and $\tau$ merely constitute one pair, the system of equations Eqs. \eqref{eq:curv} and \eqref{eq:tors} is closed. This suggests that the Euler-scale dynamics of the vector $\partial_x\mathbf{\Omega}$ is decoupled from the higher conserved charges, which is not too unexpected in a local, classical field theory.

\subsubsection{Coarse-graining and KPZ}

Let us now move away from the idealized scaling limit of the previous section and consider the effect of gauge dynamics at finite length scales $\ell \ll \ell_{\Omega} < \infty$. Since there are no useful analytical results in the latter regime, we will proceed in the spirit of nonlinear fluctuating hydrodynamics and coarse-grain the hydrodynamics of the pseudovacuum mode, Eqs. \eqref{eq:curv} and \eqref{eq:tors}, in a phenomenological manner, following the method outlined in Sec.~\ref{nlfhd}.

The first (exact) step is to linearize the Euler equations about a background stationary state. By the above results, states with constant average energy density and torsion, $\mathcal{E}(x) = \mathcal{E}_0, \, \tau(x)=\tau_0,$ are stationary with respect to the Euler-scale hydrodynamic evolution. Note that the torsion in a thermal equilibrium state would vanish by inversion symmetry; it will be instructive to work instead in a background that breaks inversion symmetry (e.g., carries an energy current), so that nonzero values of $\tau_0$ are allowed.
%
%
Linearizing the Euler equations about such a state, we obtain the dynamics
\begin{equation}
\label{eq:linH}
\partial_t \begin{pmatrix} \delta \mathcal{E} \\ \delta \tau \end{pmatrix} + \lambda \begin{pmatrix} 2\tau_0 & 2\mathcal{E}_0 \\ -1 & 2\tau_0\end{pmatrix} \partial_x \begin{pmatrix} \delta \mathcal{E} \\ \delta \tau \end{pmatrix} = 0.
\end{equation}
One immediately runs into the problem that the linearized hydrodynamics is \emph{unstable}; the eigenvalues of the velocity matrix are given by
\begin{equation}
\label{eq:normalmodes}
c_{\pm} = 2\tau_0 \pm i\sqrt{2\mathcal{E}_0},
\end{equation}
and for $\mathcal{E}_0>0$ predict a normal mode that grows exponentially in time. This pathology is telling us that the hydrodynamic description Eq. \eqref{eq:linH} is inconsistent for general values of $\mathcal{E}_0$. Let us therefore set $\mathcal{E}_0=0$ and see what the theory predicts. In this case, the velocity matrix is similar to a 2-by-2 Jordan block and therefore has a \emph{single} eigenvector, that corresponds to the torsional mode $\delta\tau$. Notice that this peculiarity cannot arise in standard nonlinear fluctuating hydrodynamics, which assumes implicitly that the velocity matrix is diagonalizable~\cite{spohn2016fluctuating}.

We deduce that $\mathcal{E}_0=0$ is the only value of the bulk energy density for which a consistent Euler-scale theory exists and that at this value, $\delta \mathcal{E}$ is not a hydrodynamic mode. (Various physical explanations for how the energy decouples in practice have been put forward in the literature~\cite{Vir20,NMKI20,Fava20}.) Restoring nonlinearities, this leaves a single Euler-scale hydrodynamic equation for $\delta\tau$:
\begin{equation}
\partial_t \delta \tau + \partial_x [2\lambda \tau_0 \delta \tau + \lambda(\delta \tau)^2] = 0.
\end{equation}

To capture the effects of dissipation on this evolution, we now introduce a phenomenological coupling to noise and diffusion terms, along the lines of the discussion in Sec.~\ref{nlfhd}. For a thermal background with $\tau_0=0$, this yields a stochastic Burgers equation for fluctuations of the torsion field,
\begin{equation}
\label{eq:KPZ}
\partial_t \delta \tau + \partial_x[\lambda(\delta \tau)^2 - D \partial_x \delta \tau + \sigma \zeta] = 0,
\end{equation}
with $\zeta$ a unit-normalized white noise as in Eq. \eqref{eq:NLFH}. Upon the standard mapping~\cite{KPZ} to a ``height field'' $\delta \tau = \partial_x\eta$, it follows that $\eta$ satisfies the Kardar-Parisi-Zhang equation, so that autocorrelation functions of $\delta \tau$ satisfy~\cite{spohn2016fluctuating}
\begin{equation}
\langle \delta \tau(x,t) \delta \tau(0,0) \rangle = \chi f_{\rm KPZ}\left(x/(\Gamma t)^{3/2}\right)/((\Gamma t)^{3/2}),
\end{equation}
where the superdiffusive spreading rate $\Gamma = 2\sqrt{2} \lambda$ and the susceptibility $\chi = \sigma^2/2D$ by the fluctuation-dissipation relation. The significance of this result for spin transport is that the torsion field measures the local magnetization imbalance~\cite{Vir20}. We deduce superdiffusive spreading of spin autocorrelation functions in the KPZ universality class, with dynamical exponent $z=3/2$.

At this point it is worth highlighting some general caveats on the use of nonlinear fluctuating hydrodynamics as a tool for understanding anomalous transport. Fluctuating hydrodynamics is by design an effective field theory; as such, Eq. \eqref{eq:KPZ} is not expected to predict the exact value of the KPZ coupling constant $\lambda$, which is renormalized from its bare value by the approximations leading to Eq. \eqref{eq:KPZ}. Similarly, the prediction of a KPZ scaling function from Eq. \eqref{eq:KPZ} holds insofar as the torsional mode is decoupled from all other hydrodynamic modes in the system at long times~\cite{Ertas}. While this is plausible in certain classes of physical states in $SU(2)$-symmetric spin chains, it would be premature to conclude that a single, decoupled, hydrodynamic mode is responsible for $z=3/2$ superdiffusion in every spin chain with nonabelian internal symmetry.

\subsection{From giant solitons to pseudovacuum fluctuations}

A formal unification of the ``microscopic'' and ``macroscopic'' perspectives on superdiffusion, described respectively in the two sections above, can be achieved by taking a particular formal limit of the GHD equations at finite $h$~\cite{NGIV20}. From the discussion above we have already identified the dominant strings at field $h$ to be those for which $s h = O(1)$. In the small-$h$ limit, these strings get very large compared with the lattice scale, so one might expect to be able to treat them as classical solitons in the continuum.
Such a correspondence between the long-wavelength spectra of an integrable quantum lattice model and the classical continuum Landau-Lifshitz theory indeed exist, as shown in Ref.~\onlinecite{NGIV20}.
Importantly, this correspondence holds \emph{at finite temperature}, at the level of the thermodynamic Bethe ansatz equations, rather than above the vacuum.

The key observation underlying this construction is that at small $h$, a very large number of strings $\sim 1/h$ contribute to transport; thus, if we define a variable $\xi = s h$, spin transport is dominated by $\xi \approx 1$ no matter what the value of $h$ is. However, for small $h$ the sums over $s$ can be replaced, using the Euler-Maclaurin formula, by \emph{integrals} over the continuous variable $\xi$ (which denotes the rescaled size of the soliton). Thus for example one can write the expression for the diffusion constant as
\begin{equation}
D = \int d\xi du D(\xi, u).
\end{equation}
Here we have also rescaled the rapidity as $u = h\lambda$. One can also write rescaled versions of the thermodynamic Bethe ansatz and GHD equations for the XXX model more generally. As an important example, we discuss the scattering kernel:
\begin{equation}
T_{\xi, \xi'}(u) = \frac{1}{\pi} \log \frac{4u^2 + (\xi + \xi')^2}{4u^2 + (\xi - \xi')^2}.
\end{equation}
This equation can be identified with the scattering phase shift of two classical solitons in the Landau-Lifshitz field theory~\cite{Takhtajan1977,Faddeev1987}. Indeed, a more careful analysis shows that \emph{all} the thermodynamic properties of the XXX spin chain and the classical soliton gas are in one-to-one correspondence~\cite{NGIV20}.

What does this correspondence teach us? Consider fixing a hydrodynamic cell size $\ell$ and taking $h \ell \ll 1$. Spin transport in this limit is dominated by giant solitons of size $s \approx 1/h \gg \ell$. These giant solitons are equivalently described by the continuum limit of the classical theory~\cite{lakshmanan1976} (i.e., by the Landau-Lifshitz model). This confirms the phenomenological identification of the Landau-Lifshitz model as the correct coarse-grained theory of pseudovacuum fluctuations, and also allows us to read off the renormalized coupling constant of this Landau-Lifshitz theory from exact Bethe ansatz data. In this sense, the scaling limit described above unifies the microscopic and macrosopic approaches to superdiffusion.



\subsection{Open questions}

Having introduced various fundamental notions that explain the emergence of an anomalous dynamical exponent $z=3/2$ in spin chains, we close this section by outlining the most pressing questions that remain unresolved.

The most ambitious goal is simply stated: a first-principles derivation of KPZ universal behaviour, including the $z=3/2$ exponent and the conjectured KPZ scaling function, from Bethe's ansatz. While the arguments outlined above provide a detailed physical intuition for how KPZ physics comes about in the Heisenberg chain, strictly speaking none of them are at a level of rigour comparable to the understanding of ballistic transport in integrable systems~\cite{YoshSpohn}. Since this seems to be primarily a technical exercise, and moreover one that is well beyond reach of present techniques, we will not dwell on it further and instead focus on more concrete challenges for future work.

Let us begin with refinements of the discussion in Sec.~\ref{pseudovacuum}. There we outlined one possible route to the fluctuating hydrodynamics of the Landau-Lifshitz equation. In fact, there does not seem to be a well-developed theory of fluctuating hydrodynamics for nonlinear target spaces, and some non-trivial geometrical questions arise in the coupling to noise and dissipation. Second, as already stressed in Sec.~\ref{pseudovacuum}, a proper treatment of the pseudovacuum dynamics seems to require a manifestly ``gauge covariant'' formulation of generalized hydrodynamics, that does not yet exist. (Concretely, a hydrodynamical initial value problem that correctly models relaxation of the weak domain wall $\hat{\rho}_{h}$ has not been written down.) Finally, we observe that first-principles derivations of coupling to a local bath, as is required for classical KPZ physics (c.f Eq. \eqref{eq:KPZ}) \emph{are} possible in closed, Hamiltonian quantum systems, but they generally require introducing a UV length-scale $\ell$ and using the Keldysh formalism to integrate over degrees of freedom at wavelengths $< \ell$. To our knowledge, this has not been attempted in integrable systems; the natural setting for such an approach is not the Heisenberg chain but rather the continuous integrable quantum field theories that have been conjectured to support anomalous spin transport, such as the $O(3)$ nonlinear sigma model on a line~\cite{ssd}.

From the kinetic-theory perspective, two important issues remain somewhat unsettled. First, the arguments we have provided so far seem incapable of establishing the form of the full dynamical structure factor, or even of the optical conductivity in general. In some ways the most physically complete of these theories is that in which we work directly with the Kubo formula and the screening-time ansatz; however, at present this approach has not been made fully quantitative, and has not been extended beyond the zero-momentum limit. Second, and more importantly, it is not clear what happens beyond linear response. If we consider domain walls between half-lines with chemical potentials $\pm h$, the KPZ scaling observed as $h \rightarrow 0$ eventually gives way to log-diffusion as $h \rightarrow \infty$ (see Sec. \ref{domainwalls}). Understanding the physics of this crossover is an important task for future work. 



\section{Models with higher-rank nonabelian symmetries}\label{SUN}

As discussed in Sec.~\ref{isoXXZ}, general theoretical arguments suggest that the phenomenon of infinite-temperature superdiffusion (with dynamical exponent $z = 3/2$) is common to all integrable models with short-range interactions that exhibit global invariance under a nonabelian Lie group. That there might be an entire universality class of integrable systems that exhibit anomalous transport was first conjectured from Bethe ansatz calculations in Ref.~\onlinecite{Ilievski18}. A persuasive case for a universal phenomenon, enabled by nonabelian symmetry and protected by integrability, was made in Ref.~\onlinecite{DupontMoore19} on the grounds of numerical tDMRG simulations, which provided evidence for $z=3/2$ spin transport in integrable chains with $SU(3)$ and $SO(5)$ symmetry. Subsequently, superdiffusion was demonstrated both numerically and analytically~\cite{Fava20} in the canonical example of an integrable system with higher-rank nonabelian symmetry, namely the one-dimensional Fermi-Hubbard model with global symmetry group $SO(4)\cong [SU(2)\times SU(2)]/\mathbb{Z}_{2}$. Recent work has unified the theory of superdiffusion in integrable systems with nonabelian symmetry~\cite{ssd}, by demonstrating that a $z=3/2$ dynamical exponent can be deduced from structural features of the thermodynamic Bethe ansatz in a wide range of models, including infinite families of integrable spin chains invariant under the classical Lie groups $SU(N)$, $SO(N)$ and $Sp(N)$.

\subsection{Higher-rank ferromagnetic spin chains}

Let us summarize here the main properties of integrable quantum spin chains with global invariance under a simple Lie group $G$~\footnote{We define a ``simple Lie group'' to be a connected Lie group whose Lie algebra is simple. Recall that a simple Lie algebra, $\mathfrak{g}$, is a nonabelian Lie algebra that does not contain non-trivial proper ideals. Simple Lie algebras, which have been classified by Killing and Cartan, comprise four infinite (classical) families that cover unitary, orthogonal and symplectic transformations, along with five extra exceptional cases.}.
We assume, in addition, that local degrees of freedom transform in an arbitrary unitary irreducible representation of the Lie algebra $\mathfrak{g}$ and that interactions have finite range. This ensures that the quasiparticle spectrum always involves $r={\rm rank}(G)$ different types of elementary excitations; one can simply picture them as magnons of distinct flavors $a=1,\ldots,r$. Due to mutual attraction, magnons of the same flavor can form bound states with $s$ constituent magnons binding into a Bethe $s$-string of arbitrary size ($s\in \mathbb{N}$).
When $\mathfrak{g}$ is a simple Lie algebra, magnons of different flavors cannot bind together; this is different in models with fermionic local degrees of freedom, such as the Fermi--Hubbard model discussed below. In $G$-invariant spin chains, the complete spectrum of thermodynamic excitations can therefore be arranged on nodes of an $a \times s$ lattice (corresponding bijectively to unitary irreducible representations of $\mathfrak{g}$, graphically represented by rectangular Young diagrams).

Magnon excitations of different flavors arrange themselves in a nested hierarchy, defined with respect to a specific choice of the continuously degenerate ferromagnetic pseudovacuum. The elementary excitations above it are normally called primary magnons (assigned $a=1$); they represent spin-waves formed out of flipped spins in a uniform background. When $r>1$ however, spin projections onto the quantization axis can be lowered multiple times. Subsequent ``flips'' lead to the notion of auxiliary magnons, which may be conveniently visualized as magnon waves of flavor $a$ that propagate in a sublattice of magnons of flavor $a-1$. In this sense, only neighboring flavors interact with one another. In any finite volume, quasimomenta for each flavor of magnons become quantized according to the Nested Bethe Ansatz equations~\cite{Hubbard_book}. While the total energy and momentum of an eigenstate are only functions of primary magnons, they are nonetheless affected by auxiliary magnons indirectly via the Bethe equations. Giant magnons refer to $s$-strings of flavor $a$ at large $s$.

\medskip

\paragraph*{Unitary quantum chains.}
We further elaborate on the structure of quasiparticle spectra for the class of $SU(N)$-symmetric quantum spin chains. For a more complete account (covering other classes of models, including higher-rank integrable QFTs), we refer the reader to Ref.~\onlinecite{ssd}.
Let us here assume, mainly for simplicity, that local degrees of freedom transform in the fundamental representation of $\mathfrak{su}(N)$. This leaves us with a family of integrable Hamiltonians (for arbitrary $N\geq 2$) of the simple form $H \simeq \sum\nolimits_i \Pi_{i, i+1}$.
Here $\Pi |\alpha \otimes \beta\rangle = |\beta \otimes \alpha \rangle$ denotes the permutation (swap gate) that exchanges the state on site $i$ with that on the adjacent site $i + 1$ in the product Hilbert space of the chain. In the simplest case of $SU(2)$, this yields (modulo an overall scale and constant shift) the familiar isotropic Heisenberg spin-$1/2$ chain.

The task at hand is then to characterize finite-temperature linear response in such $SU(N)$-symmetric integrable quantum chains, focusing again on the Noether charges. We thus introduce grand-canonical equilibrium states by
including the full set of $N-1$ Cartan charges in the $G$-invariant Gibbs state. This breaks the global $G$-invariance, reducing it down to the abelian (torus) subgroup $U(1)^{\otimes N-1}$. The effect of such an explicit breaking of symmetry
is that the conductivity tensor in the sector of the Noether charges (of dimension $N-1$) becomes non-trivial. Its structure is nevertheless still dictated by the full $SU(N)$ symmetry group of the Hamiltonian. (This is discussed in greater detail in Sec.~\ref{cosetLL} for the case of classical Landau--Lifshitz ferromagnets on coset spaces $G/H$.)

This complication aside, in the unbroken symmetry sectors the outlined arguments for the Heisenberg model generalize quite directly to $SU(N)$-symmetric integrable chains and also to other models that possess higher-rank symmetries.
In this respect it is once again crucial that all integrable $G$-invariant ferromagnets feature continuously degenerate ground states (i.e. pseudovacua) whose low-lying excitations are type-II ($z_0 = 2$) Goldstone modes.
Since these can grow into interacting giant magnons (yielding a finite contribution to static susceptibilities),
the formal requirements for $z = 3/2$ superdiffusion remain fulfilled. Indeed, in Ref.~\onlinecite{ssd} the TBA equations for the $SU(N)$ spin chains are solved explicitly, confirming that the dynamical scaling properties inferred from the Heisenberg model are robust to the choice of global nonabelian Lie group symmetry.

There is nevertheless one major difference relative to the $SU(2)$ case worth highlighting. We remind the reader that in Sec.~\ref{simplesec}, we computed the diffusion constant in the presence of a  finite chemical potential $h$, which was subsequently determined in a self-consistent manner. In the $SU(N)$ case (and more generally), where multiple chemical potentials enter the description, one can find special surfaces of enhanced symmetry in the parameter space, where the residual symmetry group still contains a nonabelian subgroup; the basic example would be an $SU(3)$ model, where there is a choice of chemical potentials that induces symmetry breaking down to $SU(2) \times U(1)$. In this scenario, there is residual superdiffusion \emph{even} in the presence of finite chemical potentials (albeit only for charges that do not acquire finite expectation values). It is not currently understood how to deal with these special lines and it thus remains an open problem to generalize the self-consistent argument for computing the KPZ coefficient to this case.

\medskip

\paragraph*{Goldstone modes and stacks.}

We have already discussed how anomalously large contributions of giant magnons in the spectrum result in divergent charge diffusion constants. At low momenta or, equivalently, large rapidities, such giant magnons admit a purely classical description and manifest themselves as interacting nonlinear waves of an integrable classical field theory. Remarkably, one finds that they correspond to soliton modes of higher-rank Landau--Lifshitz ferromagnets~\cite{Stefanski04,MatrixModels,ssd}. Below we briefly outline two complementary viewpoints on this correspondence.

Most directly, and with no recourse to integrability, the effective classical Landau-Lifshitz action can be obtained by projecting the Hamiltonian dynamics of a $G$-invariant quantum spin chain onto the many-body coherent-state manifold~\cite{MatrixModels}. Taking the continuum limit, one ends up with a purely classical field theory that lives on a coset space $G/H$, whose points are in a bijective correspondence with Perelomov $G$-coherent states. The subgroup $H\subset G$ leaves the pseudovacuum invariant (modulo a phase). For example, in the case of unitary Lie group symmetry $G=SU(N+1)$, one finds a family of integrable Landau-Lifshitz models on complex projective target spaces $\mathbb{CP}^{N}\cong SU(N+1)/[SU(N)\times U(1)]$, that are described in more detail in Sec.~\ref{cosetLL}. Remarkably, the resulting classical, mean-field equations of motion turn out to be integrable (integrability is unfortunately not manifest in this approach).

An alternative approach is to perform all computations within the framework of Algebraic Bethe Ansatz.
This approach provides a direct means to derive the (classical) Lax pair which realizes an auxiliary linear problem, along with the associated monodromy matrix. From this viewpoint, classical spin-wave solutions emerge as solutions to the Bethe equations in the low-energy sector (above the ferromagnetic pseudovacuum), describing eigenstates with long wavelengths and vanishingly small momentum $P\sim O(1/L)$ at large length $L$. In this way, one can retrieve the entire spectrum of classical (finite-gap) solutions~\cite{Kazakov2004,BKS06,BKSZ06,Vicedo07}; individual nonlinear modes can be understood, from the viewpoint of a quantum chain, as macroscopic coherent condensates of magnons. We emphasize that giant magnons are rather special in that their constituent magnons (i.e. the Bethe roots) are equidistant, implying that their associated condensates have a constant, unit spectral density~\cite{Bargheer08,MIG20}.
At the classical level, such uniform condensates describe solitons. Solitons are particular localized nonlinear field configurations that behave (asymptotically) as quasiparticles. Scattering of solitons is purely elastic and governed by a completely factorizable scattering matrix, just as for other Bethe Ansatz solvable models. In classical systems invariant under the action of a nonabelian Lie group, solitons can be thought of nonlinear analogues of Goldstone modes stabilized by integrability.

The above picture suggests there is a simple correspondence between giant Bethe strings in quantum chains and classical soliton modes
of Landau--Lifshitz-type models on the corresponding coset spaces. However, in attempting to establish this correspondence directly, one encounters the following puzzling fact: the number of distinct flavors $r$ does not match the number of Goldstone modes. As discussed in detail in Ref.~\onlinecite{ssd}, this discrepancy is resolved by the presence of \emph{emergent} classical degrees of freedom. For an illustration, let us consider an integrable quantum chain with $SO(5)$ symmetry. Since the group has rank $2$, one would naively expect the low-energy fluctuations above the ferromagnetic pseudovacuum to comprise two magnon branches. On the other hand, by the general Goldstone theorem~\cite{WM12,WM13,Hidaka13} there should instead be three Goldstone modes in the spectrum, corresponding to half the real dimension of the classical target space $SO(5)/[SO(3)\times SO(2)]$. This mismatch can be elegantly resolved by introducing additional types of excitations called ``stacks'', which are emergent excitations of mixed flavor where auxiliary magnons glue onto the primary one~\cite{ssd} (also found in a supersymmetric sigma model~\cite{BKS06}). Even though in many-body quantum eigenstates, such stacks are prohibited simply by the Pauli exclusion principle, at the classical level they become allowed and should be treated as independent modes. Physically speaking, such stack degrees of freedom represent all possible polarizations in which a classical field (with internal structure) can oscillate.

Stacks are present also in $SU(N)$ chains, despite the fact that in this simpler setting, $r=N-1$ correctly describes the number of Goldstone modes associated with complex projective manifolds $\mathbb{CP}^{N-1}\cong SU(N)/[SU(N-1)\times U(1)]$. 
Even in this case, the Goldstone modes correspond to stacks and not to auxiliary magnons (we remind the reader that auxiliary magnons cannot be excited without first having primary magnons in the state, and thus cannot correspond to proper Goldstone modes, which are physical excitations). A graphical algorithm for inferring the total number of such physical stacks (based on paths on a Hasse diagram of the underlying root lattice) is described in Ref.~\onlinecite{ssd}.

\subsection{Fermi-Hubbard model}

We now discuss the most experimentally relevant integrable system with nonabelian higher-rank symmetry, namely
the Fermi--Hubbard model \cite{Hubbard_book}
\begin{eqnarray}
    H = -\sum_{i=1}^{L}\sum_{\sigma \in \{\uparrow,\downarrow\}}
    (c^{\dagger}_{i,\sigma}c_{i+1,\sigma} + c^{\dagger}_{i+1,\sigma}c_{i,\sigma}) +
    U \sum_{i=1}^{L}\Big(n_{i,\uparrow}-\frac{1}{2}\Big)
    \Big(n_{i,\downarrow}-\frac{1}{2}\Big).
\end{eqnarray}
The model exhibits an $SU(2)$ spin-rotation symmetry; moreover, when the chemical potential is tuned precisely to the particle-hole symmetric point, the model has an $SU(2)$ particle-hole symmetry. These two symmetries combine to yield the global Lie-group symmetry $[SU(2) \times SU(2)]/\mathbb{Z}_{2} \cong SO(4)$ (for even length $L$) when both charge and spin degrees of freedom are at half filling.
By virtue of integrability, this symmetry gets elevated to that of an infinite-dimensional Yangian $Y[SO(4)]$ \cite{Uglov1994}.

The Hubbard model can be diagonalized by Nested Bethe Ansatz \cite{Takahashi72,Hubbard_book}. Its thermodynamic quasiparticle content is however, owing to its intrinsically fermionic character, markedly different from the magnonic spectrum of $G$-invariant ferromagnetic chains described above. A closer look shows that the model is closely related (from an algebraic viewpoint) to another integrable model of interacting spinful fermions exhibiting global invariance under the $\mathbb{Z}_{2}$-graded Lie algebra (also known as a Lie superalgebra) $SU(2|2)$, introduced in Ref.~\onlinecite{EKS92}. The local Hilbert space of the Fermi--Hubbard model is the four-dimensional fundamental representation of $\mathfrak{su}(2|2)$. The latter contains two copies of a bosonic $\mathfrak{su}(2)$ subalgebra, with Cartan generators $S^{z}=\tfrac{1}{2}(n_{\uparrow}-n_{\downarrow})$ and $N_{e}=n_{\uparrow}+n_{\downarrow}$, corresponding to spin and electron charge, respectively.

Elementary momentum-carrying charge excitations above an empty vacuum represent unbound spin-up electrons. There is however an infinite tower of additional (auxiliary) spin excitations that correspond to flipping spins of individual electrons. Much like in other interacting spin chains, quasiparticles of the Fermi--Hubbard model include bound states. This time however, spin degrees of freedom not only form Bethe strings of their own (which do not carry electron charge), but also combine with electrons into spin-neutral compounds that carry finite electron charge~\cite{Takahashi72}.

As far as spin and charge transport are concerned, the phenomenology of the $SU(2)$ Heisenberg chain survives with no substantive differences: away from half-filling one finds ballistic transport with finite spin and charge Drude weights, which both vanish precisely at their respective half-fillings as a consequence of particle-hole symmetry \cite{IN_Drude,IN17}. At the level of quasiparticles, this is manifested through vanishing of dressed magnetization and charge respectively. Moreover, in approaching half-filling the respective diffusion constants diverge~\cite{Ilievski18}. It was shown in Ref.~\onlinecite{Fava20} that the arguments developed in the previous section for the Heisenberg chain carry through with little modification for the Hubbard model, and predict KPZ universal behaviour for both spin and charge transport at their respective half-fillings. The associated $z = 3/2$ dynamical exponent is clearly visible in numerical tDMRG simulations~\cite{Fava20}. We remark that some earlier numerical papers had seen evidence for diffusion~\cite{prosen2012diffusive, steinigeweg2017charge}, but the charge (spin) diffusion constant is analytically known to diverge in the half-filling (zero-magnetization) limit~\cite{Ilievski18}.

\subsection{Matrix models in discrete space-time}

In another recent paper~\cite{MatrixModels} the authors studied charge transport in a family of nonabelian integrable models by constructing a class of classical ``circuit models'' with matrix-valued local degrees of freedom that propagate on a discrete space-time lattice (generalizing the construction of Ref.~\onlinecite{KP20}). The matrices takes values on certain compact manifolds: complex Grassmannian manifolds ${\rm Gr}_{\mathbb{C}}(k,N)=SU(N)/S[U(N-k)\times U(k)]$ (including complex projective spaces $\mathbb{CP}^{N-1}$ as a special case for $k=1$), or Lagrangian Grassmannians ${\rm L}(N)\equiv USp(2N)/U(N)$. The coset manifolds ${\rm Gr}_{\mathbb{C}}(k,N)$ consist of Hermitian complex matrices of dimension $N$ which square to one,
whereas ${\rm L}(N)$ are manifolds of complex $2N$-dimensional anti-symplectic unitary matrices $M^{\rm T}JM+J=0$ (with symplectic unit $J$) that satisfy $M^{2}=MM^{\dagger}=1$. These families of models are none other than \emph{integrable} discretizations (or Trotterizations) of classical nonrelativistic Landau--Lifshitz-type field theories that govern the low-energy sector of integrable ferromagnets~\cite{ssd}. In particular, classical field theories with $\mathbb{CP}^{N-1}$ target spaces (see also Sec.~\ref{undular}) describe the long-wavelength solutions to integrable $SU(N)$ quantum chains with fundamental degrees of freedom.

One particular advantage of working in a purely discrete setting is that it allows for an explicit integration scheme that consists of decomposing the many-body propagator into a sequence of two-body symplectic maps $\Phi_{\tau}:\mathcal{M}\times \mathcal{M}\rightarrow \mathcal{M}\times \mathcal{M}$, with $\tau$ playing the role of a discrete time-step. For the class of models considered in Ref.~\onlinecite{MatrixModels}, the map $\Phi_{\tau}$ maps a pair of input matrices $M_{1}$ and $M_{2}$ to an output pair $M^{\prime}_{1}$ and $M^{\prime}_{2}$ via~\cite{MatrixModels}
\begin{equation}
    M^{\prime}_{1}={\rm Ad}_{F\,S_{\tau}}(M_{2}),\quad
    M^{\prime}_{2}={\rm Ad}_{F\,S_{\tau}}(M_{1}),\quad {\rm with}
    \quad S_{\tau}(M_{1},M_{2}) = M_{1} + M_{2} + \ii \tau,
\end{equation}
where the twisting matrix $F$ is a constant (invertible) $GL(N;\mathbb{C})$ matrix representing an external field and ${\rm Ad}_{X}(M)\equiv X^{-1}\,M\,X$ is the adjoint action of $X$ on $M$. As shown in Ref.~\onlinecite{MatrixModels}, such maps arise from a fully discrete zero-curvature principle as a compatibility condition for linear transport around an elementary plaquette on a lightcone square lattice, with a linear Lax connection $L(\lambda)=\lambda+\ii\,M$; from this perspective, integrable matrix models in discrete space-time have an exceptionally simple integrable structure.

Together with their $G$-valued Noether charges, integrable matrix models possess infinitely many conserved quantities in involution. One might therefore expect that they share universal features of transport in common with other integrable systems with nonabelian symmetry~\cite{Ilievski18,DupontMoore19,ssd}. Numerical simulations~\cite{MatrixModels} confirm that this is indeed the case. In fact, the long-time behavior of dynamical correlation functions not only exhibits the anticipated KPZ dynamical exponent $z=3/2$, but also provides some of the best numerical evidence for KPZ scaling profiles in this entire (quantum and classical) class of models.

\section{Nonthermal initial states}\label{domainwalls}

In previous sections, we have discussed various types of spin transport that take place in thermal equilibrium, where spreading of fluctuations on a hydrodynamic scale can be quantified by linear-response transport coefficients. More broadly, hydrodynamics captures the non-linear response of physical states that are globally far from equilibrium, provided such states are \emph{locally} in equilibrium.

One of the most common settings for studying such far-from-equilibrium behaviour within generalized hydrodynamics is the Riemann problem, as discussed above around Eq. \eqref{eq:twores}. The initial state of a system is prepared in a factorizable form, with each half initialized in its own local equilibrium state; the evolution is stationary everywhere except at the interface of finite width in the middle where there a mismatch in temperature or chemical potentials. On the ballistic (Euler) scale, the Hamiltonian evolution generates dynamics within a light-cone that spreads from the interface. Studies of such initial conditions in spin chains date back at least twenty years~\cite{Antal1999,Gobert2005}. On the theoretical side, the Riemann problem for the easy-plane XXZ model is well-understood by now within the framework of generalized hydrodynamics~\cite{Fagotti,PhysRevB.96.020403,IN_Drude,Bulchandani18,PhysRevB.97.081111}. For the Riemann problem in the easy-axis XXZ model, a recent study~\cite{PhysRevB.96.115124} found that the magnetization exhibits abrupt jumps that lie outside the scope of conventional GHD. These jumps were explained in terms of the motion of giant quasiparticles. However, the general Riemann problem for the easy-axis and isotropic XXZ models still remains largely unresolved (cf. Sec. \ref{pseudovacuum}). We remark in passing that the Riemann problem was recently generalized to the case of two half-chains with different Hamiltonians~\cite{PhysRevB.93.205121, biella2019ballistic}, for systems with ballistic transport; it would be interesting to generalize these ideas to systems where one or both half-chains exhibit anomalous transport, but this remains to be done.

In this section we review recent progress on the nonequilibrium dynamics of two simple and experimentally accessible examples of non-thermal initial states, namely magnetic domain walls and spin spirals.

\medskip

\subsection{Domain walls}

The melting of a magnetic domain wall provides one of the cleanest problems in out-of-equilibrium dynamics, at least from a conceptual standpoint. This problem can be understood as an extreme case of the Riemann problem, where each of the two infinite partitions is initialized in a distinct ground or anti-ground state. In the following, we review the current status of this problem, focusing on the best studied example of the anisotropic spin-$1/2$ Heisenberg chain and its classical counterpart -- the Landau--Lifshitz classical field theory with uniaxial anisotropy.

Depending on the interaction anisotropy along the $z$-axis, the following three dynamical regimes can be discerned:
\begin{itemize}
    \item the easy-plane regime, showing ballistic transport,
    \item the easy-axis regime, showing absence of transport,
    \item the isotropic ``critical point", showing anomalous diffusion that is conjectured to be logarithmically enhanced diffusion.
\end{itemize}
In the rest of this section, we discuss the underlying microscopic mechanisms and offer some basic physical intuition for each of these dynamical laws.

The central point to address first is how to quantify magnetization transport far away from equilibrium, where linear response can no longer be assumed. An intuitive choice is to quantify melting by measuring the growth of net magnetization in one half of the system~\cite{Gobert2005}
\begin{equation}
    m(t) = \int^{\infty}_{0}\dd x \, S^{z}(x).
\end{equation}
Equivalently, one can compute the time-integrated spin current density at the midpoint $x=0$. A spreading exponent $\alpha$ can then be inferred from the asymptotic long-time behaviour $m(t)\sim t^{\alpha}$. Exponents $\alpha = 1$ and $\alpha = 1/2$ correspond to ballistic and diffusive transport respectively, while $\alpha=0$ signifies lack of transport.

\bigskip

\subsubsection{Quantum Heisenberg spin-1/2 chain}

The first systematic study of the domain wall dynamics in the Heisenberg spin chain was carried out in Ref.~\onlinecite{Gobert2005}, which reported three distinct behaviours: ballistic dynamics in the easy-plane regime (with linear magnetization profiles resembling the free-fermion point~\cite{Antal1999}), an apparent lack of transport in the easy-axis regime, and anomalous transport with a numerically estimated superdiffusive dynamical exponent $\alpha \approx 3/5$ at the isotropic point. This problem was revisited a decade later~\cite{Ljubotina2017-diff,Misguich2017}, using tDMRG techniques that represent the state-of-the-art at the time of writing.

Ref.~\onlinecite{Ljubotina2017-diff} studied a more general
initial-value problem, consisting of a one-parameter family
of initial product states $\ket{\Psi} = (\cos{(\theta_{0}/2)}\ket{\uparrow}+\sin{(\theta_{0}/2)}\ket{\downarrow})^{\otimes L/2}\otimes (\sin{(\theta_{0}/2)}\ket{\uparrow}+\cos{(\theta_{0}/2)}\ket{\downarrow})^{\otimes L/2}$, which interpolates between two tilted ferromagnetic domains (recovering $\ket{\Psi_{\rm DW}}$ for $\theta_{0}=0$). By measuring the asymptotic growth of $m(t)=\int^{t}_{0}\dd t'\,j(0,t')$ for $j(0,t)=\bra{\Psi(t)}j\ket{\Psi(t)}$, it was estimated numerically that $m(t)\sim t^{\alpha}$ with $\alpha \approx 1/2$ for values of the tilting angle $\theta_{0}$, except close to $\theta_{0}=0$ where a pronounced drift towards $\alpha \approx 3/5$ was observed. Despite this anomaly, the authors conclude in favor of normal diffusive spreading. In addition, they propose a classical mean-field equation in the space of product spin-coherent states which, in the continuum limit, yields a single-particle Schr\"{o}dinger equation. The approximation becomes exact in the small-bias (linear response) regime, yielding an explicit formula for the resulting magnetization profile in terms of Fresnel functions.

A follow-up study~\cite{Misguich2017}, focused on the fully-polarized domain wall, investigated the possibility of modified diffusive spreading by fitting numerical data to various types of correction terms. Even though Ref.~\onlinecite{Misguich2017} could not rule out normal spin diffusion, it nevertheless provided convincing evidence for a marginal enhancement of diffusion, in the form of a multiplicative logarithmic correction, while concluding that \emph{``\ldots time is ripe for analytical investigations of this vexing problem, either by using integrability or by studying some effective and simplified models\ldots''.}

\bigskip

\paragraph*{Easy-plane regime.}

In the easy-plane regime, magnetization transport is ballistic at the linear-response level. As explained in Sec.~\ref{gaplessxxzsec}, magnetization is transported through the system by two distinguished quasiparticle species that (in distinction to other quasiparticles in the spectrum) remain polarized (i.e. carry finite dressed magnetization) even when the background state has no net magnetization. The presence of such special excitations is intimately tied to the existence of quasilocal conservation laws that break $\mathbb{Z}_{2}$ invariance of the Hamiltonian~\cite{IN_Drude}. This picture remains valid even far from equilibrium.

The domain-wall problem in the easy-plane phase has been formulated in the language of GHD~\cite{PhysRevB.97.081111}. Remarkably, the corresponding dressing equations 
are algebraic and allow for a closed-form solution. Indeed, the evolution of the domain wall profile only involves the special pair of quasiparticles at roots of unity, with rapidity-independent Fermi occupation functions. Moreover, when the anisotropy parameter corresponds to a rational multiple of $\pi$ (such values densely fill the easy-plane interval) one again encounters the discontinuities that were discussed above in the context of the spin Drude weight. Nevertheless, by taking the denominator $\ell \to \infty$ while keeping the parameter $\gamma$ finite, the discontinuities gradually diminish and the magnetization profile smears out into a linear slope.

One distinguished property of the domain-wall state is that in the easy-plane regime, quasiparticles that constitute the domain wall experience no dressing~\cite{PhysRevB.97.081111}. This allows for a curious reinterpretation of the dynamics in terms of noninteracting fermions confined to a rescaled Brillouin zone. Based on this picture, it was further argued in Ref.~\onlinecite{PhysRevB.97.081111} that in the neighborhood of the quasiparticle front the characteristic scaling is no longer of the Tracy--Widom type $t^{1/3}$, in contrast to what happens for free fermions~\cite{Eisler2013}. In addition, they suggested that magnetization profiles near the isotropic point $\Delta \to 1^{-}$ exhibit diffusive characteristics $x/t^{1/2}$. The nature of quasiparticle fronts in the domain wall problem was clarified in subsequent studies~\cite{VBCK,Stephan_front}. The main point is that \emph{two} sharply defined fronts emerge as the domain wall melts: the interacting front, $x_i/t = \sqrt{1-\Delta^2}$, which defines the edge of the bulk ballistic profile, and the free front, $x_f/t = 1$, which defines a kind of (ballistic) Lieb-Robinson bound for magnon dynamics~\cite{Stephan_front}. Away from the free-fermion point, $\Delta=0$, whose front-scaling is well-understood~\cite{Eisler2013}, we have $x_i < x_f$, leading to front-scaling distinct from the non-interacting case.

Diffusive scaling at the interacting front $x_i$ is supported by stationary phase calculations, and shows reasonably good agreement with numerics~\cite{PhysRevB.97.081111,Stephan_front}. However, it is worth stressing~\cite{VBCK} that such scaling does not arise from diffusion in the conventional sense, as there is no microscopic mechanism that can generate a local Markovian bath~\cite{Boldrighini1997,Cao2018,DeNardis2018,Gopalakrishnan18}.

The free front arises due to an inhomogeneity in energy, $\delta E = -J\Delta/2$ at the domain wall itself, that forms a subextensive, lattice-scale correction to the standard hydrodynamic initial condition Eq. \eqref{eq:twores}. The resulting excitation is maximally localized and therefore possesses maximal uncertainty in velocity, generating excitations that travel faster than the bulk ballistic front. Such excitations travel \emph{in vacuo}, and thus exhibit free-fermionic behaviour in the vicinity of the free front $x_f$, characterized by subdiffusive $t^{1/3}$ spreading~\cite{VBCK}. Although this behaviour can be understood in terms of Tracy--Widom statistics (magnetization profiles near the free front are proportional to a spatial derivative of the Airy kernel)~\cite{VBCK}, it differs from the free fermionic case insofar as the spatial probability distribution function of the fastest quasiparticle is no longer Tracy--Widom
~\cite{PhysRevB.97.081111,Stephan_front}.

\medskip

\paragraph*{Easy-axis regime.}

For gapped domain walls $(\Delta>1)$ in the $XXZ$ model, the bulk magnetization dynamics freezes at long times~\cite{Gobert2005,Mossel2010}. This effect is caused by stable kink configurations, which become additional degenerate ground states~\cite{Koma1998,Dijkgraaf2009} in the thermodynamic limit. In the regime of weak anisotropy $\Delta \to 1^+$, the asymptotically frozen magnetization profiles were found to be well-described by static soliton profiles of the classical anisotropic Landau-Lifshitz equation, with no fitting parameters \cite{Misguich2019}. By exploiting integrability, this correspondence can actually be made precise \cite{MIG20}. The classical Landau-Lifshitz dynamics even predicts the relaxation time to such profiles, which has been numerically verified~\cite{Misguich2019} to scale as $\tau \sim (\Delta-1)^{-1}$. Modelling both this relaxation and the frozen steady state is currently beyond generalized hydrodynamics, for the reasons discussed at length in Sec.~\ref{pseudovacuum}.

As in the easy-plane regime, corrections to this bulk profile arise from the inhomogeneity in energy at the domain wall \cite{VBCK}, and again yield a non-interacting ballistic front at $x_f=t$ with subdiffusive $t^{1/3}$ broadening. The non-interacting front can be described analytically as $\Delta \to \infty$ by writing down an effective Hamiltonian for the single-magnon hopping problem~\cite{pcomm}. The appropriate generalized hydrodynamic initial condition in this limit consists of a delta function in the single-magnon sector at the origin, and is given to leading order in $1/\Delta$ by
\begin{equation}
\label{eq:1magnon}
\rho_{1,k}(x) = \frac{1}{4\Delta^2} \delta(x), \quad \rho_{n,k}(x)=0, \quad n=2,3,\ldots.
\end{equation}
Note that since the quasiparticles travel in a bulk pseudovacuum, there is no dressing. By the results obtained for a single magnon in Ref. \onlinecite{VBCK}, the mapping Eq. \eqref{eq:1magnon} predicts profiles of energy, spin and entanglement entropy near $x_f$ that are asymptotically exact as $\Delta \to \infty$.

\medskip

\paragraph*{Isotropic point.}

There are few analytical results available for the isotropic point at present and it would be premature to offer any definite conclusions on this topic. However, there exist two notable exact results that seem to support (albeit indirectly) the conjectured logarithmically amplified diffusion law~\cite{Misguich2017}. The first is an analytical solution of the domain wall dynamics in the classical isotropic Landau--Lifshitz model, covered below in Sec.~\ref{DWLL}; in this problem, the logarithmic correction is due to a logarithmic divergence in the density of states. On the grounds of a conjectured quantum-classical correspondence~\cite{Gamayun19}, it is expected that this anomalous feature survives quantization. Another hint in favor of a multiplicative logarithmic correction comes from an exact expression for the return probability amplitude computed in Ref.~\onlinecite{Stephan17}, reading $\mathcal{R}(t)\sim \sqrt{t}\exp{(-c\sqrt{t})}$ with $c=\pi^{-1/2}\zeta(3/2)$. (Once again, there is an additional non-interacting front at $x_f=t$ that exhibits subdiffusive $t^{1/3}$ broadening~\cite{VBCK,Weiner2019}.)

\subsubsection{Domain wall in the Landau--Lifshitz ferromagnet}\label{DWLL}

Obtaining an exact solution to the quantum domain wall problem (or even a hydrodynamic description on a sub-ballistic scale) remains out of reach, even for the integrable spin-$1/2$ XXZ chain. This motivates addressing the classical analogue of the same problem first. A universal description of long-wavelength ferromagnetic fluctuations in extended systems is given by the classical Landau--Lifshitz equation~\cite{lakshmanan2011fascinating}, which in one spatial dimension takes the form (cf. Sec. \ref{pseudovacuum})
\begin{equation}
\label{anisoLL}
    {\bf S}_{t} = {\bf S}\times {\bf S}_{xx} + {\bf S}\times \mathcal{J}{\bf S},
\end{equation}
where ${\bf S}$ is a spin field (${\bf S}\cdot {\bf S}=1$)
on the unit $2$-sphere and $\mathcal{J}={\rm diag}(0,0,\delta)$ diagonal an anisotropy tensor parametrized by $\delta$. The Landau--Lifshitz equation is a canonical example of a completely integrable PDE~\cite{Takhtajan1977,Sklyanin1979,Faddeev1987}. It is solvable by the classical inverse scattering method, which represents one of the high points of twentieth-century mathematics~\cite{GGKM1967,Lax1968} and was elevated to an art form by the Leningrad and Kyoto schools of mathematical physics~\cite{Faddeev1987}.

Even though the one-dimensional Landau--Lifshitz equation has been studied extensively~\cite{Faddeev1987,lakshmanan2011fascinating}, the domain-wall initial value problem was not attempted until very recently~\cite{Gamayun19}. What is perhaps most remarkable is that the classical version of the melting domain wall problem features the same dynamical behavior as the quantum Heisenberg chain~\cite{Gamayun19}. The quantum domain wall problem thus provides one example of a strongly non-equilibrium quantum problem that can be understood at the level of its (continuum) classical mean-field dynamics. The validity of the classical Landau--Lifshitz description was subsequently investigated in greater detail in a comprehensive study~\cite{Misguich2019}. This phenomenology seems to be independent of integrability; for example it applies equally to spin-$1$ chains~\cite{MedenjakSpinOne}. An intuitive way to understand this physics is via the observation that the domain wall is a very low-entanglement initial state, whose dynamics is therefore well-approximated by a time-dependent variational principle\cite{TDVP} with low bond dimension $\chi$, which recovers mean-field dynamics in the limit $\chi=1$. Meanwhile, the continuum approximation becomes justified at long times~\cite{Misguich2019}. This combination of low entanglement and long wavelengths yields time evolution that is qualitatively (and in some cases quantitatively~\cite{Misguich2019,MIG20}) similar to the continuum mean-field description Eq. \eqref{anisoLL}.


Before turning our attention to magnetization dynamics, we wish to quickly familiarize the reader with a few central ingredients of the inverse scattering transform~\cite{AblowitzSegur_book} (IST) for classical integrable systems. Intuitively, the IST can be thought of as a non-linear analogue of the Fourier transform: every phase-space configuration allows for a unique decomposition into decoupled, nonlinear harmonics. More precisely, the IST allows for a systematic construction of action-angle variables. This can be achieved via a geometrical formulation proposed by Lax~\cite{Lax1968}, in which the nonlinear evolution equation to be solved is recast as a linear equation for an auxiliary field, whose space-time evolution is encoded as parallel transport with a flat connection. In this scheme, the original nonlinear equation becomes a consistency (zero-curvature) condition for the auxiliary parallel-transport problem~\cite{AKNS1973,AKNS1974}. With the aid of the connection, one can define a monodromy operator corresponding to parallel transport around a closed, constant-time loop in space (periodic boundary conditions are usually assumed). This operator turns out to evolve in time via an isospectral flow; its characteristic equation therefore defines a time-independent algebraic object, known as the ``spectral curve'', which encodes the full set of local conservation laws of the model in question. From the monodromy operator, it is relatively straightforward to construct action-angle variables; in this way, the dynamics of classically integrable systems can be reduced explicitly to quasiperiodic motion on Liouville-Arnol'd tori. A remarkable finding~\cite{GGKM1967} was that such variables exist for certain nonlinear classical PDEs, such as the Landau-Lifshitz equation~\cite{Faddeev1987}, to which we now return.



The Cauchy problem corresponding to an initial domain wall in the classical Landau-Lifshitz magnet is defined over a non-compact spatial domain and therefore requires an appropriate boundary condition, in the form of asymptotic fields. One of the simplest possible choices is the following one-parameter family of smooth domains of width $x_{0}$, of the form
\begin{equation}
    \mathbf{S}_{0}(x) = ({\rm sech}(x/x_{0}),0,{\rm cosh}(x/x_{0})),
\label{eqn:domain_wall_initial}
\end{equation}
connecting two asymptotic ferromagnetic states oriented
in opposite directions, that is $S^{z}(x\to \pm \infty) = \pm 1$.
Moreover, with no loss of generality one can set $x_{0}=1$ by an appropriate rescaling of space, time and interaction parameter $\delta$. It is worth stressing that despite integrability there are only rare instances~\cite{Satsuma}
when the direct scattering problem admits a closed-form solution. This is why it is customary to provide the initial data in the spectral representation. Remarkably, profiles given by Eq.~\eqref{eqn:domain_wall_initial} are another exception to the rule and do allow for a closed-form solution \cite{Gamayun19}. The associated classical monodromy matrix is fully determined by the spectral data $\{a(\lambda),b(\lambda)\}$, which consists of two complex-valued functions with simple time evolution, $a(\lambda,t) = a(\lambda,0)$, and $b(\lambda,t) = b(\lambda,0)\,e^{\ii \omega t}$,
where the magnon dispersion $\omega(\lambda)=\lambda^{2}-\varepsilon^{2}$. Here, $a(\lambda)$ encodes the local charges which pertain to moments of the density of states $\varrho(\lambda)=\log|a(\lambda)|^{2}$; the precise analytic structure of $a(\lambda)$ and $b(\lambda)$ will not concern us further. The time-evolved spin field can be retrieved by finally performing the inverse mapping (from the spectral plane to the phase space of physical fields) which takes the form of a Fredholm-type integral equation.
This is unfortunately where one encounters another practical challenge, as general solutions to the inverse transform appear to be possible only for purely discrete (reflectionless) potentials corresponding to multisoliton states, when they can be expressed in the form of determinants. The trouble with the initial conditions \eqref{eqn:domain_wall_initial} is that they involve a continuous spectrum. One is therefore obliged to solve the inverse scattering equations numerically (with time $t$ entering simply as a parameter in the kernel).

We now review the conclusions of Ref.~\onlinecite{Gamayun19}. Parametrizing the interaction parameter in terms of $\varepsilon = \ii \sqrt{\delta}$, there are three regimes to be distinguished: (i) the easy-plane regime with $\varepsilon^{2}>0$, (ii) the easy-axis regime with $\varepsilon^{2}<0$ and (iii) the isotropic point at $\varepsilon=0$. 
\medskip

\paragraph*{Easy-plane regime.}

In the easy-plane regime, the function $a(\lambda)$ is free of zeros in the upper-half complex plane, indicating that the spectrum of modes consists of purely dispersive radiation. As noted above, this precludes obtaining a solution in explicit form. However, if one is merely interested in explaining the ballistic character of the expanding interface, it is sufficient to operate in the ballistic scaling limit. Upon discarding a dispersive term, one is left with the Euler-scale equations \cite{Gamayun19}
\begin{equation}
    S^{z}_{t} - [(1-(S^{z})^{2})v]_{x} = 0,\qquad
    v_{t} - [(\varepsilon^{2}-v^{2})S^{z}]_{x} = 0,
\end{equation}
where the magnetization $S^{z}$ and phase $v=-\ii [\log S^{+}]_{x}$ (pertaining to canonical coordinates $S^{z}=\cos{(\theta)}\to p$ and $\phi \to q$) are
regarded as slow variables. One then seeks the simplest solution invariant under ballistic space-time scaling (i.e. depending only on the ray coordinate $\xi = x/t$) that connects two asymptotic vacuum regions at the fronts $\xi_{\pm}$, set by $S^{z}(\xi_{\pm})=\pm 1$. The solution yields $S^{z}(\xi)=\xi/(2|\varepsilon|)$ and $v=|\varepsilon|\equiv v_{0}$ with $\xi_{\pm}=\pm 2|\varepsilon|$, implying ballistic asymptotic growth $m(t)\simeq t\int^{v_{0}}_{0}\dd \xi (1-S^{z}(\xi))=|\varepsilon|t$. Further details can be found in Refs. \onlinecite{Gamayun19,Misguich2019}.

\medskip

\paragraph*{Easy-axis regime.}

In the easy-axis regime with $\delta>0$, the spectrum acquires discrete modes which physically correspond to solitary waves. Their precise number depends on the value of $\delta$. For generic $\delta$, the spectrum assumes in addition a continuous component consisting of dispersive radiation modes. Curiously, at the `resonances' $\varepsilon=\ii(2n+1)$ for $n\in \mathbb{N}$ the continuous spectrum entirely disappears. A distinctive feature of the discrete spectrum is the presence of (anti)kinks; in contrast with ordinary (asymptotically free) solitons, kinks represent \emph{static} topologically non-trivial configurations. Kinks have precisely the form \eqref{eqn:domain_wall_initial}, with characteristic width set by the anisotropy, $x_{0}(\delta)=\pm 1/\sqrt{\delta}$. More importantly, kinks persist in the spectrum at any $\delta > 0$. This fact alone implies that asymptotically $m(t)\sim t^{0}$, as the kink is stable against decay. For general $\delta>0$, one finds dispersive ballistically propagating magnon modes superimposed on a kink background, and possibly additional (discrete) breather modes that remain localized at the interface.
Kinks have been (semiclassically) quantized and reconciled with the Bethe Ansatz description in Ref.~\onlinecite{MIG20}.

\bigskip

\paragraph*{Isotropic point.}

Finally we discuss the subtle case of isotropic interactions. At $\varepsilon=0$, the kinks are no longer stable and the spectrum of the domain wall profile is once again a radiative continuum. Another peculiarity is that as $\lambda \to 0$, the density of states $\varrho(\lambda)=\log |a(\lambda)|^{2}$ develops a logarithmic singularity~\cite{Gamayun19}. Curiously, this can be seen as an artefact of the boundary condition under consideration;
if one considers instead a family of ``twisted'' domain walls, $\mathbf{S}=(\cos{\Phi},0,\sin{\Phi})$ with $\Phi=(\gamma/\pi){\rm arcsin}({\rm tanh}(x))$ and $\gamma \in [0,\pi)$, the singularity at $\lambda=0$ disappears for any $\gamma>0$, rendering $a(0)$ finite~\cite{Gamayun19}.
As we explain in turn, for general $\gamma \in [0,\pi)$ one has $\mathcal{D}(\gamma)<\infty$, whereas $\mathfrak{D}(\gamma \to \pi)$ is singular.

It turns out that rotational symmetry of the isotropic Landau--Lifshitz equation allows for a special class of self-similar solutions that depend on the diffusive scaling variable $\zeta = x/\sqrt{t}$ \cite{lakshmanan1976} and satisfy the ODE $-2\zeta \mathbf{S}_{\zeta}=\mathbf{S}\times \mathbf{S}_{\zeta \zeta}$. In this scaling limit, the domain wall initial profile \eqref{eqn:domain_wall_initial} gets mapped to a singular initial condition with a jump discontinuity at the origin. Due to the absence of a scale in the problem, it is natural to assume that at large times the solution converges to a self-similar form. One thus expects asymptotic scaling of the form $m(t)\simeq \mathfrak{D}(\gamma)t^{1/2}$, with the coefficient $\mathfrak{D}(\gamma)$ understood as the ``diffusion constant''. By numerical integration of the inverse scattering equations, a plot of $m(t)/\sqrt{t}$ versus time $t$
shows~\cite{Gamayun19} a robust signature of a mildly divergent diffusion constant as $\sim \log(t)$. Plotting $\mathfrak{D}$ as a function of twist angle $\gamma$ reveals a singularity upon removing the twist $\gamma \to \pi$.

The observed breakdown of normal diffusion can be partially reconciled with the aforementioned logarithmic divergence in the density of states in Ref.~\onlinecite{Gamayun19} by studying self-similar solutions~\cite{Gamayun_self}. The Landau-Lifshitz equation is known to appear in a different incarnation, where it governs the evolution of a vortex filament in the local induction approximation, $\mathbf{f}_{t}=\mathbf{f}_{x}\times \mathbf{f}_{xx}$, sometimes known as the Da Rios equations~\cite{Ricca1991}. Self-similar solutions to the latter equation were studied in Ref. \onlinecite{GRV03}. In the vortex filament picture, the parameter $\gamma$ defines the angle of a wedge initial profile, while energy $E=\mathbf{S}^{2}_{\zeta}$ is a conserved quantity, with filament curvature proportional to $\sqrt{E}$. A twist angle $\gamma$ maps to energy via the relation $\exp{(-\pi E/2)}=\cos{(\gamma/2)}$, signalling a divergence
of $\mathfrak{D}(\gamma)$ in the $\gamma\to \pi$ limit. Despite this suggestive physical picture, a fully rigorous justification for emergent anomalous diffusive scaling at the isotropic point is still lacking at present. One possible route forward would be an asymptotic analysis of the associated Riemann--Hilbert problem \cite{Deift1993,DIZ93}.

\subsection{Spin spirals}

Spin spirals provide another conceptually simple and experimentally natural class of far-from-equilibrium initial states in spin chains, and their relaxation dynamics has been successfully probed in optical lattices.
In Ref.~\onlinecite{hild2014far}, the authors measured the far-from-equilibrium relaxation of magnetization in an isotropic Heisenberg ferromagnet realized by ultracold $^{87}\mathrm{Rb}$ atoms. Initializing the system in ``transverse spin spiral states'' (at finite energy density),
of the form $\ket{\chi({\bf k})}=\prod_{i}(\ket{\uparrow}_{i} + e^{\ii {\bf k}\cdot {\bf x}}\ket{\downarrow})$,
(where ${\bf k}$ denotes the imprinted wavevector and ${\bf x}$ location on the lattice) and using single-site spectroscopy to measure the decay of spin correlations, they detected normal spin diffusion in one dimension (relating the diffusion constant $D$ to the characteristic decay time via $1/\tau = D|{\bf k}|^{2}$), whereas on a two-dimensional lattice (where the model is not integrable) they reported having found traces of superdiffusive dynamics. We note that a theoretical program for studying relaxation of spin spirals in the 2D Heisenberg model has been initiated very recently~\cite{rodrigueznieva2020transverse}.

Concurrently, experimental interest in spin spirals has also been resurrected. By exploiting the capacity of $^7\mathrm{Li}$ ions to realize effective \emph{anisotropic} Heisenberg spin couplings within the usual two-component Bose--Hubbard model for the first time, Ref. ~\onlinecite{jepsen2020spin} was able to study far-from-equilbrium relaxation dynamics in the spin-$1/2$ XXZ chain over a large range of anisotropies. The experiment focused on probing relaxation dynamics of spin spiral states of wavelength $\lambda$ (in the XZ-plane) by performing imaging in the perpendicular $x$-direction, thereby measuring time dependence of the expectation value of the local magnetization $\hat{S}^{z}$. By measuring the decay time $\tau(k) \sim 1/k^{\alpha}$, they observed a continuous range of dynamical exponents $\alpha$ as a function of quasimomentum $k$, spiral wavelength $\lambda$ and interaction anisotropy $\Delta$. Remarkably, while in the easy-plane regime $|\Delta |<1$ (and throughout the entire antiferromagnetic phase $\Delta < -1$) they encountered the expected ballistic behavior ($\alpha \approx 1$), they detected an abrupt change at approximately $\Delta = 1$ upon crossing into the easy-axis phase with sub-diffusive exponents $\alpha \gtrsim 2$. Such behavior, which is markedly different from predictions within linear-response theory, has not been theoretically explained yet.

Although the theoretical situation is somewhat obscure at present, Ref.~\onlinecite{jepsen2020spin} offers one intriguing clue. The authors consider the evolution of the spin modulation $S^z_k \equiv \sum_i \cos(k i) S^z_i$ in the short-time expansion. By time-reversal invariance of the XXZ Hamiltonian, one can argue that $\langle S^z_k(t) \rangle \approx \langle S^z_k(t = 0) \rangle + t^2 \langle [H, [H, S^z(k)]] \rangle \equiv \langle S^z_k(t = 0) \rangle + (t/\tau_k)^2$, where the expectation values are taken in the spin spiral. This double commutator therefore captures the short-time relaxation rate $\tau_k$ of the spiral. Remarkably, the easy-plane and isotropic regimes differ in the asymptotics of this quantity as $k \rightarrow 0$: in the easy-plane phase, $\tau_k \sim (k \sqrt{1 - \Delta})^{-1}$, and at the isotropic point $\tau_k \sim 1/k^2$. It is unclear whether this change in the short-time behavior also persists to late times (or at least to long enough times that it captures the experimentally observed behavior).

\section{Anomalous transport from emergent or approximate integrability}\label{lowT}

\subsection{Anomalous diffusion of energy in Luttinger liquids}

We now discuss an example of anomalous diffusion that can occur in generic, chaotic many-body systems, in contrast to the integrable examples that have been considered so far. This is the free expansion of a localized heated region in a one-dimensional metal into a bulk ground state, which exhibits anomalous, nonlinear diffusion of energy, of the type discussed in Sec.~\ref{NLD}.

Recall that the universal theory of interacting fermions in one dimension is the Luttinger liquid, which is the analogue of the Fermi liquid in $d>1$, with several basic differences arising from its reduced dimensionality~\cite{Giamarchi}. The low-energy limit of the Luttinger liquid is a free bosonic field theory, which is well-known to exhibit ballistic transport of its conserved charges, including energy. By contrast, realistic metals include interactions, which break integrability and lead to thermalization and conventional, diffusive linear response, even in the absence of disorder. Nevertheless, the proximity of non-integrable Luttinger liquids to a conformally invariant critical point at $T=0$ implies the divergence of linear response transport coefficients as $T \to 0$. It is this singular dependence of the thermal conductivity on temperature (specifically, the ``dangerous irrelevance'' of perturbations to the $T=0$ critical point~\cite{DSTransport}) that allows for anomalous energy diffusion in one-dimensional metals.

We first illustrate the general principles leading to such anomalous behaviour, before describing an explicit example for which anomalous diffusion has been observed numerically~\cite{Bulchandani12713}. Therefore consider the effective action for a non-integrable Luttinger liquid with an irrelevant density-wave-type instability,
\begin{align}
H = \frac{u}{2}\int_0^L dx\, (\Pi^2 + (\partial_x \phi)^2) + h \int_0^L \dd x \, \cos{\alpha \phi} + \ldots,
\end{align}
where the momentum and phase degrees of freedom satisfy canonical commutation relations $[\Pi(x),\phi(x')] = i\delta(x-x')$ and the condition $\alpha^2 > 8 \pi$ ensures irrelevance of the cosine perturbation. Here the ellipsis denotes additional, less relevant interaction terms that break integrability (note that at least two cosine terms are required to break integrability of the field theory, since perturbing the one-dimensional free boson by a single cosine term yields the integrable sine-Gordon model).

The charge conductivity of $H$ can be obtained via the Kubo formula~\cite{Kubo1,Sirker2011}
\begin{equation}
\sigma(q,\omega) = \frac{K}{2\pi} \ii \omega \langle \phi \phi \rangle^{\rm ret}(q,\omega).
\end{equation}
Assuming that $h$ is the dominant irrelevant coupling, a timescale for relaxation of charge can be extracted from the boson self-energy, via the relation~\cite{Sirker2011,OshikawaAffleck,Luther74,LutherPRB74}
\begin{equation}
\label{eq:reltime}
\tau \propto \lim_{\omega \to 0} \frac{\omega}{\mathrm{Im}[\Pi_{h}(q=0,\omega)]} \sim h^{-2}T^{3-\alpha^2/2\pi},
\end{equation}
where the scaling with staggered field and temperature comes from the leading (in $h$) contribution to the self-energy. (Note that the scaling form Eq. \eqref{eq:reltime} can also be obtained from a simple scaling argument~\cite{Huang2013}.)

Implicit in the analytical derivation of this result for the charge conductivity is a non-perturbative summation of terms via the Dyson series, which is possible because the Kubo formula for the (bosonized) charge conductivity involves only a two-point function. Let us now consider the thermal conductivity. Modulo the well-known subtleties of using the Kubo formula to compute thermal conductivities\cite{Kubo2,Luttinger}, linear response theory predicts that the a.c. thermal conductivity is related to a four-point function, schematically $\kappa(q,\omega) \sim \langle \Pi \phi \, \Pi \phi \rangle^{\rm ret}(q,\omega)$, and hence inaccessible via the Dyson series. Since there are no comparably rigorous non-perturbative methods for evaluating four-point functions, we shall treat the value of $\kappa(q,\omega)$ as unknown and attempt to proceed on general grounds.

First, by emergent conformal invariance of the low energy theory, it is expected that the DC thermal conductivity behaves as a power-law in temperature, $\kappa_{\rm d.c.} \sim T^\lambda$. Since thermal transport in interacting Luttinger liquids becomes ballistic as $T \to 0$, it follows that the exponent $\lambda < 0$. A na{\"i}ve estimate for the exponent $\lambda$ comes from Wiedemann-Franz scaling, which assumes that charge and thermal transport are controlled by the same relaxation time $\tau$, so that by dimensional analysis $\kappa_{\rm d.c.} \sim T \sigma_{\rm d.c.} \sim T^{4-\alpha^2/2\pi}$. (Notice that for $\alpha^2 > 8 \pi$ irrelevant, this is guaranteed to diverge as $T \to 0$.) The key point is that even if the Wiedemann-Franz law $\lambda = 4 - \alpha^2/2\pi$ does not hold, a low-temperature, power-law divergence of $\kappa_{\rm d.c.}(T)$, that is expected on general grounds, will give rise to low-temperature superdiffusion of energy.

To see this, notice that the linear response heat current, evaluated at leading order in temperature and derivatives about $T=0$, is given by
$j_Q \sim -T^{\lambda} \partial_x T$. In flows that are driven solely by temperature gradients (we assume particle-hole symmetry) the heat current coincides with the energy current. Combined with the equation of state for the local energy density in a low-temperature Luttinger liquid, $\rho_E \sim T^2$, that follows by conformal invariance of the $T=0$ fixed point, we deduce a non-linear diffusion equation for the local energy density,
\begin{equation}
\label{eq:heatNLDE}
\partial_t \rho_E = D \partial^2_x(\rho^{m}_E),
\end{equation}
with exponent $m = (1+\lambda)/2$ and $D$ a non-universal prefactor. Since $\lambda <0$, this is a fast diffusion equation in the terminology of Sec.~\ref{NLD}, and its space-time scaling $x \sim t^\alpha$ is determined by the superdiffusive dynamical exponent
\begin{equation}
\label{eq:exponent}
\alpha = \frac{1}{m+1} = \frac{2}{3+\lambda} \in (2/3,1).
\end{equation}

As discussed in Sec.~\ref{NLD}, at any finite bulk temperature $T$, the linear response behaviour of Eq.~\eqref{eq:heatNLDE} is diffusive. However, we expect a crossover from transient anomalous behaviour on a timescale $t_D(T) \sim T^{\lambda-1}$ (the effective diffusion time), that diverges as $T \to 0$. In particular, free expansion of a localized heated region into a bulk ground state will be anomalous, and characterized by the spreading exponent in Eq. \eqref{eq:exponent}. This is true \emph{even under the assumption} that thermalization is strong enough in the low temperature regime for linear response theory to be applicable.

\begin{figure}[t]
\includegraphics[width=0.7\linewidth]{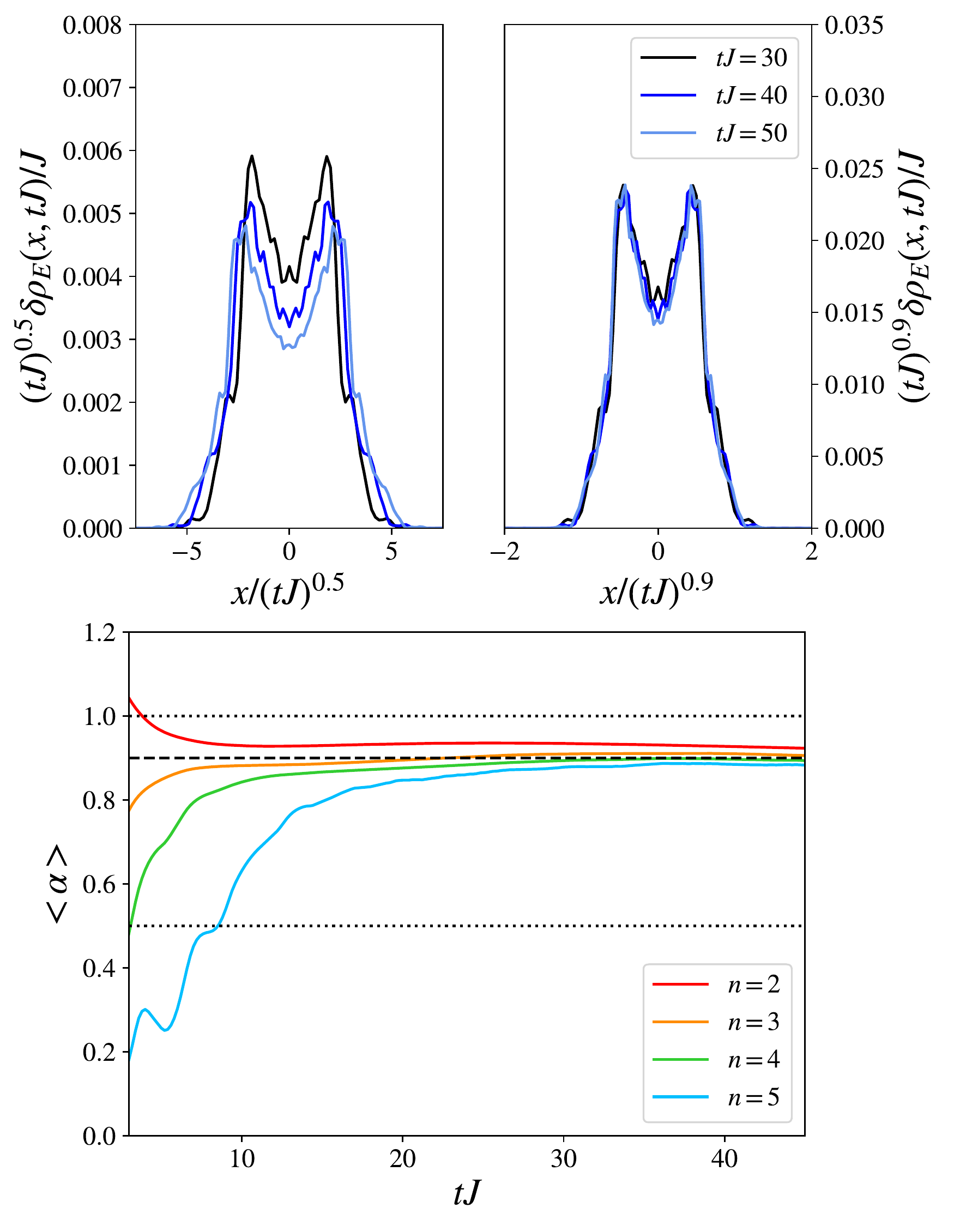}
\caption{Anomalous scaling for a spreading thermal wavepacket in the interacting Luttinger liquid phase of the spin chain Hamiltonian Eq. \eqref{eq:stagfield}. The initial condition is given by Eq. \eqref{eq:initcond} and model parameters are $\Delta=-0.85$, $h=0.2$ $\beta J=12$, $\beta_MJ=8$, $l=2$. \textit{Top}: Diffusive scaling of the spreading wavepacket (\textit{left}) versus superdiffusive scaling (\textit{right}) with exponent $\alpha \approx 0.9$. Scaling collapse to a superdiffusive exponent shows a marked improvement compared to normal diffusion. \textit{Bottom}:
rates of change of log absolute moments, as in Eq. \eqref{eq:1pscaling}, exhibit a good scaling collapse to a single exponent $\alpha \approx 0.9$ (dashed line), that is neither ballistic ($\alpha=1$) nor diffusive ($\alpha=0.5$) (dotted lines). Figure based on tDMRG simulations performed by C. Karrasch and reported in Ref. \onlinecite{Bulchandani12713}.}
\label{fig:NLDE}
\end{figure}

These predictions were corroborated~\cite{Bulchandani12713} by tDMRG numerical simulations of a non-integrable spin chain that realizes Luttinger liquid physics, namely the spin-$1/2$ XXZ chain perturbed by an integrability-breaking staggered field,
\begin{equation}
\label{eq:stagfield}
H = \frac{J}{4}\sum_{j=1}^L \left(\sigma_j^x\sigma_{j+1}^x+\sigma_j^y\sigma_{j+1}^y + \Delta \sigma_j^z \sigma_{j+1}^z\right) +\frac{h}{2} \sum_{j=1}^L (-1)^j \sigma_j^z.
\end{equation}
For $\Delta \neq 0$ and $h>0$, this model can be verified to exhibit a crossover to Wigner--Dyson level statistics via exact diagonalization~\cite{Huang2013}. Its low-energy effective action for infinitesimal $h$ is given by standard bosonization techniques~\cite{Giamarchi,Lukyanov1998}, and has the form
\begin{equation}
H = \frac{u}{2} \int_0^L \dd x \, (\Pi^2 + (\partial_x \varphi)^2) + ch \int_0^L \cos {(2\sqrt{\pi K} \phi)} + \ldots
\end{equation}
where $u,\,c$ are non-universal constants and the ellipsis includes other irrelevant terms (Umklapp, band curvature, etc.). For $K>2$, i.e. $-1 <\Delta<-\sqrt{2}/2$, the band curvature term is irrelevant and the Hamiltonian Eq. \eqref{eq:stagfield} remains in a gapless Luttinger liquid phase for $h>0$. It was found numerically~\cite{Huang2013} that this phase exhibits charge transport with a charge conductivity $\sigma \sim h^{-2}T^{3-2K}$ in agreement with analytical bosonization predictions, and suggesting that the staggered field is the dominant irrelevant perturbation for sufficiently large $L$. Subsequent work~\cite{Bulchandani12713} studied heat transport in the model Eq. \eqref{eq:stagfield}, by numerically simulating time evolution from locally thermal initial conditions
\begin{equation}
\label{eq:initcond}
\beta(x) = \beta - (\beta-\beta_M)e^{-(x/l)^2}
\end{equation}
with a central ``hot spot'' $\beta_M < \beta$, and examining the time evolution of the absolute moments of the resulting thermal wavepacket. For nonlinear diffusion, these should obey one-parameter scaling
\begin{align}
\label{eq:1pscaling}
\frac{1}{n} \frac{d \log \langle|x|^n\rangle}{d \log{t}} \to \alpha, \quad t \to \infty,
\end{align}
in time. Some sample simulations ($\Delta=-0.85$, $h=0.2$ $\beta J=12$, $\beta_MJ=8$) are recorded in Fig. \ref{fig:NLDE}. There is a clear collapse to an anomalous space-time scaling exponent $\alpha \approx 0.9$ that is neither ballistic ($\alpha=1$) nor diffusive ($\alpha=0.5$). Such single-parameter scaling of the moments is consistent with the simple anomalous diffusion model Eq. \eqref{eq:heatNLDE}, although the shape of the scaling profiles is non-monotonic away from $x=0$ and thus differs from the Barenblatt--Pattle solutions depicted in Fig. \ref{fig:BPsolns}. One possible explanation for the discrepancy at $x=0$ is that the initial spreading of the wavepacket occurs so fast that the system does not have time to reach local thermodynamic equilibrium, invalidating the hydrodynamic description Eq. \eqref{eq:heatNLDE} at short times. 

To summarize, the expansion of energy into a bulk ground state of a one-dimensional metal yields a simple and generic example of anomalous diffusion in a quantum many-body system. Since materials exhibiting interacting Luttinger liquid physics range from spin ladders~\cite{spinLL} to carbon nanotubes~\cite{Bockrath1999}, we expect that such behaviour should be observable experimentally. Indeed, one established physical realization of ``hot spot'' initial conditions occurs in experiments that use laser irradiation of small, localized regions to study the dynamics of outward heat flux~\cite{henseldynes,photothermalmic,HotSpot1,HotSpot2}.

On the theoretical side, the most pressing open question related to this anomalous diffusion phenomenon is whether there exists an accurate, non-perturbative theory of the thermal conductivity in interacting Luttinger liquids, which does not seem to have been attempted so far. Although there are good reasons to expect continuous power law scaling, $\kappa \sim T^{\lambda(K)}$, as for the charge conductivity, detailed analytical or numerical justification for this ansatz would be desirable. We also note there are certain physical regimes in Luttinger liquids for which the linear response heat conductivity exhibits an essential singularity as $T \to 0$, for example, the case of thermally activated momentum relaxation~\cite{Rosch} $\kappa \sim e^{\Delta E/T}$. In this case, an analysis in terms of fast diffusion ceases to be consistent as $T \to 0$, possibly indicating a crossover to ballistic behaviour. Another interesting open question is how the picture sketched above is modified in Luttinger liquids away from half-filling, for which thermopower effects become important and theory~\cite{Matveev} again predicts unusually slow relaxation of energy as $T \to 0$.

\subsection{Anomalous low-temperature transport in Haldane antiferromagents and nonlinear sigma models}\label{NLSM}

The study of $S = 1$ spin chains at low but nonzero temperatures has been of longstanding interest, given both the experimental relevance of the topic and Haldane's remarkable prediction~\cite{Haldane83} of a spectral gap $\varDelta$ opening only in chains of integer spin $S$. Haldane's argument invokes an effective field theory computation at large $S$, yielding a classical nonlinear sigma model which, upon quantization, includes a topological $\varTheta$-term that is sensitive to whether $2S$ is even or odd.

There have been several experiments with inelastic neutron scattering on quasi-1D Haldane-gap materials~\cite{PhysRevLett.69.3571,PhysRevB.38.543,Steiner,PhysRevB.50.15824,PhysRevLett.96.047210}, which support the Haldane gap conjecture. On the theoretical side, various competing effective descriptions have been proposed to compute spin transport and the relaxation rate of nuclear magnetic resonance (NMR) in the low-temperature regime. Unfortunately, these various methods lead to mutually inconsistent conclusions. Spin transport at long wavelengths $k\sim 0$ has recently been revisited~\cite{NMKI19} using GHD.

We begin by briefly reviewing the earlier findings. 
Various theories of NMR for Haldane chains have been developed and compared in Refs.~\onlinecite{PhysRevB.50.9265,SagiAffleck96}. The computations are largely based on computing matrix elements at zero temperature, which are then extended to the regime of interest $\omega_{N}\ll T\ll \varDelta$ (with $\omega_{N}$ being the NMR frequency) by simply inserting the Boltzmann weights and subsequently integrating over the entire momentum range.

An alternative semiclassical approximation was put forward in Refs.~\onlinecite{SD97,DS00},
in which quantum excitations are treated as (interacting) classical trajectories carrying spin projections $m\in \{-1,0,1\}$) (with a \emph{momentum independent} two-particle scattering matrix $S^{m_{1}m_{2}}_{m^{\prime}_{1}m^{\prime}_{2}}=(-1)\delta_{m_{1}m^{\prime}_{2}}\delta_{m_{2}m^{\prime}_{1}}$),
yielding a purely classical statistical model that had been solved decades earlier by Jepsen. In particular, the semi-classical theory predicts the spin diffusion constant
\begin{equation}
    D^{\rm cl} = \frac{c^{2}e^{\beta \varDelta}}{\varDelta(1+2\cosh{(\beta h)})},
    \label{eqn:semiclassical_diffusion}
\end{equation}
and the NMR relaxation time $T_{1}$ scaling as $T_{1}\sim 1/\sqrt{D_{s}h}$, consistent with
experimental observations \cite{Takigawa}. As argued in Ref.~\onlinecite{SD97}, one might expect such a semi-classical description to be effectively exact at sufficiently low temperatures, since the density of quasiparticles is suppressed
as $\sim e^{-\beta \varDelta}$, and therefore the spacing between quasiparticles greatly exceeds the thermal de Broglie wavelength of each quasiparticle. The semi-classical picture of Ref.~\onlinecite{SD97} was expected to hold in the regime $T\ll \varDelta$ at timescales $t\gg 1/T$ (with $h/T$ arbitrary). These predictions were nonetheless difficult to confirm as the regime of interest is difficult to access in numerical simulations.

A more refined computational approach makes use of the truncated form-factor expansion~\cite{Konik2003}
(see Ref.~\onlinecite{EK09} for a review), which unlike the semi-classical approximation uses the \emph{full} scattering matrix of the $O(3)$ nonlinear sigma model (NLSM). It is widely believed that computations with bare (as opposed to dressed) form factors should accurately describe spectral functions at low temperatures. Accordingly, Ref.~\onlinecite{Konik2003} rests on an assumption that one can truncate the spectral sum for the dynamical correlation functions $\langle \hat{m}^{a}(x,t)\hat{m}^{a}(0,0) \rangle^{c}_{\beta}$ can be efficiently truncated at $n$-particle states (which are weighted by a factor $e^{-n\beta \varDelta}$). This approach yields a divergent relaxation time $T_{1}\sim \log{h}$, with thermally activated temperature-dependence $T_{1}\sim e^{\beta \varDelta}$, in agreement with earlier form-factor computations~\cite{SagiAffleck96}, but in clear disagreement with the semi-classical analysis, which instead yields $T_{1}\sim e^{3\beta \varDelta/2}$. Moreover, Ref.~\onlinecite{Konik2003} also includes a computation of the d.c. spin conductivity, reporting a finite spin Drude weight even in the limit of half filling, that is $\lim_{h\to 0}\mathcal{D}_{\rm spin}(h)\neq 0$, and no diffusive subleading correction (the same statement, based on a TBA computation, indeed appeared earlier~\cite{Fujimoto99}), again conflicting with the semi-classical result predicting a finite spin diffusion constants and zero Drude weight.

We now turn to a recent study~\cite{NMKI19} which leverages the capabilities of the GHD toolbox. From a technical standpoint, the crucial improvement
was to account for the thermal dressing of quasiparticles in an exact non-perturbative fashion, in contrast to earlier approaches.
The low-temperature limit can only be safely taken at the very end of the computation. The findings
of Ref.~\onlinecite{NMKI19}, to which we now turn, radically differ from the previous conclusions reviewed above.

\medskip

\paragraph*{Nonlinear sigma model.}

The quantum $O(3)$ nonlinear sigma model is a Lorentz invariant QFT with Euclidean action
\begin{equation}
    \mathcal{S}[\hat{\bf n}] = \frac{1}{2g}\int \dd x\, \dd t\,
    \left((\partial_{t}{\bf n})^{2}-(\partial_{x}{\bf n})^{2}\right),
    \label{eqn:O3_action}
\end{equation}
where $\hat{\bf n}=(\hat{n}^{x},\hat{n}^{y},\hat{n}^{z})$ is a three-component vector field obeying the unit normalization constraint $\hat{\bf n}^{2}=1$, and $g$ is the coupling constant. At second order in derivatives, there is additionally an allowed topological $\varTheta$-term:
\begin{equation}
    \mathcal{S}_{\varTheta}[\hat{\bf n}] = \ii\frac{\varTheta}{4\pi}\int \dd x\, \dd t\,
    \hat{\bf n}\cdot \partial_{t}\hat{\bf n}\times \partial_{x}\hat{\bf n}.
    \label{eqn:O3_topological}
\end{equation}
The $O(3)$ NLSM is the effective low-energy theory of one-dimensional $SU(2)$-symmetric antiferromagnetic chains~\cite{Haldane83}, $\hat{H} \simeq J\sum_{j}\hat{\bf S}_{j}\cdot \hat{\bf S}_{j+1}$, with spin exchange coupling $J>0$. By assuming $S$ is large, and making the continuum approximation, fluctuations of lattice spins can be split into staggered and ferromagnetic fields~\cite{Haldane83}, $\hat{\bf S}_{j}\approx S(-1)^{j}\hat{\bf n}+\hat{\bf m}$, where $\hat{\bf m}=g^{-1}{\bf n}\times {\bf p}$ is the angular momentum generating rotations about $\hat{\bf n}$, i.e., $\hat{\bf m}\cdot \hat{\bf n}=0$, and ${\bf p}=g^{-1}\partial_{t}\hat{\bf n}+(\varTheta/4\pi)\hat{\bf n}\times \partial_{x}\hat{\bf n}$ is the canonical momentum conjugate to $\hat{\bf n}$.
In the Hamiltonian description, one finds~\cite{Haldane83}
\begin{equation}
    H_{O(3)} = \frac{v}{2}\int \dd x
    \left[g\left(\hat{\bf m} + \frac{\varTheta}{4\pi}\partial_{x}\hat{\bf n}\right)^{2}+g^{-1}(\partial_{x}\hat{\bf n})^{2}\right],
\end{equation}
with spin-dependent coupling constant $g=2/S$ [to match Eq.~\eqref{eqn:O3_action} one should put $v=2JS\to 1$].
In antiferromagnetic $SU(2)$ spin-$S$ chains, the topological angle depends on whether $S$ is integer or half-integer, $\varTheta = 2\pi S$.
Most remarkably, the spectrum of the quantized $O(3)$ sigma model acquires a gap (via dynamical mass generation) for $\varTheta = 0$, whereas $\varTheta = \pi$ precludes opening a gap.

At the classical level, the $O(3)$ NLSM describes three real scalar fields that transform in the fundamental (vector) representation under $O(3)$ rotations. The vacuum state is continuously degenerate and can point in any direction on the $2$-sphere $S^{2}\cong O(3)/O(2)$.
In a state with finite vacuum expectation value, $\langle {\bf n}\rangle \neq 0$, the full $O(3)$ symmetry is said to be spontaneously broken down to $O(2)$. The low-energy spectrum comprises gapless type-I (i.e., linearly dispersing) Goldstone modes, whose
number equals ${\rm dim}\,O(3)-{\rm dim}\,O(2) = 2$, matching the dimension of the $S^{2}$ target space.
Quantization however radically affects this picture and, in the absence of the topological term ($\varTheta = 0$),
one ends up having instead \emph{three} gapped (massive) elementary excitations. 
%
This nonperturbative mechanism, by which the ``would-be massless'' Goldstone modes rearrange themselves and acquire a mass $\varDelta$, can be understood using standard renormalization-group calculations~\cite{weinbergtext}. 

\medskip

\paragraph*{Factorizable scattering matrices.}

Another remarkable fact about the $O(3)$ NLSM is that it is integrable for both $\varTheta=0$ and $\varTheta=\pi$. By rotational invariance, the model possesses conserved Noether currents. Owing to integrability \cite{Wiegmann1985}, there is, in addition, an infinite sequence of local (and non-local) conservation laws in the system. Local conservation laws can be thought as a consequence of complete two-body reducibility of the underlying many-body scattering matrix \cite{ZZ78,ZZ79}. As just noted, in the non-topological version of the model, $\varTheta = 0$, the elementary excitations form a spin triplet of massive bosons with dispersion $e(k)=\sqrt{k^{2}+\varDelta^{2}}$ (absent an external field), which are associated with a pair of creation and annihilation operators $Z_{a}(\theta)$ ($a\in \{x,y,z\}$), with rapidity variable $\theta$ parametrizing their bare momentum $k(\theta)=\varDelta\sinh{(\theta)}$ and energy $e(\theta)=\varDelta\cosh{(\theta)}$. Multiparticle states are of the form $\ket{\theta_{1},\ldots,\theta_{n}}=\prod_{i=1}^{n}Z^{\dagger}_{a_{n}}(\theta_{n})\cdots Z^{\dagger}_{a_{2}}(\theta_{2})Z^{\dagger}_{a_{1}}(\theta_1)\ket{0}$,
where the Fock vacuum $\ket{0}$ satisfies $Z_{a}(\theta)\ket{0}=0$. Quasiparticle excitations belong to noncommuting operators obeying the Faddeev--Zamolodchikov algebra \cite{ZZ92}, whose `structure constants' are prescribed by the $S$-matrix.

Even though quasiparticle collisions are purely elastic, the many-body scattering matrix is still non-diagonal in the quasi-particle basis
due to non-trivial exchange of spin degrees of freedom upon collisions. This is overcome by adjoining to the elementary excitations a set of auxiliary degrees of freedom, interpreted as magnons propagating in a fictitious system of physical quasiparticles. Having done that, the equilibrium partition sum can be subsequently computed in the framework of TBA, yielding an infinite coupled system of Fredholm-type integral equations. To highlight the structural similarity with the isotropic Heisenberg spin chain, it is instructive to display these equations explicitly. By assigning each quasiparticle species a spectral density (in rapidity space), the total density of available states satisfies Bethe--Yang integral equations of the following type~\cite{ZZ92,NMKI19}
\begin{eqnarray}
\rho^{\rm tot}_{0} = \frac{\partial_{\theta}k}{2\pi} + s \star \overline{\rho}_{2},\qquad
\rho^{\rm tot}_{s} = \delta_{s,2}(s\star \rho_{0})
+ s\star (\overline{\rho}_{s-1}+\overline{\rho}_{s+1}),\qquad s\in \mathbb{Z}_{\geq 0},
\end{eqnarray}
with convolution kernel $s(\theta)=1/[2\cosh{(\theta)}]$. The subscript $0$ refers here to physical relativistic particles, while $s\geq 1$ pertain to the aforementioned auxiliary magnonic excitations (the bar indicates the density of holes). By conventionally introducing thermodynamic $\mathcal{Y}$-functions $\mathcal{Y}_{s\geq 0}(\theta)=\overline{\rho}_{s}(\theta)/\rho_{s}(\theta)$ (encoding dressed energies through $\log{\mathcal{Y}_{s\geq 0}(\theta)}=\beta\,\varepsilon_{s}(\theta)$), the TBA equations can be presented in a compact group-theoretic form (after performing
the particle-hole transformation, $\rho_{0}\leftrightarrow \overline{\rho}_{0}$)
\begin{eqnarray}
    \log \mathcal{Y}_{s} = -\delta_{s,0}\beta e + s\star I^{D_{\infty}}_{s,s'}\log(1+\mathcal{Y}_{s'}),
\end{eqnarray}
where $I^{D_{\infty}}$ stands for the incidence (adjacency) matrix of the $D_{\infty}$ Dynkin diagram with a bifurcation at node $s=2$ (that is $I_{s,s'}=1$ when node $s'$ is adjacent to $s$, and otherwise zero) that captures the morphology of interparticle interactions. At the same time, this reveals that an infinite tower of internal magnonic modes attached to the momentum-carrying fundamental bosonic excitation indeed correspond to the magnons of the isotropic Heisenberg chain.
We mention in passing that TBA equations with similar structure occur in entire infinite families of integrable QFTs with isotropic degrees of freedom~\cite{Fendley99,Fendley01,Fendley_topo}; the $O(3)$ NLSM is thus a canonical example of an integrable QFT displaying anomalous transport properties (see the discussion in Ref.~\onlinecite{ssd}).

\medskip

\paragraph*{Spin transport.}

By Noether's theorem, the $O(3)$ NLSM model possesses a conserved Lorentz two-current
\begin{equation}
    \partial_{\mu}\hat{\bf j}_{\mu} = 0,\qquad
    \hat{\bf j}_{\mu} = g^{-1}\hat{\bf n}\times \partial_{\mu}\hat{\bf n}.
\end{equation}
The conserved Noether charge is therefore the ferromagnetic order parameter ${\bf m}(x,t)$ which governs fluctuations at large wavelengths $k\sim 0$.
We stress that GHD can only access $k\sim 0$ but not $k\sim \pi$, which instead pertains to antiferromagnetic fluctuations governed by ${\bf n}(x,t)$ (which is not a density of a local conserved current).

We now finally turn to the nature of magnetization transport in the quantum $O(3)$ NLSM within GHD~\cite{NMKI19}. Although we shall focus exclusively on low temperatures, mainly to allow for comparison with Haldane spin chains, the arguments presented here apply to the pure $O(3)$ NLSM field theory at any temperature. We wish to emphasize that thermodynamic equilibrium states are characterized by a non-zero density of energy and entropy. This fact alone tells us that computations based on a dilute gas of excitations above the Fock vacuum (used in the bare form factor technique) are bound to fail as they are not able to capture the effective renormalization of charges due to interparticle (elastic) interactions. This is true (at least for asymptotic hydrodynamic properties) down to arbitrary low temperatures; one has to first wait for long enough (possibly exponentially long in inverse temperature) for the quasiparticle to get dressed by the surrounding thermal environment and only then compute the thermodynamic quantities of interest. To be concrete, consider any local charge density $\hat{q}$. The dominant matrix elements at large wavelengths are those coming from particle-hole type of excitations (for all quasiparticle species, including bound states), reading
\begin{equation}
    \langle \{\rho_{s}(\theta)\} |\hat{q}(x)| \{\rho_{s}(\theta);\theta^{p}_{s},\theta^{h}_{s}\} \rangle_{\beta,h} = e^{\ii \Delta k_{s}x}q^{\rm dr}_{s},
\end{equation}
where $\theta^{p}_{s}$ and $\theta^{h}_{s}$ denote rapidities of the added and removed quasiparticles (labelled by integers $s\geq 0$),
respectively, whereas $q^{\rm dr}_{s}(\theta)$ denote the dressed values of $q_{s}(\theta)$ associated to a local density $\hat{q}$.
Hydrodynamics on the Euler scale corresponds to the regime of vanishingly small momentum difference $\Delta k_{s} = k_{s}(\theta^{p})-k_{s}(\theta^{h})$.
The leading higher-order contributions, which come from two particle-hole excitations or single particle/hole excitations, become suppressed at large times as $\sim 1/t$ \cite{cubero2020generalized}. For instance, the dressed magnetization $m^{\rm dr}_{s}$ that enters in the mode resolution of the d.c. conductivity does not equal simply the number of magnetization quanta $m^{\rm bare}_{s} = s$ but depends both on temperature and also (quite abruptly, in fact) on the chemical potential $h$ coupling to magnetization. The effect of dressing becomes most pronounced near half filling where (in close analogy with integrable ferromagnetic chains) $m^{\rm dr}_{s}\sim s^{2}h$ for $s\ll 1/h$, crossing over to $m^{\rm dr}_{s}\approx s$ at large $s\gg 1/h$ for ``giant magnons''.

We now summarize the main statements of Ref.~\onlinecite{NMKI19} concerning the nature of spin transport. The spin Drude weight $\mathcal{D}_{s}$ is finite for any $T>0$ and $h>0$ and exactly vanishes as $h\to 0$. This is merely a manifestation of particle-hole symmetry at $h=0$ which indeed ensures that $\lim_{h\to 0}m^{\rm dr}_{s}(h)=0$.
We remark that a previous computation carried out in Ref.~\onlinecite{Fujimoto99} is entirely correct, except at the last step where some sloppiness occurred in analyzing the low-temperature behavior (assuming a large ratio $h/T$ for $T,h\ll \varDelta$, instead of demanding $h\ll T$). An explanation for the finite spin Drude weight at half filling which invokes quantum bilocal conserved currents~\cite{Konik2003} is likewise not valid; the Yangian symmetry generators~\cite{Luscher78} do not constitute local thermodynamic degrees of freedom (spanning the mode space of GHD). Indeed, analogous bilocal charges which become exact thermodynamic conserved quantities exist in the Heisenberg XXX spin chain, where similarly they play no essential role.

On the sub-ballistic scale, the Drude weight is generically accompanied by a finite diffusive correction of the form~\cite{NMKI19}
\begin{equation}
    D(T,h) = \sum_{s}\rho_{s,\theta}(1-n_{s,\theta})
    |v^{\rm eff}_{s,\theta}| \mathcal{W}^{2}_{s,\theta},\qquad
    \mathcal{W}_{s,\theta} = \lim_{s'\to \infty}
    \frac{\mathcal{K}^{\rm dr}_{s\,\theta,s'\theta^{\prime}}}{\rho^{\rm tot}_{s',\theta^{\prime}}},
\end{equation}
which yields a finite value as long as $h \neq 0$. In the $h\to 0$ limit, explicit computation of dressed two-body scattering kernels
$\mathcal{K}^{\rm dr}_{s\,\theta,s'\,\theta^{\prime}}$ can be circumvented by exploiting the ``magic formula'' (see Eq.~\eqref{eqn:magic_formula}). This allows one to compactly express $D_{s}(T,h)$ at half-filling $h=0$ through the curvature of the Drude self-weight \cite{DS17} with respect to the filling parameter $\nu = 4T\langle \hat{S}^{z} \rangle_{T,h} = 4T\chi(T,0)h+\mathcal{O}(h^{2})$,
\begin{equation}
   \lim_{h\to 0}D_{s}(T,h) =
   \frac{\partial^{2}\mathcal{D}^{\rm self}_{s}(T,\nu)}{\partial \nu^{2}}\Big|_{\nu=0}.
\end{equation}
Most importantly, in approaching half-filling $D_{\rm spin}(T,h)$ \emph{diverges} as $\sim 1/h$ at any finite temperature.
Upon restoring global $O(3)$ invariance at half filling, spin dynamics becomes \emph{superdiffusive},
$\langle m^{a}(x,t)m^{a}(0,0) \rangle^{c}_{\beta,h=0}\sim t^{-1/z}$, with dynamical exponent $z=3/2$, precisely as in Sec.~\ref{isoXXZ}.
Regarding behavior at low temperatures, there are two regimes with markedly different behavior to be distinguished.
In the experimentally relevant regime $h\ll T\ll \varDelta$ one has \cite{NMKI19}
\begin{equation}
    D = \sum_{s\geq 0}D_{s} \sim \frac{e^{\varDelta/T}}{3\varDelta|h|}
    + \mathcal{O}(h^{0}), 
\label{eqn:spin_diffusion_lowT}
\end{equation}
while the corresponding d.c. conductivity reads
$\sigma(T,h)=\chi(T,h)D(T,h)=\kappa(T)/|h|+\mathcal{O}(h^{0})$, with $\kappa(T)\sim T^{-1/2}$. Notice that the semi-classical result, cf. Eq.~\eqref{eqn:semiclassical_diffusion}, coincides precisely with the first term $D_{0}$ in an infinite sum over quasiparticle species in Eq.~\eqref{eqn:spin_diffusion_lowT}. Terms with $s\geq 1$ are contributions of spin waves which cannot be safely neglected; indeed, in approaching $h\to 0$ they become amplified and cause $D_{\rm spin}$ to diverge! The NMR relaxation time of (longitudinal) spin correlators behaves as $1/T_{1} \sim e^{3\beta \varDelta/2}/\sqrt{h}$ at small $h$, consistent with experimental findings~\cite{Takigawa}. While this result somewhat surprisingly matches the semi-classical prediction, its particular temperature dependence has a different origin that can be traced to internal magnon degrees of freedom. In contrast, for $T\ll \varDelta$ but with $h/T\gg 1$, contributions due to magnons ($s\geq 1$) become suppressed and (to leading order) the GHD predictions agree with the semi-classical theory.

\medskip

\paragraph*{Including the topological term.}

As already mentioned, for $\varTheta = \pi$ no dynamical transmutation takes place and the spectrum of excitations of the $O(3)$ NLSM retains two gapless modes from its classical counterpart; they are the left and right movers which transform as an $SU(2)$ doublet and are assigned labels $a\in \pm$, with chiral bare dispersion relations $e_{\pm}(\theta)=(\Lambda/2)e^{\pm \theta}$. Integrability of the $O(3)$ model survives the addition of the topological term \cite{ZZ92}.
In contrast to CFTs, the left and right movers interact among themselves and with one another, and the scattering matrix of the internal spin degrees of freedom can be diagonalized with aid of the nested Bethe Ansatz~\cite{ZZ92}. The Bethe--Yang and TBA equations can be found in Refs. \onlinecite{ZZ92,Konik2003,NMKI19}, while the low-temperature expansion is given in Ref. \onlinecite{NMKI19}.

Intuition may perhaps suggest that presence or absence of the spectral gap will have an impact on transport properties of the model at low temperatures. This is however not the case (at least not on a qualitative level) and the phenomenology of the non-topological ($\varTheta=0$) model persists also in the topological version of the $O(3)$ NLSM. Specifically, divergence of the spin diffusion constant as $\sim 1/h$ in approaching half filling $h=0$ can once again be attributed to the particular anomalous thermal dressing of interacting bound states.

\medskip

\paragraph*{Irrelevant operators and integrability breaking terms.}
To conclude this section, some general remarks are in order on the applicability of effective field theories in condensed matter physics. For understanding the low-energy physics of generic condensed matter systems, effective field theory remains an invaluable tool; the problem of spin transport in Haldane chains is no exception. However, the derivation of the $O(3)$ NLSM as an effective action for low-energy spin chains relies on a number of non-trivial assumptions; beyond the usual neglect of irrelevant terms that is strictly valid only at $T=0$, one must also assume~\cite{Haldane83,auerbach2012interacting} a semiclassical regime of large $S$. Thus there is no guarantee that for small, finite $S$ (and especially for $S=1$) the phenomenology of the large-$S$ effective theory remains qualitatively unaltered. While numerical studies (and experiments) confirm the presence of a gap for spin chains with small integer $S$, the regime of validity for modelling more complicated observables with the effective field theory  is \emph{a priori} unclear.


Transport coefficients can be especially fragile since they concern the long-time behavior of correlation functions. See, for instance, a recent numerical study \cite{PhysRevB.100.094411} where certain discrepancies (e.g., the form of the NMR relaxation rate) have been observed. Incorporating additional realistic effects that are present in spin-chain compounds, such as interchain couplings and on-site interaction anisotropy, is a separate question but might also affect thermodynamic properties. More importantly, we have tacitly assumed that the effective theory continues to provide a reasonable approximation even at non-zero (albeit low) temperatures. However, the effective theory is completely integrable and its quasiparticle excitations are infinitely long lived. In view of standard RG arguments, one expects dangerously irrelevant operators to cause finite quasiparticle lifetimes at $T>0$ (see e.g. Ref. \onlinecite{VM_review}), and the central question in this regard is to reliably estimate the timescales on which RG irrelevant terms and other integrability-breaking perturbations have an appreciable effect on relaxation. Recent progress on modelling various scenarios of integrability breaking~\cite{PhysRevB.101.180302,lopez2020hydrodynamics,durnin2020non,hutsalyuk2020integrability,bouchoule2020effect} may shed light on this question.

\subsection{Persistence of anomalous behavior in nonintegrable isotropic chains}\label{sec:persistence}

We now turn to the (largely still open) problem of transport in spin chains where integrability is weakly broken. In the absence of strict integrability, one might expect conventional hydrodynamics to govern the behavior at asymptotically late times, with the crossover timescale set by some form of Golden Rule calculation~\cite{essler2014quench, bertini2015prethermalization, PhysRevX.8.021030, PhysRevX.9.021027, PhysRevB.101.180302, PhysRevB.102.161110, durnin2020non, leblond2019entanglement, lopez2020hydrodynamics, bouchoule2020effect, hutsalyuk2020integrability}. On timescales short compared with this crossover, the dynamics will follow the integrable behavior; on timescales that are much longer, one might naively expect integrability to be irrelevant to the dynamics. However, the analysis of Refs.~\onlinecite{PhysRevB.101.180302, PhysRevB.102.161110} is also consistent with the idea that there might be a broad spectrum of relaxation timescales, because different quasiparticles couple differently to the integrability-breaking perturbation. If some of these relaxation timescales are slow enough, anomalous transport can persist away from the integrable limit. A separate possibility is that the conventional hydrodynamics of systems with nonabelian symmetries might be rich enough to permit anomalous transport; however, recent field-theoretical studies do not seem to support this possibility~\cite{glorioso2020hydrodynamics}. 

One might suspect that such broad distributions of timescales occur for the isotropic Heisenberg model subject to \emph{isotropic} integrability-breaking perturbations, for a few distinct reasons. First, there is now considerable numerical evidence~\cite{NMKI19, richter1, richter2, NMKI20, NGVW21} that spin transport even in isotropic spin chains quite far from integrability is not diffusive on the accessible timescales (but see also Refs.~\onlinecite{karadamoglou2004diffusive, steinigeweg2016heat, richter2019magnetization}). Depending on the perturbation, one either sees KPZ scaling persist to the numerically accessible times or sees a large temporal regime with length-time scaling $x^2 \sim t \log t$. By contrast, energy diffusion can be seen on much shorter timescales, as can spin diffusion in the presence of perturbations that break the nonabelian symmetry~\cite{NGVW21}. Taken together, these results strongly suggest that there is at least a strong quantitative suppression of the relaxation rate in the presence of the nonabelian symmetry. 

Perturbation theory supports this conclusion~\cite{NGVW21}. The perturbative calculation for the Heisenberg chain is simplest for the case of spatially correlated noise, with a large correlation length, coupling to the energy density. In this limit, one can show~\cite{DeNardis_SciPost} that the coupling of the perturbation to an $s$-string scales as $1/s$ (i.e., as the dressed energy of the $s$-string). One would expect the lifetime of an $s$-string to scale as $s^2$ by the Golden Rule. Plugging this scaling for the string lifetime into the Kubo formula~\eqref{kuboheis} gives a logarithmically diverging diffusion constant. 

This phenomenon can also be understood somewhat more generally, as follows. The giant quasiparticles of the Heisenberg model are made up of Goldstone modes, and look locally like vacuum rotations. Thus, isotropic local integrability-breaking perturbations cannot effectively couple to them: the matrix element for scattering an $s$-string must generally include factors of $1/s$. Thus, within low-order perturbation theory in the integrability-breaking parameter, one will generically find that the giant solitons persist in having long lifetimes (precisely how long is sensitive to how the density of final states scales with $s$, which is nonuniversal). Finally, even if the decay rates vanish sufficiently rapidly in perturbation theory, higher-order or nonperturbative processes might restore diffusion. At present, no complete theoretical analysis of this problem exists, and it remains one of the fundamental open questions in this area.

\section{Undular diffusion in nonabelian systems}\label{undular}

Now that we have navigated our way through an array of anomalous transport laws, the question arises of whether there are any possibilities for
anomalous transport in clean, short-ranged, Hamiltonian quantum chains that we have left unexplored. In this section we review yet another distinctive type of anomalous dynamics in systems with nonbelian symmetries, which is due to discernible consequences of symmetry breaking \emph{at non-zero temperature}.

Thus far we have focused on exactly (or approximately) integrable models. We have explained how nonabelian symmetries play a surprising role in enabling anomalous transport, allowing for the possibility of anomalous spin or charge transport in integrable lattice systems, that can be precisely quantified through divergences in the associated diffusion constants. Introducing chemical potentials (coupling to the Cartan charges) will in general restore normal diffusion, with its associated Gaussian fluctuations and dynamical exponent $z=2$. As we shall see shortly, such explicit breaking of a global Lie symmetry has a profound effect on transport properties, regardless of microscopic integrability and the temperatures studies. In particular, signatures of symmetry breaking persist to \emph{infinite} temperature. As demonstrated in a recent study \cite{undular}, two-point dynamical correlation functions amongst different components of $G$-invariant Noether charges show two different types of behavior: while longitudinal correlators lying in the unbroken symmetry sector exhibit normal diffusive behavior (with Gaussian asymptotic scaling profiles) as expected, the transverse (i.e. symmetry-broken) sector reveals another type of anomalous transport law characterized by (i) diffusive dynamical exponent $z=2$ and (ii) oscillatory stationary scaling profiles---hence the name undular diffusion~\cite{undular}. 

An intuitive interpretation of undular diffusion is that it arises from an interplay of normal diffusive transport and the dispersive physics of ``Type-II'' Goldstone modes. Both types of dynamics have the same dynamical exponent $z=2$, and thus at the level of linear response they must be linearly superposed to yield a complex diffusion constant. This reasoning has been confirmed by a systematic derivation of fluctuating hydrodynamics for systems with nonabelian global symmetry, within the language of effective field theory~\cite{glorioso2020hydrodynamics}. This phenomenon is counterintuitive in the sense that Goldstone physics is not usually believed to hold sway at high temperatures; that Goldstone modes \emph{can} be important for high-temperature transport is nevertheless consistent with the results described in Secs. \ref{isoXXZ} and \ref{SUN}. An intriguing question that immediately arises is whether undular diffusion can be understood in terms of a microscopic random walk, since a complex diffusion constant seems to invalidate the usual central-limit-theorem arguments for normal transport (see Sec. \ref{griffiths}).

\medskip

\paragraph*{Goldstone modes at finite density.}
The phenomenon of undular diffusion is a generic feature of Hamiltonian dynamical systems that exhibit symmetry under the action of a nonabelian Lie group.
For definiteness, we confine ourselves to nonrelativistic systems and defer comments about Lorentz invariant theories to the end of this section. Importantly, we do not demand integrability or impose any other constraints on the microscopic evolution law.

Undular diffusion is, in a nutshell, a manifestation of soft (Goldstone) modes at the level of equilibrium states, namely states with a non-zero density of background charge. In what follows, we explain the underlying mechanism and recapitulate the main conclusions of a recent paper Ref.~\onlinecite{undular}, where the phenomenon is introduced and examined at the level of classical nonlinear sigma models.

Nonlinear sigma models are generally understood as field theories whose target spaces are compact coset manifolds $G/H$. Recall that coset spaces are equivalence classes with respect to right multiplication by element of a stabilizer subgroup $H$ (also known as the little group), namely $G/H = \{g\sim gh;g\in G,h\in H\}$. Nonlinear sigma models thus exhibit global invariance under the action of the group $G$ (assumed to be simple) and local gauge invariance under the action of the subgroup $H \subset G$. The class of non-relativistic coset sigma models considered in Ref.~\onlinecite{undular} can be viewed as ferromagnets with a $G/H$-valued order parameter.

\subsection{Isotropic Landau--Lifshitz model and complex diffusion constant}

The physics behind undular diffusion is best explained using, as a basic example, the isotropic Landau--Lifshitz magnet, a sigma model whose target space is simply a two-sphere, that is $S^{2} \cong SU(2)/U(1)$, with nonabelian isometry group $G=SU(2)$ and local gauge group $H=U(1)$. The two-sphere is most commonly parametrized by a three-component spin field ${\bf S}$ subjected to unit normalization ${\bf S}\cdot {\bf S} = 1$. The latter can be thought of as a source of nonlinearity which ultimately renders time-evolution non-trivial. By virtue of global invariance under $G$, the (ferromagnetic) vacuum $\boldsymbol{\Omega}$ is continuously degenerate and can point anywhere on the 2-sphere; spontaneous breaking of symmetry amounts to choosing a particular orientation. Here we make the conventional choice and adopt (with no loss of generality) the vacuum polarization to be aligned with the positive $z$-axis. The associated stability subgroup $H$, which generates rotations about this polarization axis, leaves $\boldsymbol{\Omega}$ intact, i.e. $h \boldsymbol{\Omega}h^{-1}=\boldsymbol{\Omega}$ for $h\in H$.
The nonlinearity constraint can be approximately relaxed only in close proximity to the vacuum. One then finds linear fluctuations (i.e., excitations of a ferromagnetic background) that consist of transverse Goldstone modes -- in low-temperature ferromagnets, these are the usual quadratically dispersing (i.e. type-II) magnons. Recall that classical magnons (in the presence of an external magnetic field of strength $b$) are governed by a linear PDE $\big(\ii \partial_{t} \pm (\partial_{xx} - b)\big)S^{\pm}(x,t) = 0$, with $S^{\pm}(x,t) = S^{x} \pm \ii S^{y}$. This represents a uniform rotation (precession) in the plane perpendicular to a reference polarization axis. The associated Green's function in Fourier space is given by $\mathcal{G}_{\pm}(k)=e^{\mp i \omega(k)t}$, with $\omega(k) = \ii(k^{2}+b)$.

\medskip

At this juncture, we are interested in the fate of Goldstone modes in thermal equilibrium at a non-zero density of background charges,i.e. not near the ferromagnetic vacuum but at higher energies. To enforce non-zero expectation values for the Cartan charges (associated with Cartan generators spanning the maximal torus $\mathfrak{t}\subset \mathfrak{g}$ with ${\rm dim}(\mathfrak{t})=r$), we couple them to the corresponding chemical potentials ${\bf \mu}$ and introduce an equilibrium stationary measure $\varrho_{\boldsymbol{\mu}}$, representing an integration measure on a compact coset manifold $G/H$ (defined below). More importantly, introducing $\boldsymbol{\mu}$ breaks $G$-invariance of equilibrium correlation functions and (assuming generic values of chemical potentials) leaves behind only the residual symmetry of the maximal torus subgroup $T=U(1)^{r}\subset H$. This reduction splits $\mathfrak{g}$ into a longitudinal sector, generated by $\mathfrak{t}$, and a transverse sector spanned by the complement of $\mathfrak{t}$, as we detail below. While in integrable models, the long-time behavior of dynamical two-point functions is characterized by (generically) finite Drude weights, in non-integrable lattice systems one expects to find normal diffusive transport on general grounds (cf. the discussion in Sec.~\ref{reminder}).

This is not quite what is observed in the transverse sector. While charge dynamics in the symmetry-broken sector exhibits diffusion-like spreading with a normal dynamical exponent $z=2$, the stationary Gaussian scaling profiles are characterized by a \emph{complex-valued} ``diffusion constant''~\cite{undular}. This law is thus to be distinguished from normal diffusion (i.e. the Fourier/Fick law), which is associated with a single real constant proportional to the variance of a microscopic Brownian motion (cf. Sec. \ref{griffiths}). In simple terms, Goldstone modes do not disappear at high temperatures; they merely acquire an extra diffusive component.

\bigskip

\paragraph*{Dynamics of transverse modes.}

To illustrate the implications of explicit symmetry breaking in equilibrium at non-zero temperature, we return to our basic example of the isotropic Landau--Lifshitz ferromagnet. In terms of the spin field ${\bf S} \equiv (S^{x},S^{y},S^{z}) \in S^{2}$, the model is governed by a nonlinear equation of motion
\begin{equation}
    {\bf S}_{t} = {\bf S}\times {\bf S}_{xx} + {\bf S}\times {\bf B},
    \label{eqn:LL_EOM}
\end{equation}
where we have included a constant longitudinal external field of magnitude $b$ pointing along the $z$-axis, ${\bf B}=b\,{\bf e}_{z}$.

\begin{figure}[tb]
\begin{center}
\includegraphics[width = 0.64\textwidth]{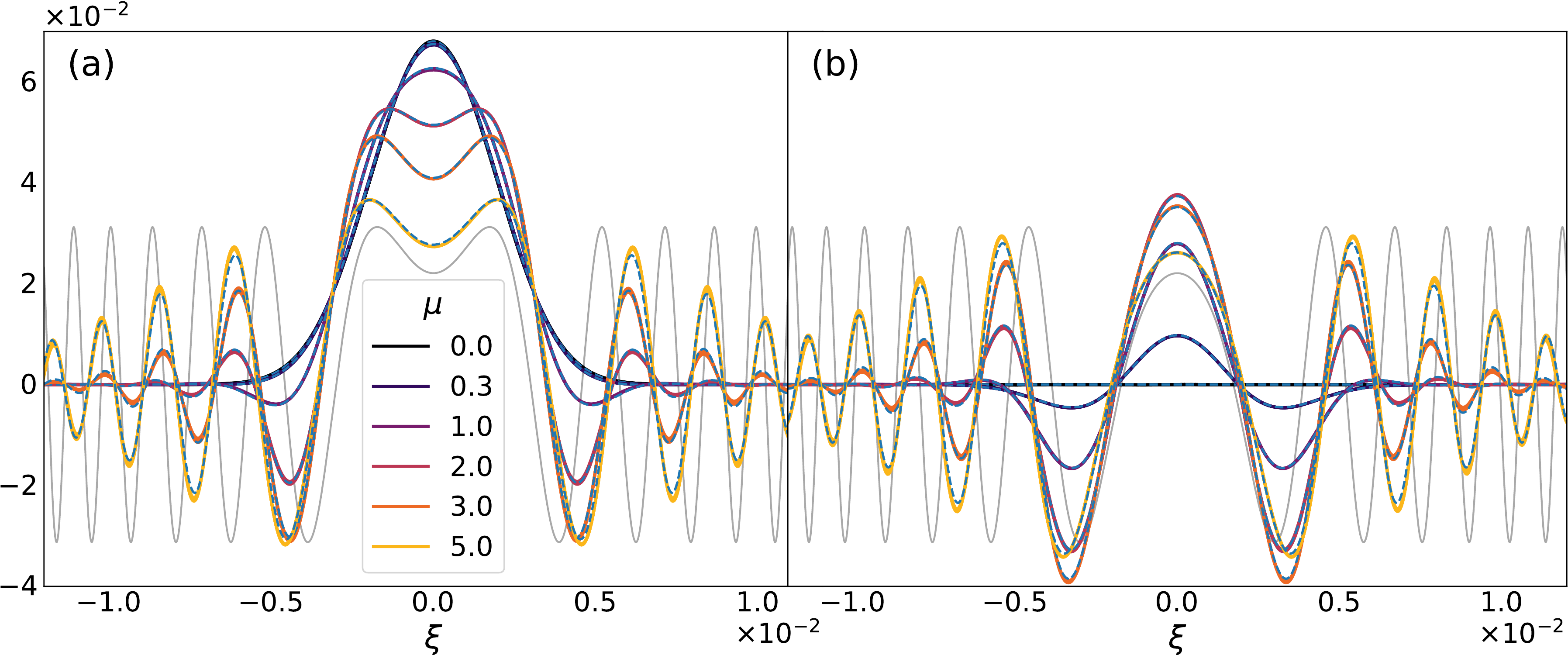}
\includegraphics[width = 0.35\textwidth]{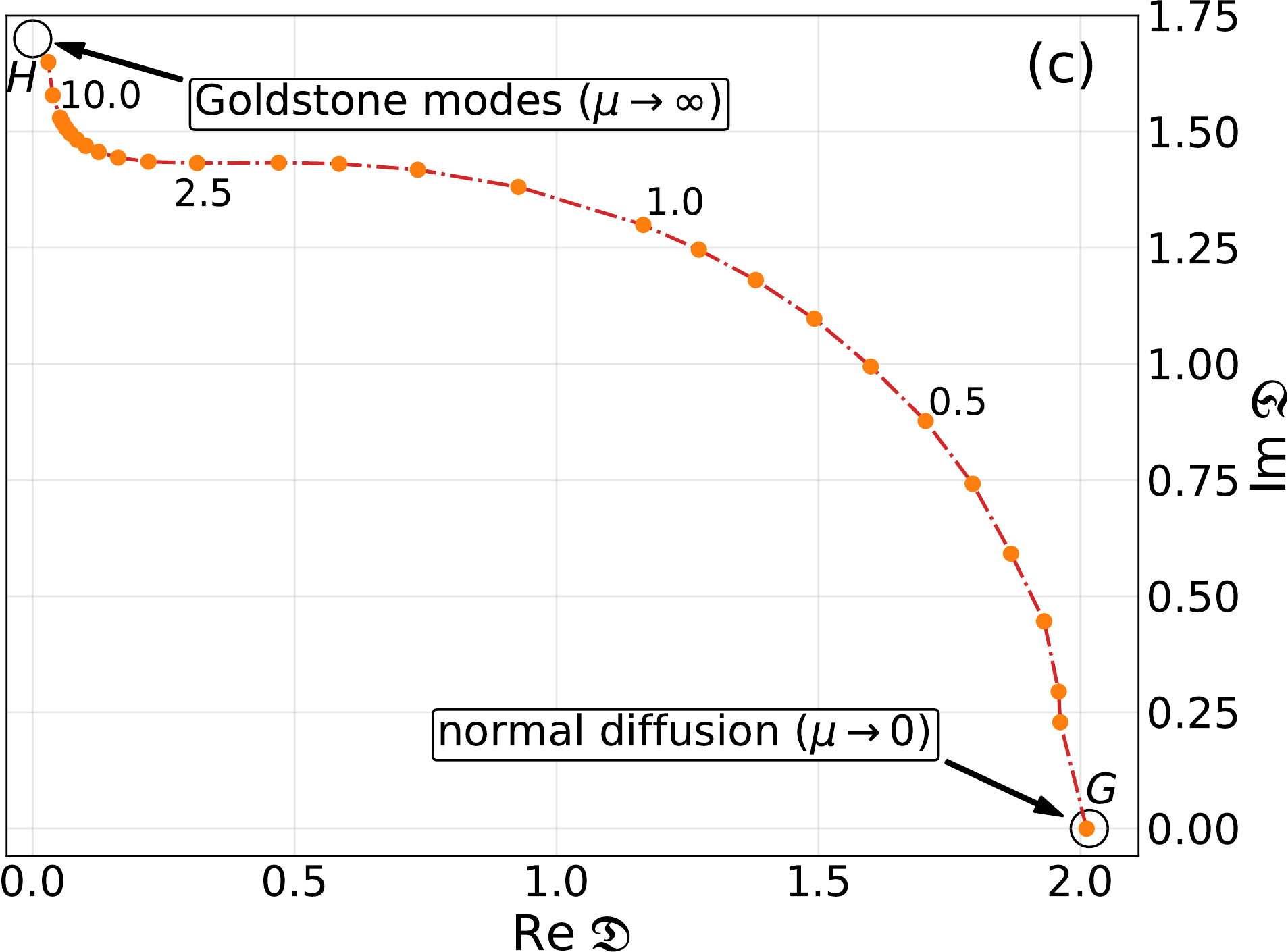}
\caption{Left: Real (a) and imaginary (b) part of asymptotic stationary profiles of transverse dynamical correlation functions $t^{1/2}\langle S^{+}(x,t)S^{-}(0,0) \rangle^{c}_{\mu}/2$ as functions of scaling variables $\xi=x/\sqrt{t}$ and chemical potential $\mu$ of an invariant stationary measure, shown for a \emph{non-integrable} lattice discretization of the Landau--Lifshitz field theory on $S^{2}$; grey lines indicate the linear dynamics of Goldstone modes (in arbitrary units) whereas dashed lines shown the best fit with a complex diffusion constant $\mathfrak{D}(\mu)$.
Right: Real and imaginary parts of $\mathfrak{D}(\mu)$ as functions of $\mu$, interpolating between the diffusive and Goldstone modes.
Both panels reproduced from \v{Z}. Krajnik et al., \textit{Phys. Rev. Lett.} {\bf 125}, 240607 (2020).}
\label{undular_plot}
\end{center}
\end{figure}

The task is to infer the structure of the full three-dimensional tensor of dynamical correlation functions $\langle S^{a}(x,t)S^{b}(0,0) \rangle^{c}_{\mu}$ ($a,b\in \{x,y,z\}$), evaluated at non-zero chemical potential $\mu$ coupling to $\int \dd x\, S^{z}$ (below we shall consider a more general equilibrium measure by including the temperature). By setting $\mu= 0$, the invariant measure corresponds to the uniform $SU(2)$-invariant Liouville measure on the phase space. For finite $\mu$ on the other hand, the symmetry is lowered to $U(1)$ rotations about the $z$-axis, and the density of the corresponding stationary measure takes the form $\varrho_{\mu}=\mathcal{Z}^{-1}_{\mu}\exp{[\mu(1-2S^{z})]}$, with partition function $\mathcal{Z}_{\mu} = \pi \sinh{(\mu)}/\mu$.
A minor comment is in order regarding our choice of illustrative example, which is non-generic in the sense that Eq.~\eqref{eqn:LL_EOM} is a completely integrable PDE~\cite{Takhtajan1977,Sklyanin1979}. It is well-understood that one consequence of integrability is that longitudinal structure factors ($a=b=z$) in polarized states at $\mu \neq 0$ will undergo ballistic spreading (characterized by non-zero Drude weights) with a finite subleading diffusive correction. Upon breaking integrability, either by lattice effects or by adjoining higher-order terms in derivatives, one nonetheless expects to recover normal spin diffusion. Our primary concern here is not the longitudinal sector but instead dynamical correlation functions in the symmetry-broken (transverse) sector, namely in the $xy$-plane perpendicular to the symmetry axis. To this end, it is convenient to introduce complex fields $S^{\pm}=S^{x}\pm \ii S^{y}$. As a direct consequence of $U(1)$ symmetry, it is clear that both $\langle S^{\pm}(x,t)S^{\pm}(0,0) \rangle_{\mu}$ and $\langle S^{z}(x,t)S^{\pm}(0,0)\rangle_{\mu}$ vanish~\cite{undular}. This leaves two non-trivial correlators, namely $\langle S^{\pm}(x,t)S^{\mp}(0,0) \rangle^{c}_{\mu}$. The corresponding stationary profiles, collapsed to the diffusive scaling variable $\xi=x/\sqrt{t}$, are illustrated in Fig.~\ref{undular_plot}. The observed oscillation patterns are governed by a single hydrodynamic $z=2$ mode with a \emph{complex-valued} ``diffusion constant'' $\mathfrak{D}$, that interpolates between normal diffusion with $\mathfrak{D}\in \mathbb{R}$ in the absence of a chemical potential ($\langle S^{z} \rangle = 0$) and a Goldstone mode with $\mathfrak{D}\in \ii \mathbb{R}$ in the limit of large polarizations $\mu \to \infty$.

\bigskip

\subsection{Coset sigma models of higher-rank symmetry}\label{cosetLL}

Goldstone modes are a generic feature in systems with spontaneously broken symmetries. We have so far outlined the basics of undular diffusion in the simplest case of $SU(2)$ symmetry and a single magnon mode. We now focus on a more general scenario where the symmetry Lie group $G$ has higher rank so that symmetry breaking patterns $G\to H$ give rise to a larger number of Goldstone modes. According to the Goldstone theorem \cite{NielsenChadha76,WM12,Hidaka13}, in the absence of linear (type-I) Goldstone modes, the total number of type-II Goldstone modes is given by one half the number of symmetry-broken generators, that is $n_{G}=\tfrac{1}{2}({\rm dim}(G)-{\rm dim}(H))$. We note that Goldstone modes are conventionally studied in low-temperature settings, where they can essentially be modelled as non-interacting waves and as such cannot exert forces on one another. Here we are instead concerned with genuinely \emph{nonlinear} equations of motion, for which different field components interact among themselves in a non-trivial fashion. In this setting, the key question is whether coupling of different transverse modes can result in a more intricate structure of dynamical correlations than normal diffusion.

This question was also addressed in Ref.~\onlinecite{undular}, for a specific class of \emph{nonrelativistic} sigma models with complex projective target spaces $\mathcal{M}_{n}\equiv \mathbb{CP}^{n}\cong SU(n+1)/[SU(n)\times U(1)]$. The latter represent K\"{a}hler manifolds of complex dimension $n$, and it is thus natural to employ a description in terms of $n$ complex (affine) coordinates $\mathbb{CP}^{n}$, ${\bf z}\equiv (z_{1},z_{2},\ldots,z_{n})^{\rm T}$. The Lagrangian density can be split in two parts,
\begin{equation}
    \mathcal{L} = \mathcal{L}_{\rm WZ} - \mathcal{L}_{\mathcal{M}_{n}},
    \label{eqn:CPn_LL}
\end{equation}
with the geometric Wess--Zumino term,
\begin{equation}
    \mathcal{L}_{\rm WZ} = \ii (1+{\bf z}^{\dagger}{\bf z})^{-1}({\bf z}^{\dagger}{\bf z}_{t}-{\bf z}_{t}{\bf z})
\end{equation}
and the kinetic term
\begin{equation}
    \mathcal{L}_{\mathcal{M}_{n}} = \sum_{a,b=1}^{n}\eta_{ab}\overline{z}^{a}_{x}z^{b}_{x},\qquad
    \eta_{ab}=\frac{(1+{\bf z}^{\dagger}{\bf z})\delta_{ab} - \overline{z}_{a}z_{b}}{(1+{\bf z}^{\dagger}{\bf z})^{2}},
\end{equation}
expressed in terms of the standard Fubini--Study metric tensor $\eta$. We have only kept the leading-order (i.e. quadratic) term in the gradient expansion; one could in principle add additional higher-order terms in spatial derivatives but this is of no significance for what follows. One disadvantage of working directly with Eqs.~\eqref{eqn:CPn_LL} is that like the Landau--Lifshitz magnet, they too represent completely integrable nonlinear PDEs. Indeed, they emerge as an effective low-energy description of certain integrable $SU(n)$-invariant quantum ferromagnetic chains~\cite{MatrixModels}. To avoid unwanted artifacts due to integrability, it is safer to work with non-integrable discretizations of these equations~\cite{undular}.

\medskip

\paragraph*{Dynamical correlation funtions and structure factors at finite temperature.}
In order to describe the complete structure of dynamical structure factors, we need to first identify the longitudinal and transverse fields. To this end, it proves convenient to switch to the Hamiltonian formulation,
\begin{eqnarray}
    H = \frac{1}{4}\int \dd x \, {\rm Tr}(M^{2}_{x}) + {\rm Tr}(B\,M),
    \qquad M \in \mathcal{M}_{n},
    \label{eqn:LL_Hamiltonian}
\end{eqnarray}
and further exploit the fact that coset manifolds $G/H$ can be realized as $G$-orbits of the stabilizer subgroup $H$, by restricting the matrices $M\in \mathcal{M}_{n}$ to the compact submanifold of $SU(n)$ singled out by the constraint $M^{2}= 1$. The higher-rank Landau--Lifshitz equations generated by Eqs.~\eqref{eqn:LL_Hamiltonian} take a compact universal form
\begin{equation}
    M_{t} = \frac{1}{2\ii}[M,M_{xx}] + \ii[B,M].
\end{equation}
To make full use of the underlying algebraic structure, one next decomposes matrices $M \in \mathbb{CP}^{n}$ as
\begin{equation}
    M = \frac{n-1}{n+1}\mathbf{1} + \sum_{j=1}^{n}\phi^{j}{\rm H}^{j}
    + \sum_{\pm \alpha}\phi^{\pm \alpha}{\rm X}^{\mp \alpha},
\end{equation}
in terms of Cartan and Weyl fields,
\begin{equation}
    \phi^{i} = \sum_{i,j=1}^{n}(\kappa^{-1})_{ij}{\rm Tr}(M\,{\rm H}^{j}),\qquad
    \phi^{\pm \alpha} = {\rm Tr}(M\,{\rm X}^{\pm \alpha}),
\end{equation}
respectively. Here we have used the Cartan--Weyl basis of $\mathfrak{g}=\mathfrak{su}(n+1)$, where ${\rm H}^{i}$ ($i \in \Delta_{0}\in \{1,\ldots,n\}$) constitute the maximal subset of commuting generators spanning the Cartan subalgebra, $[{\rm H}^{i},{\rm H}^{j}]=0$, whereas $X^{\pm \alpha}$ label the Weyl generators associated with one-dimensional complex root spaces spanned by root vectors $\pm \alpha \in \Delta_{\pm}$. The root lattice of $\mathfrak{g}$ is given by $\Delta = \Delta_{+}\cup \Delta_{-}$ of $\mathfrak{g}$. For reference, the Cartan--Weyl commutation relations read
\begin{equation}
    [{\rm H}^{i},{\rm X}^{\pm \alpha}] = \alpha^{i}{\rm X}^{\pm \alpha},\quad
    [{\rm X}^{\alpha},{\rm X}^{-\alpha}] = \sum_{i,j=1}^{n}\alpha^{i}(\kappa^{-1})_{ij}{\rm H}^{j},\quad
    [{\rm X}^{\alpha},{\rm X}^{\gamma \neq \alpha}] = C_{\alpha \gamma}{\rm X}^{\alpha +\gamma},
\end{equation}
where we have adopted the non-diagonal Killing metric $\kappa_{ij} = {\rm Tr}({\rm H}^{i}{\rm H}^{j})$ and normalization ${\rm Tr}({\rm X}^{\alpha}{\rm X}^{-\alpha})=1$ for the Weyl generators.

\medskip

Next we introduce the grand-canonical Gibbs measure. Since complex projective spaces $\mathbb{CP}^{n}$ are compact manifolds, they can be equipped with a stationary $G$-invariant measure. As already announced above, we shall switch to lattice Hamiltonians $H_{\rm lattice}$ which describe nonintegrable lattice discretizations of Eqs.~\eqref{eqn:CPn_LL}, see Ref.~\onlinecite{undular} for details. The motivation for doing so is, apart from breaking integrability, two-fold: on the one hand, the use of functional integral techniques can be entirely avoided, while on the other hand the benefit is that final results are more general. For a lattice of length $L$, the phase space consists of an $L$-fold Cartesian product of $\mathcal{M}_{n}$, and the equilibrium expectation value of any observable $\mathcal{O}$ (at inverse temperature $\beta$ and $U(1)$ chemical potentials $\boldsymbol{\mu}$) is given by
\begin{equation}
    \langle \mathcal{O} \rangle_{\beta,\boldsymbol{\mu}} = \frac{1}{\mathcal{Z}_{\beta,\boldsymbol{\mu}}}\int_{\mathcal{M}^{\times L}_{n}}
    \prod_{\ell=1}^{L}\dd \Omega^{(n)}_{\ell}\varrho_{\boldsymbol{\mu},\ell}\,
    e^{-\beta H_{\rm lattice}}\mathcal{O},
    \label{eqn:CPn_exp_value}
\end{equation}
with normalization $\mathcal{Z}_{\beta,\boldsymbol{\mu}}$ in Eq.~\eqref{eqn:CPn_exp_value} corresponding to the grand-canonical partition function. In the above expression $\dd \Omega^{(n)}_{\ell}$ denotes the $\mathbb{CP}^{n}$ volume element of the local phase-space attached to the lattice site $\ell$, and the corresponding equivariant (Duisermaat--Heckmann) measure $\varrho^{(n)}_{\boldsymbol{\mu}}\dd \Omega^{(n)}$ has density
\begin{equation}
    \varrho^{(n)}_{\boldsymbol{\mu}}(M) = e^{{\rm Tr}(H^{T}_{\boldsymbol{\mu}}M)}.
\end{equation}
prescribed by the ``torus Hamiltonian''
$H^{\rm T}_{\boldsymbol{\mu}}=\sum_{i=i}^{n}\mu_{i}{\rm H}^{i}=
-\tfrac{1}{2}{\rm diag}(\mu_{0},\mu_{1},\ldots,\mu_{n})$, where the parameters $\{\mu_{i}\}$ with $\sum_{i=0}^{n} \mu_i = 0$ define $n$ distinct chemical potentials. Specifically, in the $\beta \to 0$ limit, dependence on $H_{\rm lattice}$ drops out and the partition sum completely factorizes
\begin{equation}
    \mathcal{Z}^{(n)}_{\boldsymbol{\mu}}
    = \int_{\mathcal{M}_{n}}\dd \Omega^{(n)}e^{{\rm Tr}(H^{T}_{\boldsymbol{\mu}}M)}.
\end{equation}

The following two general statements have been proven in Ref.~\onlinecite{undular}:
\begin{itemize}
    \item {\tt (neutrality rule)} All \emph{non-neutral} static $N$-point correlation functions evaluated in grand-canonical Gibbs states vanish,
    \begin{equation}
        \sum_{j \in \{1,\ldots,N\};j \notin \Delta_{0}}^{N}\sigma_{j} \neq {\bf 0}
        \qquad \Longrightarrow \qquad \left\langle \prod_{j=1}^{N}
        \phi^{\sigma_{j}}_{\ell_{j}} \right\rangle^{c}_{\beta,\boldsymbol{\mu}} = 0,
    \end{equation}
    where vector indices $\sigma_{j}$ take values in the root lattice $\Delta$. 
    \item {\tt (imaginary part of intersectoral correlators)} Assuming that Hamiltonian dynamics is invariant under spatial reflection, $\ell \mapsto L-\ell+1$, the imaginary components of all two-point static correlation functions corresponding to conjugate pairs of Weyl fields vanish
    \begin{equation}
        {\rm Im}\left\langle \phi^{\pm \alpha}_{\ell_{1}}\phi^{\mp \alpha}_{\ell_{2}} \right\rangle^{c}_{\beta,\boldsymbol{\mu}} = 0.
    \end{equation}
\end{itemize}
The above statements are in fact valid for arbitrary $G$-invariant lattice Hamiltonians $H_{\rm lattice}$ which are invariant under spatial inversion~\cite{undular}. This can be seen by directly performing the phase-space integrals, which is most conveniently done
in the basis of $2n$ real canonical (Darboux) coordinates $\{p^{i},\varphi^{i}\}_{i=1}^{n}$ of $\mathbb{CP}^{n}$, in terms of which both the volume element, $\dd \Omega^{(n)}=2^{-n}\prod_{i=1}^{n}\dd p^{i}\dd \varphi^{i}$, and the stationary equivariant density, $\varrho^{(n)}_{\boldsymbol{\mu}}=\prod_{i=1}^{n}e^{\mu_{i}p^{i}}$, factorize.

The main result regarding \emph{dynamical} two-point correlation functions of $\phi$-fields is that all intersectoral correlators identically vanish
\begin{equation}
    \left\langle \phi^{\pm \alpha}_{\ell_{1}}\phi^{\gamma}_{\ell_{2}} \right\rangle^{c}_{\beta,\boldsymbol{\mu}} = 0 ,\qquad {\rm for} \quad \gamma \nparallel \alpha,
\end{equation}
a property which has been called \emph{dynamical decoupling}~\cite{undular}. This statement follows from the fact (i) at initial time the correlation amongst the Weyl field from different sectors are zero and (ii) that $G$-invariant time-evolution (expressed formally as a series of nested commutators) cannot produce non-trivial correlations as a corollary of the neutrality rule.

Finally, in the table below we list the complete formal structure of equilibrium dynamical two-point correlations functions in
\emph{non-integrable} non-relativistic Landau--Lifshitz field theories (including lattice discretizations thereof) with $\mathbb{CP}^{n}$ target spaces.

\begin{center}
  \begin{tabular}{| c | c |}
    \hline
    {\tt correlators} & {\tt transport} \\
    \hline \hline
    $\langle \phi^{i}(x,t)\phi^{j}(0,0) \rangle_{\boldsymbol{\mu}}$ & normal diffusion \\ \hline
    $\langle \phi^{\pm \alpha}(x,t)\phi^{\mp \alpha}(0,0) \rangle_{\boldsymbol{\mu}}$ & undular diffusion \\
    \hline
    $\langle \phi^{i}(x,t)\phi^{\pm \alpha}(0,0) \rangle_{\boldsymbol{\mu}}$ &  \\
    $\langle \phi^{\pm \alpha}(x,t)\phi^{\pm \alpha}(0,0) \rangle_{\boldsymbol{\mu}}$ & trivial \\
    $\langle \phi^{\pm \alpha}(x,t)\phi^{\gamma \nparallel \alpha}(0,0) \rangle_{\boldsymbol{\mu}}$ & \\
    \hline
  \end{tabular}
\end{center}

\bigskip

\paragraph*{Open directions.}
We conclude with some open questions. We have seen that undular diffusion crucially depends on the presence of type-II Goldstone modes, i.e. soft modes with dynamical exponent $z=2$. In contrast, classical \emph{relativistic} nonabelian systems possess only type-I Goldstone modes, which may suggest that undular diffusion is incompatible with Lorentz invariance. Although this has not yet been thoroughly examined, we wish to point out that in relativistic nonlinear sigma models (whose target spaces are either Lie groups $G$ or coset manifolds $G/H$) densities of the Noether charges are not the Goldstone modes themselves but rather fields associated with the generators of $G$ (cf. the example of the $O(3)$ NLSM in Sec.~\ref{NLSM}).

From a practical standpoint, it would be valuable to develop an efficient computational scheme to extract complex diffusion coefficients. Given the discussion in Sec. \ref{pseudovacuum}, it seems significant that even in integrable systems, the phenomenon of undular diffusion is beyond the present capabilities of GHD.

Lastly, it would be interesting to address whether the outlined algebraic structure of dynamical structure factors is a universal feature of sigma models beyond complex projective manifolds, and to extend the formalism presented above to other Lie group symmetries and coset spaces.

\section{Summary and outlook}\label{conclusion}

In this review we have summarized recent progress on understanding superdiffusive transport in spin chains. We now briefly review where the problem stands at present, and list what seem to us to be the most pressing open questions.  GHD offers a relatively clear understanding of linear-response transport in integrable XXZ spin chains away from the isotropic limit. In the easy-axis case, the phenomenon of diffusion at half filling is both qualitatively and quantitatively understood~\cite{GV19, NMKI19}, and so is the dynamic structure factor at general values of the magnetization~\cite{GVW19}. In the easy-plane regime, a variety of approaches have converged on the conclusion that the spin Drude weight really is a nowhere-continuous function of the anisotropy. The nature of the low-frequency response has also been calculated using a combination of GHD and arguments from spectral sum rules; this offers a very nontrivial consistency check on the fractal Drude weight, and also agrees with numerical simulations of the finite-time response. The prediction of a quasiparticle L\'evy flight has not yet been confirmed at the level of scaling functions, and requires higher-resolution numerical studies of the dynamic structure factor. 

For isotropic spin chains, the situation is less clear at present. At the level of scaling, the exponent $z = 3/2$ can be understood on rather simple and general grounds; moreover, a quantitative theory exists for the time-dependent diffusion constant~\cite{NGIV20}. In addition, the role of nonabelian symmetries---and the giant solitonic wavepackets composed of Goldstone modes---in giving rise to anomalous transport is also relatively clear~\cite{Vir20, NGIV20, ssd}. A nontrivial prediction of this theory, which seems to be borne out numerically, is that the $z = 3/2$ exponent should be superuniversal for integrable spin chains with short-range interaction that are invariant under the action of simple Lie groups~\cite{ssd}. 

Deriving the underlying scaling function, which is conjectured to be the KPZ scaling function in several models~\cite{Ljubotina19,PhysRevE.100.042116,Weiner2019,KP20,MatrixModels}, is more difficult. We emphasize that the emergence of $z=3/2$ dynamical exponents and their associated KPZ scaling functions is generic in chaotic one-dimensional systems, where it is understood in terms of an effective theory of noisy, coupled hydrodynamic modes~\cite{KPZ,PhysRevLett.108.180601,spohn2016fluctuating}. At this level of detail, the emergence of KPZ scaling functions in integrable spin chains with isotropic symmetry is understood~\cite{Vir20,NMKI20,Fava20}. What is lacking at present is a microscopic derivation of these scaling functions that leverages integrability. The gulf in difficulty between these two problems is roughly the difference between the original Kardar-Parisi-Zhang proposal~\cite{KPZ} in 1986 and the exact results obtained for the polynuclear growth model by Pr{\"a}hofer and Spohn in 2004~\cite{Prahofer2004}. Short of reproducing the Pr{\"a}hofer-Spohn calculation for an integrable spin chain, a desirable goal would be to obtain the KPZ scaling function within GHD. This seems to require a further technical advance, namely a quantitative prescription for including pseudovacuum fluctuations within GHD, which has proved elusive so far.



Much of the experimental evidence for anomalous transport comes from experiments with ultracold atomic gases. In these systems, it is more natural to consider the relaxation of far-from-equilibrium initial states than to compute linear response about thermal states. The relaxation of an initial spin helix, in particular, does not seem accessible at present within the framework presented here---despite the recent progress, which we reviewed here, on the simpler related problem of the relaxation of a domain wall. (Very heuristically, one can think of a long-wavelength helix as a widely-spaced array of domain walls.) In some ways, the difficulties here are analogous to those mentioned above: in particular, the lack of a satisfactory framework for dealing with pseudovacuum dynamics within GHD. The difficulties in the present context might also go deeper, however: it is not clear, e.g., on what timescale such a far-from-equilibrium state approaches a local generalized equilibrium state that is describable within GHD. (This issue is, of course, a very general one with the GHD framework applied far from equilibrium.) 

Finally, a question of great importance both conceptually and in practice is the fate of anomalous transport when integrability is weakly broken while the nonabelian symmetry is preserved. It would be fair to say that the situation here is extremely unclear. Numerically, there is evidence that the crossover to normal transport (if it occurs) is much slower than a naive dimensional argument would predict. Also there is numerical evidence for a logarithmic divergence of the diffusion constant (though arguably~\cite{glorioso2020hydrodynamics} a similar effect could arise from long-time tails). Theoretically, there have been many recent attempts to incorporate integrability-breaking perturbations into GHD~\footnote{These developments will be reviewed in a companion article by A. Bastianello, A. De Luca, and R. Vasseur.}; so far, however, we do not fully understand how to apply these theories to systems with anomalous behavior in the integrable limit. Qualitatively, it seems that the giant quasiparticles responsible for superdiffusion should approximately decouple from local integrability-breaking perturbations that preserve the underlying nonabelian symmetry. However, actually computing these decay rates seems to require one to sum over a large number of decay channels, and neither the matrix elements nor the kinematics regulating these decay processes is understood.

\begin{acknowledgments}
We are indebted to Utkarsh Agrawal, Vincenzo Alba, Sounak Biswas, Xiangyu Cao, Jacopo De Nardis, Benjamin Doyon, Michele Fava, Aaron Friedman, Oleksandr Gamayun, Paolo Glorioso, David Huse, Christoph Karrasch, Vedika Khemani, Michael Knap, Robert Konik, \v{Z}iga Krajnik, Javier Lopez-Piqueres, Marko Ljubotina, Marko Medenjak, Yuan Miao, Joel Moore, Vadim Oganesyan, Sid Parameswaran, Toma\v{z} Prosen, Tibor Rakovszky, Marcos Rigol, Subir Sachdev, Shivaji Sondhi, Herbert Spohn, Vipin Varma, Romain Vasseur, Curt Von Keyserlingk, Brayden Ware, Marko \v{Z}nidari\v{c}, and many others for helpful discussions and collaborations on the topics addressed here. S.G. acknowledges support from NSF DMR-1653271, E.I. is supported by the program P1-0402 of Slovenian Research Agency. 
\end{acknowledgments}

\section*{References}

\bibliography{SSD}

\end{document}